\documentclass{article}%
\usepackage{graphicx}
\usepackage{amsmath}
\usepackage{amsfonts}
\usepackage{amssymb}%
\newtheorem{theorem}{Theorem}
{}

\newtheorem{lemma}{Lemma}
{}

\newtheorem{remark}{Remark}

\newenvironment{proof}[1][Proof]{\textbf{#1.} }{\ \rule{0.5em}{0.5em}}

\begin{document}

\author{O. A. Veliev\\{\small \ Dept. of Math, Fen-Ed. Fak, Dogus University.,}\\{\small Acibadem, Kadikoy, Istanbul, Turkey,}\\{\small \ e-mail: oveliev@dogus.edu.tr}}
\title{\textbf{Perturbation Theory for the Multidimensional Schrodinger Operator with
a Periodic Potential}}
\date{}
\maketitle

\begin{abstract}
In this paper we obtain asymptotic formulas of arbitrary order for the Bloch
eigenvalue and the Bloch function of the periodic Schrodinger operator
$-\Delta+q(x),$ of arbitrary dimension, when corresponding quasimomentum lies
near a diffraction hyperplane. Moreover, we estimate the measure of the
isoenergetic surfaces in the high energy region.

Bisides, writing the asymptotic formulas for the Bloch eigenvalue and the
Bloch function, when corresponding quasimomentum lies far from the diffraction
hyperplanes, obtained in my previous papers in improved and enlarged form, we
obtain the complete perturbation theory for the multidimensional Schrodinger
operator with a periodic potential.

\end{abstract}

\section{Introduction}

\bigskip In this paper we consider the operator%
\begin{equation}
L(q(x))=-\Delta+q(x),\ x\in\mathbb{R}^{d},\ d\geq2
\end{equation}
with a periodic (relative to a lattice $\Omega$) potential $q(x)\in W_{2}%
^{s}(F),$ where

$s\geq s_{0}=\frac{3d-1}{2}(3^{d}+d+2)+\frac{1}{4}d3^{d}+d+6,$ $F\equiv
\mathbb{R}^{d}/\Omega$ is a fundamental domain of $\Omega.$ Without loss of
generality it can be assumed that the measure $\mu(F)$ of $F$ is $1$ and
$\int_{F}q(x)dx=0.$ Let $L_{t}(q(x))$ be the operator generated in $F$ by (1)
and the conditions:%
\begin{equation}
u(x+\omega)=e^{i(t,\omega)}u(x),\ \forall\omega\in\Omega,
\end{equation}
where $t\in F^{\star}\equiv\mathbb{R}^{d}/\Gamma$ and $\Gamma$ is the lattice
dual to $\Omega$, that is, $\Gamma$ is the set of all vectors $\gamma
\in\mathbb{R}^{d}$ satisfying $(\gamma,\omega)\in2\pi Z$ for all $\omega
\in\Omega.$ It is well-known that ( see [2]) the spectrum of the operator
$L_{t}(q(x))$ consists of the eigenvalues

$\Lambda_{1}(t)\leq\Lambda_{2}(t)\leq....$The function $\Lambda_{n}(t)$ is
called $n$th band function and its range $A_{n}=\left\{  \Lambda_{n}(t):t\in
F^{\ast}\right\}  $ is called the $n$th band of the spectrum $Spec(L)$ of $L$
and $Spec(L)=\cup_{n=1}^{\infty}A_{n}$. The eigenfunction $\Psi_{n,t}(x)$ of
$L_{t}(q(x))$ corresponding to the eigenvalue $\Lambda_{n}(t)$ is known as
Bloch functions. In the case $q(x)=0$ these eigenvalues and eigenfunctions are
$\mid\gamma+t\mid^{2}$ and $e^{i(\gamma+t,x)}$ for $\gamma\in\Gamma$.

This paper consists of 6 section. First section is the introduction, where we
describe briefly the scheme of this paper and discuss the related papers.

In papers [13-17] for the first time the eigenvalues $\left\vert
\gamma+t\right\vert ^{2}$, for big $\ \gamma\in\Gamma,$ were divided into two
groups: non-resonance ones and resonance ones and for the perturbations of
each group various asymptotic formulae were obtained. Let the potential $q(x)$
be a trigonometric polynomial%

\[
\sum_{\gamma\in Q}q_{\gamma}e^{i(\gamma,x)},
\]
where $q_{\gamma}=(q(x),e^{i(\gamma,x)})=\int_{F}q(x)e^{-i(\gamma_{1},x)}dx,$
and $Q=\{\gamma\in\Gamma:q_{\gamma}\neq0\}$ consists of a finite number of
vectors $\gamma$ from $\Gamma.$ Then the eigenvalue $\left\vert \gamma
+t\right\vert ^{2}$ \ is called a non-resonance eigenvalue if $\gamma+t$ does
not belong to any of the sets

$W_{b,\alpha_{1}}=\{x\in\mathbb{R}^{d}:\mid\mid x\mid^{2}-\mid x+b\mid^{2}%
\mid<\mid x\mid^{\alpha_{1}}\},$ that is, if $\gamma+t$ lies far from the
diffraction hyperplanes $D_{b}=\{x\in\mathbb{R}^{d}:\mid x\mid^{2}=\mid
x+b\mid^{2}\},$ where $\alpha_{1}\in(0,1),$ $b\in Q$ (see [15-17]). The idea
of the definition of the non-resonance eigenvalue $\left\vert \gamma
+t\right\vert ^{2}$ \ is the following. If \ $\gamma+t\notin W_{b,\alpha_{1}}$
then the influence of $q_{b}e^{i(b,x)}$ to the eigenvalue $\left\vert
\gamma+t\right\vert ^{2}$ is not significant. If $\gamma+t$ does not belong to
any of the sets $W_{b,\alpha_{1}}$ for $b\in Q$ then the influence of the
trigonometric polynomial $q(x)$ to the eigenvalue $\left\vert \gamma
+t\right\vert ^{2}$ is not significant. Therefore the corresponding eigenvalue
of the operator $L_{t}(q(x))$ is close to the eigenvalue $\mid\gamma+t\mid
^{2}$ of $L_{t}(0).$

If $q(x)\in W_{2}^{s}(F),$ then to describe the non-resonance and resonance
eigenvalues $\left\vert \gamma+t\right\vert ^{2}$ of the order of $\rho^{2}$ (
written as $\left\vert \gamma+t\right\vert ^{2}\sim\rho^{2}$) for big
parameter $\rho$ we write the potential $q(x)\in W_{2}^{s}(F)$ in the form
\begin{equation}
q(x)=\sum_{\gamma_{1}\in\Gamma(\rho^{\alpha})}q_{\gamma_{1}}e^{i(\gamma
_{1},x)}+O(\rho^{-p\alpha}),
\end{equation}
where $\Gamma(\rho^{\alpha})=\{\gamma\in\Gamma:0<$ $\mid\gamma\mid
<\rho^{\alpha})\}$, $p=s-d,$ $\alpha=\frac{1}{q},$ $q=3^{d}+d+2,$ and the
relation $\left\vert \gamma+t\right\vert ^{2}\sim\rho^{2}$ means that
$c_{1}\rho<\left\vert \gamma+t\right\vert <c_{2}\rho$. Here and in subsequent
relations we denote by $c_{i}$ ($i=1,2,...)$ the positive, independent of
$\rho$ constants whose exact values are inessential. Note that $q(x)\in
W_{2}^{s}(F)$ means that $\sum_{\gamma}\mid q_{\gamma}\mid^{2}(1+\mid
\gamma\mid^{2s})<\infty.$ If $s\geq d,$ then
\begin{equation}
\sum_{\gamma}\mid q_{\gamma}\mid<c_{3},\text{ }\sup\mid\sum_{\gamma
\notin\Gamma(c_{1}\rho^{\alpha})}q_{\gamma}e^{i(\gamma,x)}\mid\leq\sum
_{\mid\gamma\mid\geq c_{1}\rho^{\alpha}}\mid q_{\gamma}\mid=O(\rho^{-p\alpha
}),
\end{equation}
i.e., (3) holds. It follows from (4) that the influence of $\sum_{\gamma
\notin\Gamma(c_{1}\rho^{\alpha})}q_{\gamma}e^{i(\gamma,x)}$ to the eigenvalue
$\left\vert \gamma+t\right\vert ^{2}$ is $O(\rho^{-p\alpha}).$ If $\gamma+t$
does not belong to any of the sets

$W_{b,\alpha_{1}}(c_{2})=\{x\in\mathbb{R}^{d}:\mid\mid x\mid^{2}-\mid
x+b\mid^{2}\mid<c_{2}\mid x\mid^{\alpha_{1}}\}$ for $b\in\Gamma(c_{1}%
\rho^{\alpha}),$ then the influence of the trigonometric polynomial
$P(x)=\sum_{\gamma\in\Gamma(c_{1}\rho^{\alpha})}q_{\gamma}e^{i(\gamma,x)}$ to
the eigenvalue $\left\vert \gamma+t\right\vert ^{2}$ is not significant. Thus
the corresponding eigenvalue of the operator $L_{t}(q(x))$ is close to the
eigenvalue $\mid\gamma+t\mid^{2}$ of $L_{t}(0).$ Note that changing the values
of $c_{1}$ and $c_{2}$\ in the definitions of $W_{b,\alpha_{1}}(c_{2})$ and
$P(x)$ we obtain the different definitions of the non-resonance eigenvalues.
However, in any case we obtain the same asymptotic formulas and the same
perturbation theory, that is, this changing does not change anything for
asymptotic formulas. Therefore we can define the non-resonance eigenvalue in
different way. In accordance with the case of the trigonometric polynomial it
is natural to say that the eigenvalue $\left\vert \gamma+t\right\vert ^{2}$ is
a non-resonance eigenvalue if $\gamma+t$ does not belong to any of the sets
$W_{b,\alpha_{1}}(c_{2})$ for $\mid b\mid<c_{1}p\left\vert \gamma+t\right\vert
^{\alpha}$. However, for simplicity, we give the definitions as follows. By
definition, put $\alpha_{k}=3^{k}\alpha$ for $k=1,2,...$ and introduce the sets

$V_{\gamma_{1}}(\rho^{\alpha_{1}})\equiv\{x\in\mathbb{R}^{d}:\mid\mid
x\mid^{2}-\mid x+\gamma_{1}\mid^{2}\mid<\rho^{\alpha_{1}}\}\cap(R(\frac{3}%
{2}\rho)\backslash R(\frac{1}{2}\rho))$%
\[
E_{1}(\rho^{\alpha_{1}},p)\equiv\bigcup_{\gamma_{1}\in\Gamma(p\rho^{\alpha}%
)}V_{\gamma_{1}}(\rho^{\alpha_{1}}),\text{ }U(\rho^{\alpha_{1}},p)\equiv
(R(\frac{3}{2}\rho)\backslash R(\frac{1}{2}\rho))\backslash E_{1}(\rho
^{\alpha_{1}},p),
\]%
\[
E_{k}(\rho^{\alpha_{k}},p)\equiv\bigcup_{\gamma_{1},\gamma_{2},...,\gamma
_{k}\in\Gamma(p\rho^{\alpha})}(\cap_{i=1}^{k}V_{\gamma_{i}}(\rho^{\alpha_{k}%
})),
\]
where $R(\rho)=\{x\in\mathbb{R}^{d}:\mid x\mid<\rho\},$ $\rho$ is a big
parameter and the intersection $\cap_{i=1}^{k}V_{\gamma_{i}}$ in the
definition of $E_{k}$ is taken over $\gamma_{1},\gamma_{2},...,\gamma_{k},$
that are linearly independent. The set $U(\rho^{\alpha_{1}},p)$ is said to be
a non-resonance domain and the eigenvalue $\left\vert \gamma+t\right\vert
^{2}$ is called a non-resonance eigenvalue if $\gamma+t\in U(\rho^{\alpha_{1}%
},p).$ The domains $V_{\gamma_{1}}(\rho^{\alpha_{1}})$ for $\gamma_{1}%
\in\Gamma(p\rho^{\alpha})$ are called resonance domains and $\mid\gamma
+t\mid^{2}$ is called a resonance eigenvalue if $\gamma+t\in V_{\gamma_{1}%
}(\rho^{\alpha_{1}}).$ In Remark 1 we will discuss the relations between sets
$W_{b,\alpha_{1}}(c_{2})$ and $V_{b}(\rho^{\alpha_{1}}).$

In section 2 we prove that for each $\gamma+t\in U(\rho^{\alpha_{1}},p)$ there
exists an eigenvalue $\Lambda_{N}(t)$ of the operator $L_{t}(q(x))$ satisfying
the following formulae
\begin{equation}
\Lambda_{N}(t)=\mid\gamma+t\mid^{2}+F_{k-1}(\gamma+t)+O(\mid\gamma
+t\mid^{-3k\alpha})
\end{equation}
for $k=1,2,...,[\frac{1}{3}(p-\frac{1}{2}q(d-1))],$ where $[a]$ denotes the
integer part of $a,$ $F_{0}=0,$ and $F_{k-1}$ ( for $k>1)$ is expressed by the
potential $q(x)$ and eigenvalues of $L_{t}(0).$ Besides, we prove that if the
conditions
\begin{align}
&  \mid\Lambda_{N}(t)-\mid\gamma+t\mid^{2}\mid<\frac{1}{2}\rho^{\alpha_{1}},\\
&  \mid b(N,\gamma)\mid>c_{4}\rho^{-c\alpha}%
\end{align}
hold, \ where $b(N,\gamma)=(\Psi_{N,t},e^{i(\gamma+t,x)}),$ $\Psi_{N,t}(x)$ is
a normalized eigenfunction of $L_{t}(q(x))$ corresponding to $\Lambda_{N}(t),$
then $\ $the following statements are valid:

(a) if $\gamma+t$ is in the non-resonance domain, then $\Lambda_{N}(t)$
satisfies (5) for $k=1,2,...,[\frac{1}{3}(p-c)]$ ( see Theorem 1);

(b) if $\gamma+t\in E_{s}\backslash E_{s+1},$ where $s=1,2,...,d-1,$ then%
\begin{equation}
\Lambda_{N}(t)=\lambda_{j}(\gamma+t)+O(\mid\gamma+t\mid^{-k\alpha}),
\end{equation}
where $\lambda_{j}$ is an eigenvalue of the matrix $C(\gamma+t)$ ( see (27)
and Theorem 2). Moreover, we prove that every big eigenvalue of the operator
$L_{t}(q(x))$ for all values of $t$ satisfies one of these formulae.

The results of section 2 ( see Theorem 1,2) is considered in [17]. However, in
that paper these results are written only briefly. The enlarged variant is
written in [19, 21] which can not be used as a reference. Here we write the
non-resonance case in an improved and enlarged form and so that it can easily
be used in the next sections. Moreover it helps to read section 3, where we
consider in detail the single resonance domains $V_{\gamma_{1}}^{^{\prime}%
}(\rho^{\alpha_{1}})\equiv V_{\gamma_{1}}(\rho^{\alpha_{1}})\backslash E_{2}$,
i.e., the part of the resonance domains $V_{\gamma_{1}}(\rho^{\alpha_{1}}),$
which does not contain the intersection of two resonance domains. Namely, for
this case we obtain asymptotic formulas of arbitrary order for the eigenvalues
of the $d$ dimensional periodic Schrodinger operator $L(q)$ for arbitrary
dimension $d$. This case is connected with Sturm-Liouville operators. In the
next papers, which use this, we will constructively determine a family of
spectral invariants by given Floquet spectrum and give an algorithm for
finding the potential $q(x)$ by these spectral invariants. Thus the new
results about asymptotic formulas for eigenvalues are the results
corresponding to the single resonance case (section 3). It follows from (5)
that the non-resonance eigenvalue of $L_{t}(q(x))$ is close to the eigenvalue
of the Laplace operator $L_{t}(0)$. So the influence of the potential $q(x)$
is not significant. To obtain the asymptotic formula for the non-resonance
eigenvalues we take the operator$L_{t}(0)$ for an unperturbed operator and
$q(x)$ for a perturbation. We use the formula (16) connecting the eigenvalues
and eigenfunctions of $L_{t}(q(x))$ and $L_{t}(0).$ We call (16) binding
formula for $L_{t}(q(x))$ and $L_{t}(0).$ In section 2 substituting the
decomposition (3) of $q(x)$ into (16) and iterating it several times we get
the asymptotic formulas for the non-resonance eigenvalues.

The resonance eigenvalues corresponds to the quasimomentum $\gamma+t$ lying
near the diffraction hyperplanes $\{x:\mid x\mid^{2}=\mid x+\delta\mid^{2}\}.$
In this case the influence of the directional potential%
\begin{equation}
q^{\delta}(x)=\sum_{n\in Z}q_{n\delta}e^{in(\delta,x)}=Q(\zeta),\text{ }%
\zeta=(\delta,x)
\end{equation}
is significant, but the influence of $q(x)-q^{\delta}(x)$ is not significant.
Therefore these eigenvalues of $L_{t}(q(x))$ is close to the eigenvalues of
$L_{t}(q^{\delta}(x)).$ Hence in section 3 for investigation of the resonance
eigenvalues we take the operator $L_{t}(q^{\delta}(x))$ for an unperturbed
operator and $q(x)-q^{\delta}(x)$ for a perturbation. To prove the asymptotic
formulas for the resonance eigenvalues first we consider the eigenvalues and
eigenfunctions of $L_{t}(q^{\delta}(x))$ ( see Lemma 2 ). Then we iterate the
formula (53) connecting the eigenvalues and the eigenfunctions of
$L_{t}(q(x))$ and $L_{t}(q^{\delta}(x)).$ We call (53) binding formula for
$L_{t}(q(x))$ and $L_{t}(q^{\delta}(x)).$ The formulas (16), (53) and their
iterations are similar. In the resonance case ( in section 3) we use the ideas
of the non-resonance case ( of section 2) and replace the perturbation $q(x)$
by the perturbation $q(x)-q^{\delta}(x)$, the eigenvalues $\mid\gamma
+t\mid^{2}$ and eigenfunctions $e^{i(\gamma+t,x)}$ of the unperturbed ( for
non-resonance case) operator $L_{t}(0)$ by the eigenvalues and eigenfunctions
of the \ unperturbed ( for resonance case) operator $L_{t}(q^{\delta}(x))$
respectively. Therefore the simple iterations of (16) helps to read the
complicated iteration of (53).

For investigation of the Bloch function in the non-resonance domain, in
section 4, we find the values of quasimomenta $\gamma+t$ for which the
corresponding eigenvalues are simple , namely we construct the subset $B$ of
$U(\rho^{\alpha_{1}},p)$ with the following properties:

Pr.1. If $\gamma+t\in B,$ then there exists a unique eigenvalue, denoted by
$\Lambda(\gamma+t),$ of the operator $L_{t}(q(x))$ satisfying (5). This is a
simple eigenvalue of $L_{t}(q(x))$. Therefore we call the set $B$ the simple
set of quasimomenta.

Pr.2. The eigenfunction $\Psi_{N(\gamma+t)}(x)\equiv\Psi_{\gamma+t}(x)$
corresponding to the eigenvalue $\Lambda(\gamma+t)$ is close to $e^{i(\gamma
+t,x)}$, namely
\begin{equation}
\Psi_{N}(x)=e^{i(\gamma+t,x)}+O(\mid\gamma+t\mid^{-\alpha_{1}}),
\end{equation}%
\begin{equation}
\Psi_{\gamma+t}(x)=e^{i(\gamma+t,x)}+\Phi_{k-1}(x)+O(\mid\gamma+t\mid
^{-k\alpha_{1}}),\text{ }k=1,2,...\text{ ,}%
\end{equation}
where $\Phi_{k-1}$ is expressed by $q(x)$ and the eigenvalues of $L_{t}(0).$

Pr.3. The set $B$ contains the intervals $\{a+sb:s\in\lbrack-1,1]\}$ such that
$\Lambda(a-b)<\rho^{2},$ $\Lambda(a+b)>\rho^{2},$\ and $\Lambda(\gamma+t)$
\ is continuous on these intervals. Hence there exists $\gamma+t$ such that
$\Lambda(\gamma+t)=\rho^{2}$ for $\rho\gg1.$ It implies that there exist only
a finite number of gaps in the spectrum of $L,$ that is, it implies the
validity of Bethe-Sommerfeld conjecture for arbitrary dimension and for
arbitrary lattice.

Construction of the set $B$ consists of two steps.

Step 1. We prove that all eigenvalues $\Lambda_{N}(t)\sim\rho^{2}$ of the
operator $L_{t}(q(x))$ lie in the $\varepsilon_{1}=\rho^{-d-2\alpha}$
neighborhood of the numbers

$F(\gamma+t)=\mid\gamma+t\mid^{2}+F_{k_{1}-1}(\gamma+t)$, $\lambda_{j}%
(\gamma+t)$ ( see (5), (8)), where $k_{1}=[\frac{d}{3\alpha}]+2.$ We call
these numbers as the known parts of the eigenvalues. Moreover, for
$\gamma+t\in U(\rho^{\alpha_{1}},p)$ there exists $\Lambda_{N}(t)$ satisfying
$\Lambda_{N}(t)=F(\gamma+t)+o(\varepsilon_{1})$.

Step 2. By eliminating the set of quasimomenta $\gamma+t$, for which the known
parts $F(\gamma+t)$ of $\Lambda_{N}(t)$ are situated from the known parts
$F(\gamma^{^{\prime}}+t),$ $\lambda_{j}(\gamma^{^{\prime}}+t)$ ($\gamma
^{^{\prime}}\neq\gamma)$ of other eigenvalues at a distance less than
$2\varepsilon_{1},$ we construct the set $B$ with the following properties: if
$\gamma+t\in B,$ then the following conditions (called simplicity conditions
for $\Lambda_{N}(t))$ hold
\begin{equation}
\mid F(\gamma+t)-F(\gamma^{^{\prime}}+t)\mid\geq2\varepsilon_{1}\text{ }%
\end{equation}
for $\gamma^{^{\prime}}\in K\backslash\{\gamma\},$ $\gamma^{^{\prime}}+t\in
U(\rho^{\alpha_{1}},p)$ and%
\begin{equation}
\mid F(\gamma+t)-\lambda_{j}(\gamma^{^{\prime}}+t)\mid\geq2\varepsilon_{1}%
\end{equation}
for $\gamma^{^{\prime}}\in K,\gamma^{^{\prime}}+t\in E_{k}\backslash E_{k+1},$
$j=1,2,...,$ where $K$ is the set of $\gamma^{^{\prime}}\in\Gamma$ satisfying
$\mid F(\gamma+t)-\mid\gamma^{^{\prime}}+t\mid^{2}\mid<\frac{1}{3}\rho
^{\alpha_{1}}$. Thus $B$ is the set of

$x\in U(\rho^{\alpha_{1}},p)\cap(R(\frac{3}{2}\rho-\rho^{\alpha_{1}%
-1})\backslash R(\frac{1}{2}\rho+\rho^{\alpha_{1}-1}))$ such that
$x=\gamma+t,$ where $\gamma\in\Gamma,t\in F^{\star},$ and the simplicity
conditions (12), (13) hold. As a consequence of these conditions the
eigenvalue $\Lambda_{N}(t)$ does not coincide with other eigenvalues. To prove
this, namely to prove the Pr.1 and (10), we show that for any normalized
eigenfunction $\Psi_{N}(x)$ corresponding to $\Lambda_{N}(t)$ the following
equality holds:
\begin{equation}
\sum_{\gamma^{^{\prime}}\in\Gamma\backslash\gamma}\mid b(N,\gamma^{^{\prime}%
})\mid^{2}=O(\rho^{-2\alpha_{1}}).
\end{equation}

For the first time in [15-17] we constructed the simple set $B$ with the Pr.1
and Pr.3., though in those papers we emphasized the Bethe-Sommerfeld
conjecture. Note that for this conjecture and for Pr.1, Pr.3. it is enough to
prove that the left-hand side of (14) is less than $\frac{1}{4}$ ( we proved
this inequality in [15-17] and as noted in Theorem 3 of [16] and in [18] the
proof of this inequality does not differ from the proof of (14)). From (10) we
got \ (11) by iteration (see [18]) . But in those papers these results are
written briefly. The enlarged variant is written in [19,22] which can not be
used as reference. In this paper we write these results in improved and
enlarged form. The main difficulty and the crucial point of papers [15-17]
were the construction of the simple set $B$ with the Pr.1.,Pr.3. This
difficulty of the perturbation theory of $L(q(x))$ is of a physical nature and
it is connected with the complicated picture of the crystal diffraction. If
$d=2,3,$ then $F(\gamma+t)=\mid\gamma+t\mid^{2}$ and the matrix $C(\gamma+t)$
corresponds to the Schrodinger operator with directional potential (9) ( see
[16]). So for construction of the simple set $B$ of quasimomenta we eliminated
the vicinities of the diffraction planes and the sets connected with
directional potential ( see (12), (13)). Besides,\ for nonsmooth potentials
$q(x)\in L_{2}(\mathbb{R}^{2}/\Omega),$ we eliminated a set, which is
described in the terms of the number of states ( see [15, 19, 20]). The simple
sets $B$ of quasimomenta for the first time is constructed and investigated (
hence the main difficulty and the crucial point of perturbation theory of
$L(q)$ is investigated) in [16] for $d=3$ and in [15,17] for the cases:

1. $d=2,$ $q(x)\in L_{2}(F);$ \ \ \ 2. $d>2,$ $q(x)$ is a smooth potential.

Then, Yu.E. Karpeshina proved ( see [7-9]) the convergence of the perturbation
series of two and three dimensional Schrodinger operator $L(q)$ with a wide
class of nonsmooth potential $q(x)$ for a set, that is similar to $B$, of
quasimomenta. In papers [3,4] the asymptotic formulas for the eigenvalues and
Bloch function of the two and three dimensional operator $L_{t}(q(x))$ were
obtained. In [5] the asymptotic formulae for the eigenvalues of $L_{0}(q(x))$
were obtained.

In section 5 we consider the geometrical aspects of the simple sets. We prove
that the simple sets $B$ has asymptotically full measure on $\mathbb{R}^{d}$.
Moreover, we construct a part of isoenergetic surfaces corresponding to
$\rho^{2},$ which is smooth surfaces and has the measure asymptotically close
to the measure of the isoenergetic surfaces of the operator $L(0).$ The
nonemptyness of the isoenergetic surfaces \ for $\rho\gg1$ implies the
validity of the Bethe-Sommerfeld conjecture. Note that one can read Section 4
and Section 5 without reading Section 3.

For the first time M.M. Skriganov [11,12] proved the validity of the
Bethe-Sommerfeld conjecture for the Schrodinger operator for dimension $d=2,3$
for arbitrary lattice, for dimension $d>3$ for rational lattice. The
Skriganov's method is based on the detail investigation of the arithmetic and
geometric properties of the lattice. B.E.J.Dahlberg \ and E.Trubowits [1]
using an asymptotic of Bessel function, gave the simple proof of this
conjecture for the two dimensional Schrodinger operator. Then in papers
[15-17] we proved the validity of the Bethe-Sommerfeld conjecture for
arbitrary lattice and for arbitrary dimension by using the asymptotic formulas
and by construction of the simple set $B,$ that is, by\ the method of
perturbation theory. Yu.E. Karpeshina ( see [7-9]) proved this conjecture for
two and three dimensional Schrodinger operator $L(q)$ for a wide class of
singular potentials $q(x),$ including Coulomb potential, by \ the method of
perturbation theory. B. Helffer and A. Mohamed [6], by investigations the
integrated density of states, proved the validity of the Bethe-Sommerfeld
conjecture for the Schrodinger operator for $d\leq4$ for arbitrary lattice.
Recently L. Parnovski and A. V. Sobelev [10] proved this conjecture for
$d\leq4.$ The method of this paper and papers [15-17] is a first and unique,
for the present, by which the validity of the Bethe-Sommerfeld conjecture for
arbitrary lattice and for arbitrary dimension is proved.

In section 6 we construct simple sets in the resonance domain and obtain the
asymptotic formulas of arbitrary order for the Bloch functions of the

$d$ dimensional Schrodinger operator $L(q(x)),$ where $q(x)\in W_{2}^{s}(F),$

$s\geq6(3^{d}(d+1)^{2})+d,$ when corresponding quasimomentum lies in these
simple sets. Note that we construct the simple sets in the non-resonance
domain so that it contains a big part of the isoenergetic surfaces of $L(q).$
However in the case of resonance domain we construct the simple set so that it
can be easily used for the constructive determination ( in next papers) a
family of the spectral invariants by given Floquet spectrum.

In this paper for the different types of the measures of the subset $A$ of
$\mathbb{R}^{d}$ we use the same notation $\mu(A).$ By $\mid A\mid$ we denote
the number of elements of the set $A\subset\Gamma$ and use the following
obvious fact. If $a\sim\rho,$ then the number of elements of the set
$\{\gamma+t:$ $\gamma\in\Gamma\}$ satisfying $\mid\mid\gamma+t\mid-a\mid<1$ is
less than $c_{5}\rho^{d-1}.$ Therefore the number of eigenvalues of $L_{t}(q)$
lying in $(a^{2}-\rho,a^{2}+\rho)$ is less than $c_{5}\rho^{d-1}.$ Besides, we
use the inequalities:%
\begin{align}
\alpha_{1}+d\alpha &  <1-\alpha\,,\ \ \ \ \ d\alpha<\frac{1}{2}\alpha
_{d},\ \ \ k_{1}\leq\frac{1}{3}(p-\frac{1}{2}(q(d-1)),\\
p_{1}\alpha_{1}  &  \geq p\alpha,\ \ \ \ \ 3k_{1}\alpha>d+2\alpha
,\ \ \ \ \ \ \alpha_{k}+(k-1)\alpha<1,\nonumber\\
\alpha_{k+1}  &  >2(\alpha_{k}+(k-1))\alpha\nonumber
\end{align}
for $k=1,2,...,d,$ which follow from the definitions $p=s-d,$ $\alpha
_{k}=3^{k}\alpha,$ $\alpha=\frac{1}{q},$ $q=3^{d}+d+2,$ $k_{1}=[\frac
{d}{3\alpha}]+2,$ $p_{1}=[\frac{p}{3}]+1$ of the numbers $p,q,\alpha
_{k},\alpha,k_{1},p_{1}.$

\section{Asymptotic Formulae for the Eigenvalues}

First we obtain the asymptotic formulas for the non-resonance eigenvalues by
iteration of the formula%
\begin{equation}
(\Lambda_{N}-\mid\gamma+t\mid^{2})b(N,\gamma)=(\Psi_{N,t}(x)q(x),e^{i(\gamma
+t,x)}),
\end{equation}
which is obtained from equation $-\Delta\Psi_{N,t}(x)+q(x)\Psi_{N,t}%
(x)=\Lambda_{N}\Psi_{N,t}(x)$ by multiplying by $e^{i(\gamma+t,x)}),$ where
$\gamma+t\in U(\rho^{\alpha_{1}},p).$ Introducing into (16) the expansion (3)
of $q(x)$, we get
\begin{equation}
(\Lambda_{N}-\mid\gamma+t\mid^{2})b(N,\gamma)=\sum_{\gamma_{1}\in\Gamma
(\rho^{\alpha})}q_{\gamma_{1}}b(N,\gamma-\gamma_{1})+O(\rho^{-p\alpha}).
\end{equation}
From the relations (16), (17) it follows that
\begin{equation}
b(N,\gamma^{^{\prime}})=\dfrac{(\Psi_{N,t}q(x),e^{i(\gamma^{^{\prime}}+t,x)}%
)}{\Lambda_{N}-\mid\gamma^{^{\prime}}+t\mid^{2}}=%
{\displaystyle\sum_{\gamma_{1}\in\Gamma(\rho^{\alpha})}}
\dfrac{q_{\gamma_{1}}b(N,\gamma^{^{\prime}}-\gamma_{1})}{\Lambda_{N}%
-\mid\gamma^{^{\prime}}+t\mid^{2}}+O(\rho^{-p\alpha})
\end{equation}
for all vectors $\gamma^{^{\prime}}\in\Gamma$ satisfying the inequality
\begin{equation}
\mid\Lambda_{N}-\mid\gamma^{^{\prime}}+t\mid^{2}\mid>\frac{1}{2}\rho
^{\alpha_{1}}.
\end{equation}
If (6) holds and $\gamma+t\in U(\rho^{\alpha_{1}},p),$ then
\begin{equation}
\mid\mid\gamma+t\mid^{2}-\mid\gamma-\gamma_{1}+t\mid^{2}\mid>\rho^{\alpha_{1}%
},\text{ }\mid\Lambda_{N}-\mid\gamma-\gamma_{1}+t\mid^{2}\mid>\frac{1}{2}%
\rho^{\alpha_{1}}\text{ }%
\end{equation}
for all $\gamma_{1}\in\Gamma(p\rho^{\alpha}).$ Hence the vector $\gamma
-\gamma_{1}$ for $\gamma+t\in U(\rho^{\alpha_{1}},p)$ and $\gamma_{1}\in
\Gamma(p\rho^{\alpha})$ satisfies (19). Therefore, in (18) one can replace
$\gamma^{^{\prime}}$ by $\gamma-\gamma_{1}$ and write
\[
b(N,\gamma-\gamma_{1})=%
{\displaystyle\sum_{\gamma_{2}\in\Gamma(\rho^{\alpha})}}
\dfrac{q_{\gamma_{2}}b(N,\gamma-\gamma_{1}-\gamma_{2})}{\Lambda_{N}-\mid
\gamma-\gamma_{1}+t\mid^{2}}+O(\rho^{-p\alpha}).
\]
Substituting this for $b(N,\gamma-\gamma_{1})$ into the right-hand side of
(17) and isolating the terms containing the multiplicand $b(N,\gamma)$, we
get
\[
(\Lambda_{N}-\mid\gamma+t\mid^{2})b(N,\gamma)=%
{\displaystyle\sum_{\gamma_{1},\gamma_{2}\in\Gamma(\rho^{\alpha})}}
\dfrac{q_{\gamma_{1}}q_{\gamma_{2}}b(N,\gamma-\gamma_{1}-\gamma_{2})}%
{\Lambda_{N}-\mid\gamma-\gamma_{1}+t\mid^{2}}+O(\rho^{-p\alpha})=
\]%
\[%
{\displaystyle\sum_{\gamma_{1}\in\Gamma(\rho^{\alpha})}}
\dfrac{\mid q_{\gamma_{1}}\mid^{2}b(N,\gamma)}{\Lambda_{N}-\mid\gamma
-\gamma_{1}+t\mid^{2}}+%
{\displaystyle\sum_{\substack{\gamma_{1},\gamma_{2}\in\Gamma(\rho^{\alpha
}),\\\gamma_{1}+\gamma_{2}\neq0}}}
\dfrac{q_{\gamma_{1}}q_{\gamma_{2}}b(N,\gamma-\gamma_{1}-\gamma_{2})}%
{\Lambda_{N}-\mid\gamma-\gamma_{1}+t\mid^{2}}+O(\rho^{-p\alpha}),
\]
since $q_{\gamma_{1}}q_{\gamma_{2}}=\mid q_{\gamma_{1}}\mid^{2}$ for
$\gamma_{1}+\gamma_{2}=0$ and the last summation is taken under the condition
$\gamma_{1}+\gamma_{2}\neq0.$ Repeating this process $p_{1}\equiv\lbrack
\frac{p}{3}]+1$ times, i.e., in the last summation replacing $b(N,\gamma
-\gamma_{1}-\gamma_{2})$ by its expression from (18) ( in (18) replace
$\gamma^{^{\prime}}$ by $\gamma-\gamma_{1}-\gamma_{2}$) and isolating the
terms containing $b(N,\gamma)$ etc., we obtain
\begin{equation}
(\Lambda_{N}-\mid\gamma+t\mid^{2})b(N,\gamma)=A_{p_{1}}(\Lambda_{N}%
,\gamma+t)b(N,\gamma)+C_{p_{1}}+O(\rho^{-p\alpha}),
\end{equation}
where $A_{p_{1}}(\Lambda_{N},\gamma+t)=\sum_{k=1}^{p_{1}}S_{k}(\Lambda
_{N},\gamma+t)$ ,
\[
S_{k}(\Lambda_{N},\gamma+t)=%
{\displaystyle\sum_{\gamma_{1},...,\gamma_{k}\in\Gamma(\rho^{\alpha})}}
\dfrac{q_{\gamma_{1}}q_{\gamma_{2}}...q_{\gamma_{k}}q_{-\gamma_{1}-\gamma
_{2}-...-\gamma_{k}}}{\prod_{j=1}^{k}(\Lambda_{N}-\mid\gamma+t-\sum_{i=1}%
^{j}\gamma_{i}\mid^{2})},
\]%
\[
C_{p_{1}}=\sum_{\gamma_{1},...,\gamma_{p_{1}+1}\in\Gamma(\rho^{\alpha})}%
\dfrac{q_{\gamma_{1}}q_{\gamma_{2}}...q_{\gamma_{p_{1}+1}}b(N,\gamma
-\gamma_{1}-\gamma_{2}-...-\gamma_{p_{1}+1})}{\prod_{j=1}^{p_{1}}(\Lambda
_{N}-\mid\gamma+t-\sum_{i=1}^{j}\gamma_{i}\mid^{2})}.
\]
Here the summations for $S_{k}$ and $C_{p_{1}}$ are taken under the additional
conditions $\gamma_{1}+\gamma_{2}+...+\gamma_{s}\neq0$ for $s=1,2,...,k$ and
$s=1,2,...,p_{1}$ respectively. These conditions and the inclusion $\gamma
_{i}\in\Gamma(\rho^{\alpha})$ for $i=1,2,...,p_{1}$ imply the relation
$\sum_{i=1}^{j}\gamma_{i}\in\Gamma(p\rho^{\alpha})$. Therefore from the second
inequality in (20) it follows that the absolute values of the denominators of
the fractions in $S_{k}$ and $C_{p_{1}}$ are greater than $(\frac{1}{2}%
\rho^{\alpha_{1}})^{k}$ and $(\frac{1}{2}\rho^{\alpha_{1}})^{p_{1}}$
respectively. Hence the first inequality in (4) and $p_{1}\alpha_{1}\geq
p\alpha$ ( see the fourth inequality in (15)) yield
\begin{equation}
C_{p_{1}}=O(\rho^{-p_{1}\alpha_{1}})=O(\rho^{-p\alpha}),\text{ }S_{k}%
(\Lambda_{N},\gamma+t)=O(\rho^{-k\alpha_{1}}),\forall k=1,2,...,p_{1}.
\end{equation}
Since we used only the condition (6) for $\Lambda_{N},$ it follows that
\begin{equation}
S_{k}(a,\gamma+t)=O(\rho^{-k\alpha_{1}})
\end{equation}
for all $a\in\mathbb{R}$ satisfying $\mid a-\mid\gamma+t\mid^{2}\mid<\frac
{1}{2}\rho^{\alpha_{1}}.$ Thus finding $N$ such that $\Lambda_{N}$ is close to
$\mid\gamma+t\mid^{2}$ and $b(N,\gamma)$ is not very small, then dividing both
sides of (21) by $b(N,\gamma),$ we get the asymptotic formulas for
$\Lambda_{N}$.

\begin{theorem}
$(a)$ Suppose $\gamma+t\in U(\rho^{\alpha_{1}},p).$ If (6) and (7) hold, then
$\Lambda_{N}$ satisfies formulas (5) for $k=1,2,...,[\frac{1}{3}(p-c)],$
where
\begin{equation}
F_{s}=O(\rho^{-\alpha_{1}}),\forall s=0,1,...,
\end{equation}
and $F_{0}=0,$ $F_{s}=A_{s}(\mid\gamma+t\mid^{2}+F_{s-1},\gamma+t)$ for
$s=1,2,....$

$(b)$ For $\gamma+t\in U(\rho^{\alpha_{1}},p)$ there exists an eigenvalue
$\Lambda_{N}$ of $L_{t}(q(x))$ satisfying (5).
\end{theorem}

\begin{proof}
$(a)$ To prove (5) in case $k=1$ we divide both side of (21) by $b(N,\gamma)$
and use (7), (22). Then we obtain
\begin{equation}
\Lambda_{N}-\mid\gamma+t\mid^{2}=O(\rho^{-\alpha_{1}}).
\end{equation}
This and $\alpha_{1}=3\alpha$ ( see the end of the introduction) imply that
formula (5) for $k=1$ holds and $F_{0}=0.$ Hence (24) for $s=0$ is also
proved. Moreover, from (23), we obtain $S_{k}(\mid\gamma+t\mid^{2}%
+O(\rho^{-\alpha_{1}}),\gamma+t)=O(\rho^{-\alpha_{1}})$ for $k=1,2,....$
Therefore (24) for arbitrary $s$ follows from the definition of $F_{s}$ by
induction. Now we prove (5) by induction on $k$. Suppose (5) holds for $k=j$,
that is,

$\Lambda_{N}=\mid\gamma+t\mid^{2}+F_{k-1}(\gamma+t)+O(\rho^{-3k\alpha}).$
Substituting this into $A_{p_{1}}(\Lambda_{N},\gamma+t)$ in (21) and dividing
both sides of (21) by $b(N,\gamma),$ we get
\begin{align*}
\Lambda_{N}  &  =\mid\gamma+t\mid^{2}+A_{p_{1}}(\mid\gamma+t\mid^{2}%
+F_{j-1}+O(\rho^{-j\alpha_{1}}),\gamma+t)+O(\rho^{-(p-c)\alpha})=\\
&  \mid\gamma+t\mid^{2}+\{A_{p_{1}}(\mid\gamma+t\mid^{2}+F_{j-1}%
+O(\rho^{-j\alpha_{1}}),\gamma+t)-\\
A_{p_{1}}(  &  \mid\gamma+t\mid^{2}+F_{j-1},\gamma+t)\}+A_{p_{1}}(\mid
\gamma+t\mid^{2}+F_{j-1},\gamma+t)+O(\rho^{-(p-c)\alpha}).
\end{align*}
To prove $(a)$ for $k=j+1$ we need to show that the expression in curly
brackets is equal to $O(\rho^{-(j+1)\alpha_{1}}).$ It can be checked by using
(4), (20), (24) and the obvious relation
\begin{align*}
&  \frac{1}{\prod_{j=1}^{s}(\mid\gamma+t\mid^{2}+F_{j-1}+O(\rho^{-j\alpha_{1}%
})-\mid\gamma+t-\sum_{i=1}^{s}\gamma_{i}\mid^{2})}-\\
&  \frac{1}{\prod_{j=1}^{s}(\mid\gamma+t\mid^{2}+F_{j-1}-\mid\gamma
+t-\sum_{i=1}^{s}\gamma_{i}\mid^{2})}\\
&  =\frac{1}{\prod_{j=1}^{s}(\mid\gamma+t\mid^{2}+F_{j-1}-\mid\gamma
+t-\sum_{i=1}^{s}\gamma_{i}\mid^{2})}(\frac{1}{1-O(\rho^{-(j+1)\alpha_{1}}%
)}-1)
\end{align*}
$=O(\rho^{-(j+1)\alpha_{1}})$ for $s=1,2,...,p_{1}.$

$(b)$ Let $A$ be the set of indices $N$ satisfying (6). Using (16) and Bessel
inequality, we obtain%
\[
\sum_{N\notin A}\mid b(N,\gamma)\mid^{2}=\sum_{N\notin A}\mid\dfrac{(\Psi
_{N}(x),q(x)e^{i(\gamma+t,x)})}{\Lambda_{N}-\mid\gamma+t\mid^{2}}\mid
^{2}=O(\rho^{-2\alpha_{1}})
\]
Hence, by the Parseval equality, we have $\sum_{N\in A}\mid b(N,\gamma
)\mid^{2}=1-O(\rho^{-2\alpha_{1}}).$ This and the inequality $\mid A\mid
<c_{5}\rho^{d-1}=c_{5}\rho^{(d-1)q\alpha}$ ( see the end of the introduction)
imply that there exists a number $N$ satisfying $\mid b(N,\gamma)\mid>\frac
{1}{2}(c_{5})^{-1}\rho^{-\frac{(d-1)q}{2}\alpha}$, that is, (7) holds for
$c=\frac{(d-1)q}{2}$ . Thus $\Lambda_{N}$ satisfies (5) due to $(a)$
\end{proof}

Theorem 1 shows that in the non-resonance case the eigenvalue of the perturbed
operator $L_{t}(q(x))$ is close to the eigenvalue of the unperturbed operator
$L_{t}(0).$ However, in Theorem 2 we prove that if $\gamma+t\in\cap_{i=1}%
^{k}V_{\gamma_{i}}(\rho^{\alpha_{k}})\backslash E_{k+1}$ for $k\geq1,$ where
$\gamma_{1},\gamma_{2},...,\gamma_{k}$ are linearly independent vectors of
$\Gamma(p\rho^{\alpha}),$ then the corresponding eigenvalue of $L_{t}(q(x))$
is close to the eigenvalue of the matrix constructed as follows. Introduce the sets:

$B_{k}\equiv B_{k}(\gamma_{1},\gamma_{2},...,\gamma_{k})=\{b:b=\sum_{i=1}%
^{k}n_{i}\gamma_{i},n_{i}\in Z,\mid b\mid<\frac{1}{2}\rho^{\frac{1}{2}%
\alpha_{k+1}}\},$%
\begin{equation}
B_{k}(\gamma+t)=\gamma+t+B_{k}=\{\gamma+t+b:b\in B_{k}\},
\end{equation}
$B_{k}(\gamma+t,p_{1})=\{\gamma+t+b+a:b\in B_{k},\mid a\mid<p_{1}\rho^{\alpha
},a\in\Gamma\}.$

Denote by $h_{i}+t$ for $i=1,2,...,b_{k}$ the vectors of $B_{k}(\gamma
+t,p_{1}),$ where

$b_{k}\equiv b_{k}(\gamma_{1},\gamma_{2},...,\gamma_{k})$ is the number of the
vectors of $B_{k}(\gamma+t,p_{1})$. Define the matrix $C(\gamma+t,\gamma
_{1},\gamma_{2},...,\gamma_{k})\equiv(c_{i,j})$ by the formulas
\begin{equation}
c_{i,i}=\mid h_{i}+t\mid^{2},\text{ }c_{i,j}=q_{h_{i}-h_{j}},\text{ }\forall
i\neq j,
\end{equation}
where $i,j=1,2,...,b_{k}.$ We consider the resonance eigenvalue \ $\mid
\gamma+t\mid^{2}$ for $\gamma+t\in(\cap_{i=1}^{k}V_{\gamma_{i}}(\rho
^{\alpha_{k}}))$ by using the following Lemma.

\begin{lemma}
If $\gamma+t\in\cap_{i=1}^{k}V_{\gamma_{i}}(\rho^{\alpha_{k}})\backslash
E_{k+1},$ $h+t\in B_{k}(\gamma+t,p_{1}),$

$(h-\gamma^{^{\prime}}+t)\notin B_{k}(\gamma+t,p_{1}),$ then
\begin{equation}
\mid\mid\gamma+t\mid^{2}-\mid h-\gamma^{^{\prime}}-\gamma_{1}^{^{\prime}%
}-\gamma_{2}^{^{\prime}}-...-\gamma_{s}^{^{\prime}}+t\mid^{2}\mid>\frac{1}%
{5}\rho^{\alpha_{k+1}},
\end{equation}
where $\gamma^{^{\prime}}\in\Gamma(\rho^{\alpha}),$ $\gamma_{j}^{^{\prime}}%
\in\Gamma(\rho^{\alpha}),$ $j=1,2,...,s$ and $s=0,1,...,p_{1}-1.$
\end{lemma}

\begin{proof}
The inequality $p>2p_{1}$ ( see the end of the introduction) and the
conditions of Lemma 1 imply that

$h-\gamma^{^{\prime}}-\gamma_{1}^{^{\prime}}-\gamma_{2}^{^{\prime}}%
-...-\gamma_{s}^{^{\prime}}+t\in B_{k}(\gamma+t,p)\backslash B_{k}(\gamma+t)$
for all $s=0,1,...,p_{1}-1.$ It follows from the definitions of $B_{k}%
(\gamma+t,p),B_{k}$ that ( see (26))

$h-\gamma^{^{\prime}}-\gamma_{1}^{^{\prime}}-\gamma_{2}^{^{\prime}}%
-...-\gamma_{s}^{^{\prime}}+t=\gamma+t+b+a,$ where
\begin{equation}
\mid b\mid<\frac{1}{2}\rho^{\frac{1}{2}\alpha_{k+1}},\mid a\mid<p\rho^{\alpha
},\text{ }\gamma+t+b+a\notin\gamma+t+B_{k}.
\end{equation}
Then (28) has the form
\begin{equation}
\mid\mid\gamma+t+a+b\mid^{2}-\mid\gamma+t\mid^{2}\mid>\frac{1}{5}\rho
^{\alpha_{k+1}}.
\end{equation}
To prove (30) we consider two cases:

Case 1. $a\in P$, where $P=Span\{\gamma_{1,}\gamma_{2},...,\gamma_{k}\}.$
Since $b\in B_{k}\subset P,$ we have $a+b\in P.$ This with the third relation
in (29) imply that $a+b\in P\backslash B_{k}$ ,i.e., $\mid a+b\mid\geq\frac
{1}{2}$ $\rho^{\frac{1}{2}\alpha_{k+1}}$. Consider the orthogonal
decomposition $\gamma+t=y+v$ of $\gamma+t,$ where $v\in P$ and $y\bot P.$
First we prove that the projection $v$ of any vector $x\in\cap_{i=1}%
^{k}V_{\gamma_{i}}(\rho^{\alpha_{k}})$ on $P$ satisfies
\begin{equation}
\mid v\mid=O(\rho^{(k-1)\alpha+\alpha_{k}}).
\end{equation}
For this we turn the coordinate axis so that $Span\{\gamma_{1,}\gamma
_{2},...,\gamma_{k}\}$ coincides with the span of the vectors $e_{1}%
=(1,0,0,...,0)$, $e_{2}=(0,1,0,...,0),...,$ $e_{k}$. Then $\gamma_{s}%
=\sum_{i=1}^{k}\gamma_{s,i}e_{i}$ for $s=1,2,...,k$ . Therefore the relation
$x\in\cap_{i=1}^{k}V_{\gamma_{i}}(\rho^{\alpha_{k}})$ implies that
\[
\sum_{i=1}^{k}\gamma_{s,i}x_{i}=O(\rho^{\alpha_{k}}),s=1,2,...,k;\text{ }%
x_{n}=\frac{\det(b_{j,i}^{n})}{\det(\gamma_{j,i})}\text{, }n=1,2,...,k,
\]
where $x=(x_{1},x_{2},...,x_{d}),\gamma_{j}=(\gamma_{j,1},\gamma
_{j,2},...,\gamma_{j,k},0,0,...,0),$ $b_{j,i}^{n}=\gamma_{j,i}$ for $n\neq j$
and $b_{j,i}^{n}=O(\rho^{\alpha_{k}})$ for $n=j.$ Taking into account that the
determinant $\det(\gamma_{j,i})$ is the volume of the parallelepiped
$\{\sum_{i=1}^{k}b_{i}\gamma_{i}:b_{i}\in\lbrack0,1],i=1,2,...,k\}$ and using
$\mid\gamma_{j,i}\mid<p\rho^{\alpha}$ ( since $\gamma_{j}\in\Gamma
(p\rho^{\alpha})$ ), we get the estimations
\begin{equation}
x_{n}=O(\rho^{\alpha_{k}+(k-1)\alpha})\text{ ,}\forall n=1,2,...,k;\text{
}\forall x\in\cap_{i=1}^{k}V_{\gamma_{i}}(\rho^{\alpha_{k}}).
\end{equation}
Hence (31) holds. Therefore, using the inequalities $\mid a+b\mid\geq\frac
{1}{2}$ $\rho^{\frac{1}{2}\alpha_{k+1}}$ ( see above), $\alpha_{k+1}%
>2(\alpha_{k}+(k-1)\alpha)$ ( see the \ seventh inequality in (15)), and the
obvious equalities $(y,v)=(y,a)=(y,b)=0,$%
\begin{equation}
\mid\gamma+t+a+b\mid^{2}-\mid\gamma+t\mid^{2}=\mid a+b+v\mid^{2}-\mid
v\mid^{2},
\end{equation}
we obtain the estimation (30).

Case 2. $a\notin P.$ First we show that
\begin{equation}
\mid\mid\gamma+t+a\mid^{2}-\mid\gamma+t\mid^{2}\mid\geq\rho^{\alpha_{k+1}}.
\end{equation}
Suppose, to the contrary, that it does not hold. Then $\gamma+t\in V_{a}%
(\rho^{\alpha_{k+1}}).$ On the other hand $\gamma+t\in\cap_{i=1}^{k}%
V_{\gamma_{i}}(\rho^{\alpha_{k+1}})$ ( see the conditions of Lemma 1).
Therefore we have $\gamma+t\in E_{k+1}$ which contradicts the conditions of
the lemma. \ So (34) is proved. Now, to prove (30) we write the difference
$\mid\gamma+t+a+b\mid^{2}-\mid\gamma+t\mid^{2}$ as the sum of $d_{1}\equiv
\mid\gamma+t+a+b\mid^{2}-\mid\gamma+t+b\mid^{2}$ and $d_{2}\equiv\mid
\gamma+t+b\mid^{2}-\mid\gamma+t\mid^{2}.$ Since $d_{1}=\mid\gamma+t+a\mid
^{2}-\mid\gamma+t\mid^{2}+2(a,b),$ it follows from the inequalities (34), (29)
that $\mid d_{1}\mid>\frac{2}{3}$ $\rho^{\alpha_{k+1}}$. On the other hand,
taking $a=0$ in (33), we have $d_{2}=\mid b+v\mid^{2}-\mid v\mid^{2}.$
Therefore (31), the first inequality in (29) and the \ seventh inequality in
(15) imply that $\mid d_{2}\mid<\frac{1}{3}$ $\rho^{\alpha_{k+1}},$ $\mid
d_{1}\mid-\mid d_{2}\mid>\frac{1}{3}\rho^{\alpha_{k+1}},$ that is, (30) holds
\end{proof}

\begin{theorem}
$(a)$ Suppose $\gamma+t\in(\cap_{i=1}^{k}V_{\gamma_{i}}(\rho^{\alpha_{k}%
}))\backslash E_{k+1},$ where $k=1,2,...,d-1.$ If (6) and (7) hold, then there
is an index $j$ such that
\begin{equation}
\Lambda_{N}(t)=\lambda_{j}(\gamma+t)+O(\rho^{-(p-c-\frac{1}{4}d3^{d})\alpha}),
\end{equation}
where $\lambda_{1}(\gamma+t)\leq\lambda_{2}(\gamma+t)\leq...\leq\lambda
_{b_{k}}(\gamma+t)$ are the eigenvalues of the matrix $C(\gamma+t,\gamma
_{1},\gamma_{2},...,\gamma_{k})$ defined in (27).

$(b)$ Every eigenvalue $\Lambda_{N}(t)$ of the operator $L_{t}(q(x))$
satisfies either (5) or (35) for $c=\frac{q(d-1)}{2}.$
\end{theorem}

\begin{proof}
$(a)$Writing the equation (17) for all $h_{i}+t\in B_{k}(\gamma+t,p_{1}),$ we
obtain%
\begin{equation}
(\Lambda_{N}-\mid h_{i}+t\mid^{2})b(N,h_{i})=\sum_{\gamma^{^{\prime}}\in
\Gamma(\rho^{\alpha})}q_{\gamma^{^{\prime}}}b(N,h_{i}-\gamma^{^{\prime}%
})+O(\rho^{-p\alpha})
\end{equation}
for $i=1,2,...,b_{k}$ ( see (26) for definition of $B_{k}(\gamma+t,p_{1})$).
It follows from (6) and Lemma 1 that if $(h_{i}-\gamma^{^{\prime}}+t)\notin
B_{k}(\gamma+t,p_{1}),$ then
\[
\mid\Lambda_{N}-\mid h_{i}-\gamma^{^{\prime}}-\gamma_{1}-\gamma_{2}%
-...-\gamma_{s}+t\mid^{2}\mid>\frac{1}{6}\rho^{\alpha_{k+1}},
\]
where $\gamma^{^{\prime}}\in\Gamma(\rho^{\alpha}),\gamma_{j}\in\Gamma
(\rho^{\alpha}),$ $j=1,2,...,s$ and $s=0,1,...,p_{1}-1.$ Therefore, applying
the formula (18) $p_{1}$ times, using (4) and $p_{1}\alpha_{k+1}>p_{1}%
\alpha_{1}\geq p\alpha$ ( see the\ fourth inequality in (15)), we see that if
$(h_{i}-\gamma^{^{\prime}}+t)\notin B_{k}(\gamma+t,p_{1}),$ then
\begin{equation}
b(N,h_{i}-\gamma^{^{\prime}})=\nonumber
\end{equation}%
\begin{equation}
\sum_{\gamma_{1},...,\gamma_{p_{1}-1}\in\Gamma(\rho^{\alpha})}\dfrac
{q_{\gamma_{1}}q_{\gamma_{2}}...q_{\gamma_{p_{1}}}b(N,h_{i}-\gamma^{^{\prime}%
}-\sum_{i=1}^{p_{1}}\gamma_{i})}{\prod_{j=0}^{p_{1}-1}(\Lambda_{N}-\mid
h_{i}-\gamma^{^{\prime}}+t-\sum_{i=1}^{j}\gamma_{i}\mid^{2})}+
\end{equation}%
\[
+O(\rho^{-p\alpha})=O(\rho^{p_{1}\alpha_{k+1}})+O(\rho^{-p\alpha}%
)=O(\rho^{-p\alpha}).
\]
Hence (36) has the form%
\[
(\Lambda_{N}-\mid h_{i}+t\mid^{2})b(N,h_{i})=\sum_{\gamma^{^{\prime}}%
}q_{\gamma^{^{\prime}}}b(N,h_{i}-\gamma^{^{\prime}})+O(\rho^{-p\alpha
}),i=1,2,...,b_{k},
\]
where the summation is taken under the conditions $\gamma^{^{\prime}}\in
\Gamma(\rho^{\alpha})$ and

$h_{i}-\gamma^{^{\prime}}+t\in B_{k}(\gamma+t,p_{1})$. It can be written in
matrix form
\[
(C-\Lambda_{N}I)(b(N,h_{1}),b(N,h_{2}),...b(N,h_{b_{k}}))=O(\rho^{-p\alpha}),
\]
where the right-hand side of this system is a vector having the norm

$\parallel O(\rho^{-p\alpha})\parallel=O(\sqrt{b_{k}}\rho^{-p\alpha})$. Now,
taking into account that

$\gamma+t\in\{h_{i}+t:i=1,2,...,b_{k}\}$ and (7) holds, we have
\begin{align}
c_{4}\rho^{-c\alpha}  &  <(\sum_{i=1}^{b_{k}}\mid b(N,h_{i})\mid^{2}%
)^{\frac{1}{2}}\leq\parallel(C-\Lambda_{N}I)^{-1}\parallel\sqrt{b_{k}}%
c_{6}\rho^{-p\alpha},\\
\max_{i=1,2,...,b_{k}}  &  \mid\Lambda_{N}-\lambda_{i}\mid^{-1}=\parallel
(C-\Lambda_{N}I)^{-1}\parallel>c_{4}c_{6}^{-1}b_{k}^{-\frac{1}{2}}%
\rho^{-c\alpha+p\alpha}.
\end{align}
Since $b_{k}$ is the number of the vectors of $B_{k}(\gamma+t,p_{1}),$ it
follows from the definition of $B_{k}(\gamma+t,p_{1})$ ( see (26)) and the
obvious relations $\mid B_{k}\mid=O(\rho^{\frac{k}{2}\alpha_{k+1}}),$

$\mid\Gamma(p_{1}\rho^{\alpha})\mid=O(\rho^{d\alpha})$ and $d\alpha<\frac
{1}{2}\alpha_{d}$ ( see the end of introduction), we get
\begin{equation}
b_{k}=O(\rho^{d\alpha+\frac{k}{2}\alpha_{k+1}})=O(\rho^{\frac{d}{2}\alpha_{d}%
})=O(\rho^{\frac{d}{2}3^{d}\alpha}),\forall k=1,2,...,d-1
\end{equation}
Thus formula (35) follows from (39) and (40).

$(b)$ Let $\Lambda_{N\text{ }}(t)$ be any eigenvalue of the operator
$L_{t}(q(x))$ such that

$\sqrt{\Lambda_{N}(t)}\in(\frac{3}{4}\rho,\frac{5}{4}\rho).$ Denote by $D$ the
set of all vectors $\gamma\in\Gamma$ satisfying (6). From (16), arguing as in
the proof of Theorem 1($b$), we obtain

$\sum_{\gamma\in D}\mid b(N,\gamma)\mid^{2}=1-O(\rho^{-2\alpha_{1}}).$ Since
$\mid D\mid=O(\rho^{d-1})$ ( see the end of the introduction), there exists
$\gamma\in D$ such that

$\mid b(N,\gamma)\mid>c_{7}\rho^{-\frac{(d-1)}{2}}=c_{7}\rho^{-\frac
{(d-1)q}{2}\alpha}$, that is, condition (7) for $c=\frac{(d-1)q}{2}$ holds.
Now the proof of $(b)$ follows from Theorem 1 $(a)$ and Theorem 2$(a),$ since
either $\gamma+t$ $\in U(\rho^{\alpha_{1}},p)$ or $\gamma+t\in$ $E_{k}%
\backslash E_{k+1}$ for $k=1,2,...,d-1$ ( see (43))
\end{proof}

\begin{remark}
Here we note that the non-resonance domain

$U(c_{8}\rho^{\alpha_{1}},p)=(R(\frac{3}{2}\rho)\backslash R(\frac{1}{2}%
\rho))\backslash\bigcup_{\gamma_{1}\in\Gamma(p\rho^{\alpha})}V_{\gamma_{1}%
}(c_{8}\rho^{\alpha_{1}}),$ where

$V_{\gamma_{1}}(c_{8}\rho^{\alpha_{1}})\equiv\{x:\mid\mid x\mid^{2}-\mid
x+\gamma_{1}\mid^{2}\mid<c_{8}\rho^{\alpha_{1}}\}\cap(R(\frac{3}{2}%
\rho)\backslash R(\frac{1}{2}\rho)),$ has an asymptotically full measure on
$\mathbb{R}^{d}$ in the sense that $\frac{\mu(U\cap B(\rho))}{\mu(B(\rho))}$
tends to $1$ as $\rho$ tends to infinity, where $B(\rho)=\{x\in\mathbb{R}%
^{d}:\mid x\mid=\rho\}.$ Clearly, $B(\rho)\cap V_{b}(c_{8}\rho^{\alpha_{1}})$
is the part of sphere $B(\rho),$ which is contained between two parallel hyperplanes

$\{x:\mid x\mid^{2}-\mid x+b\mid^{2}=-c_{8}\rho^{\alpha_{1}}\}$ and $\{x:\mid
x\mid^{2}-\mid x+b\mid^{2}=c_{8}\rho^{\alpha_{1}}\}.$ The distance of these
hyperplanes from origin is $O(\frac{\rho^{\alpha_{1}}}{\mid b\mid}).$
Therefore, the relations $\mid\Gamma(p\rho^{\alpha})\mid=O(\rho^{d\alpha}),$
and $\alpha_{1}+d\alpha<1-\alpha$ ( see the first inequality in (15)) imply
\begin{align}
\mu(B(\rho)\cap V_{b}(c_{8}\rho^{\alpha_{1}}))  &  =O(\frac{\rho^{\alpha
_{1}+d-2}}{\mid b\mid}),\text{ }\mu(E_{1}\cap B(\rho))=O(\rho^{d-1-\alpha}),\\
\mu(U(c_{8}\rho^{\alpha_{1}},p)\cap B(\rho))  &  =(1+O(\rho^{-\alpha}%
))\mu(B(\rho)).
\end{align}
If $x\in\cap_{i=1}^{d}V_{\gamma_{i}}(\rho^{\alpha_{d}}),$ then (32) holds for
$k=d$ and $n=1,2,...,d.$ Hence we have $\mid x\mid=O(\rho^{\alpha
_{d}+(d-1)\alpha}).$ It is impossible, since $\alpha_{d}+(d-1)\alpha<1$ ( see
the \ sixth inequality in (15)) and $x\in B(\rho).$ It means that $(\cap
_{i=1}^{d}V_{\gamma_{i}}(\rho^{\alpha_{k}}))\cap B(\rho)=\emptyset$ for
$\rho\gg1$. Thus for $\rho\gg1$ we have
\begin{equation}
R(\frac{3}{2}\rho)\backslash R(\frac{1}{2}\rho)=(U(\rho^{\alpha_{1}}%
,p)\cup(\cup_{s=1}^{d-1}(E_{s}\backslash E_{s+1}))).
\end{equation}

Note that everywhere in this paper we use the big parameter $\rho.$ All
considered eigenvalues $\left\vert \gamma+t\right\vert ^{2}$ of $L_{t}(0)$
satisfy the relations $\frac{1}{2}\rho<\left\vert \gamma+t\right\vert
<\frac{3}{2}\rho.$ Therefore in the asymptotic formulas instead of $O(\rho
^{a})$ one can take \ $O(\left\vert \gamma+t\right\vert ^{a}).$ For
simplicity, we often use $O(\rho^{a}).$ It is clear that the asymptotic
formulas hold true if we replace $U(\rho^{\alpha_{1}},p)$ by $U(c_{8}%
\rho^{\alpha_{1}},p).$ Since

$V_{b}(\frac{1}{2}\rho^{\alpha_{1}})\subset(R(\frac{3}{2}\rho)\backslash
R(\frac{1}{2}\rho))\cap W_{b,\alpha_{1}}(1)\subset$ $V_{b}(\frac{3}{2}%
\rho^{\alpha_{1}}),$\ in all considerations the resonance domain $V_{b}%
(\rho^{\alpha_{1}})$ can be replaced by $W_{b,\alpha_{1}}(1)\cap(R(\frac{3}%
{2}\rho)\backslash R(\frac{1}{2}\rho)).$
\end{remark}

\begin{remark}
Here we note some properties of the known parts

$\mid\gamma+t\mid^{2}+F_{k}(\gamma+t)$ (see Theorem 1) and $\lambda_{j}%
(\gamma+t)$ ( see Theorem 2) of the non-resonance and resonance eigenvalues of
$L_{t}(q(x)).$ Denoting $\gamma+t$ by $x$ , where $\gamma+t\in U(\rho
^{\alpha_{1}},p),$ we prove
\begin{equation}
\frac{\partial F_{k}(x)}{\partial x_{i}}=O(\rho^{-2\alpha_{1}+\alpha}),\forall
i=1,2,...,d;\forall k=1,2,...
\end{equation}
by induction on $k.$ For $k=1$ the formula (44) follows from (4) and
\begin{equation}
\frac{\partial}{\partial x_{i}}(\dfrac{1}{\mid x\mid^{2}-\mid x-\gamma_{1}%
\mid^{2}})=\dfrac{-2\gamma_{1}(i)}{(\mid x\mid^{2}-\mid x-\gamma_{1}\mid
^{2})^{2}}=O(\rho^{-2\alpha_{1}+\alpha}),
\end{equation}
where $\gamma_{1}(i)$ is the $i$th component of the vector $\gamma_{1}%
\in\Gamma(p\rho^{\alpha})$ hence is equal to $O(\rho^{\alpha}).$ Now suppose
that (44) holds for $k=s.$ Using this and (24), replacing $\mid x\mid^{2}$ by
$\mid x\mid^{2}+F_{s}(x)$ in (45) and evaluating as above we obtain
\[
\frac{\partial}{\partial x_{i}}(\dfrac{1}{\mid x\mid^{2}+F_{s}-\mid
x-\gamma_{1}\mid^{2}})=\dfrac{-2\gamma_{1}(i)+\frac{\partial F_{s}%
(x)}{\partial x_{i}}}{(\mid x\mid^{2}+F_{s}-\mid x-\gamma_{1}\mid^{2})^{2}%
}=O(\rho^{-2\alpha_{1}+\alpha}).
\]
This formula together with the definition of $F_{k}$ give (44) for $k=s+1.$

Now denoting $\lambda_{i}(\gamma+t)-\mid\gamma+t\mid^{2}$ by $r_{i}(\gamma+t)$
we prove that
\begin{equation}
\mid r_{i}(x)-r_{i}(x^{^{\prime}})\mid\leq2\rho^{\frac{1}{2}\alpha_{d}}\mid
x-x^{^{\prime}}\mid,\forall i.
\end{equation}
Clearly $r_{1}(x)\leq r_{2}(x)\leq...\leq$ $r_{b_{k}}(x)$ are the eigenvalue
of the matrix

$C(x)-\mid x\mid^{2}I\equiv C^{^{\prime}}(x),$ where $C(x)$ is defined in
(27). By definition, only the diagonal elements of the matrix $C^{^{\prime}%
}(x)$ depend on $x$ and they are

$\mid x+a_{i}\mid^{2}-\mid x\mid^{2}=2(x,a_{i})+\mid a_{i}\mid^{2},$ where
$a_{i}=h_{i}+t-\gamma-t$ and $h_{i}+t\in B_{k}(\gamma+t,p_{1}).$ It follows
from the definitions of $B_{k}(\gamma+t,p_{1})$ for $k<d$ ( see (26)), and
$\alpha_{d}$ ( see introduction) that $\mid a_{i}\mid<\frac{1}{2}\rho
^{\frac{1}{2}\alpha_{k}}+p_{1}\rho^{\alpha}<\rho^{\frac{1}{2}\alpha_{d}}$.
Using this and taking into account that $C^{^{\prime}}(x)-C^{^{\prime}%
}(x^{^{\prime}})=(a_{i,j}),$ where $a_{i,i}=2(x-x^{^{\prime}},a_{i}),$
$a_{i,j}=0$ for $i\neq j,$ we obtain $\parallel C^{^{\prime}}(x)-C^{^{\prime}%
}(x^{^{\prime}})\parallel\leq2\rho^{\frac{1}{2}\alpha_{d}}\mid x-x^{^{\prime}%
}\mid$ from which follows (46).
\end{remark}

\section{ Bloch Eigenvalues near\ the Diffraction Planes}

In this section we obtain the asymptotic formulae for the eigenvalues
corresponding to the quasimomentum $\gamma+t$ lying near the diffraction hyperplane

$D_{\delta}=\{x\in\mathbb{R}^{d}:\mid x\mid^{2}=\mid x-\delta\mid^{2}\}$. In
section 2 to obtain the asymptotic formula for the non-resonance eigenvalues (
that is, for the eigenvalues corresponding to the quasimomentum $\gamma+t$
lying far from the diffraction planes) we considered the operator
$L_{t}(q(x))$ as perturbation of the operator $L_{t}(0)$ with $q(x).$ As a
result the asymptotic formulas for these eigenvalues of $L_{t}(q(x))$ is
expressed in the term of the eigenvalues of $L_{t}(0)$. To obtain the
asymptotic formulae for the eigenvalues corresponding to the quasimomentum
$\gamma+t$ lying near the diffraction plane $D_{\delta}$ we consider the
operator $L_{t}(q(x))$ as the perturbation of the operator $L_{t}(q^{\delta
}(x)),$ where the directional potential $q^{\delta}(x)$ is defined in (9),
with $q(x)-q^{\delta}(x).$ So it is natural that the asymptotic formulas,
which will be obtained in this section, for these eigenvalues will be
expressed in the term of the eigenvalues of the operator $L_{t}(q^{\delta
}(x)).$ Therefore first of all we need to investigate the eigenvalues and
eigenfunctions of the operator $L_{t}(q^{\delta}(x)).$ Without loss of
generality it can be assumed that $\delta$ is the maximal element of $\Gamma$,
that is, $\delta$ is the element of $\Gamma$ of minimal norm belonging to the
line $\delta\mathbb{R}$, since it is easy to verify that $V_{k\delta}%
(\rho^{\alpha_{1}})\subset V_{\delta}(\rho^{\alpha_{1}})$ for $k=\pm
2,\pm3,....$ Note that $\delta$ is the maximal element of $\Gamma$ if
$\{(\delta,\omega):\omega\in\Omega\}=2\pi Z.$ Let $\Omega_{\delta\text{ }}$be
the sublattice $\{h\in\Omega:(h,\delta)=0\}$ of $\Omega$ in the hyperplane
$H_{\delta}=$ $\{x\in\mathbb{R}^{d}:(x,\delta)=0\}$, and $\Gamma_{\delta
}\equiv\{a\in H_{\delta}:(a,k)\in2\pi Z,\forall$ $k\in\Omega_{\delta}\}$ be
the dual lattice of $\Omega_{\delta}.$ Denote by $F_{\delta}$ the fundamental
domain $H_{\delta}/\Gamma_{\delta}$ of $\Gamma_{\delta}.$ Then $t\in F^{\ast}$
has a unique decomposition
\begin{equation}
t=a+\tau+\mid\delta\mid^{-2}(t,\delta)\delta,
\end{equation}
where $a\in\Gamma_{\delta},$ $\tau\in F_{\delta}.$ Define the sets
$\Omega^{^{\prime}}$ and $\Gamma^{^{\prime}}$ by $\Omega^{^{\prime}%
}=\{h+l\delta^{\ast}:h\in\Omega_{\delta},l\in Z\},$ and by $\Gamma^{^{\prime}%
}=\{b+(p-(2\pi)^{-1}(b,\delta^{\ast}))\delta:b\in\Gamma_{\delta},p\in Z\},$
where $\delta^{\ast}$ is the element of $\Omega$ satisfying $(\delta^{\ast
},\delta)=2\pi.$

\begin{lemma}
$(a)$ The following relations hold: $\Omega=\Omega^{^{\prime}},$
$\Gamma=\Gamma^{^{\prime}}.$

$(b)$ The eigenvalues and eigenfunctions of the operator $L_{t}(q^{\delta
}(x))$ are
\[
\lambda_{j,\beta}(v,\tau)=\mid\beta+\tau\mid^{2}+\mu_{j}(v(\beta,t)),\text{
}\Phi_{j,\beta}(x)=e^{i(\beta+\tau,x)}\varphi_{j,v(\beta,t))}(\zeta)
\]
for $j\in Z,$ $\beta\in\Gamma_{\delta},$ where $v(\beta,t)$ is the fractional
part of

$\mid\delta\mid^{-2}(t,\delta)-(2\pi)^{-1}(\beta-a,\delta^{\ast}),$ $\tau$ and
$a$ are uniquely determined from decomposition (47), $\zeta=(\delta,x)$, and
$\mu_{j}(v(\beta,t)),\varphi_{j,v(\beta,t)}(\zeta)$ are eigenvalues and
corresponding normalized eigenfunctions of the operator $T_{v(\beta
,t)}(Q(\zeta))$ (see (9)) generated by the boundary value problem
\[
-\mid\delta\mid^{2}y^{\prime\prime}(\zeta)+Q(\zeta)y(\zeta)=\mu y(\zeta
),\text{ }y(\zeta+2\pi)=e^{i2\pi v}y(\zeta),
\]
where, for simplicity of the notation, instead of $v(\beta,t)$ we write
$v(\beta)$ (or $v)$ if $t$ (or $t$ and $\beta$), for which we consider
$v(\beta,t),$ is unambiguous.
\end{lemma}

\begin{proof}
$(a)$ For each vector $\omega$ of the lattice $\Omega$ assign $h=\omega
-(2\pi)^{-1}(\omega,\delta)\delta^{\ast}.$ Using the relations $(\omega
,\delta)\equiv2\pi l\in2\pi Z,$ and $(\delta^{\ast},\delta)=2\pi$ we see that
$h\in\Omega$ and $(h,\delta)=0$ ,i.e., $h\in\Omega_{\delta}.$ Hence
$\Omega\subset\Omega^{^{\prime}}.$ Now for each vector $\gamma$ of the lattice
$\Gamma$ assign $b=\gamma-\mid\delta\mid^{-2}(\gamma,\delta)\delta.$ It is not
hard to verify that $b\in H_{\delta}$ and $(b,\omega)=(\gamma,\omega)\in2\pi
Z$ for $\omega\in\Omega_{\delta}$ $\subset\Omega.$ Therefore $b\in
\Gamma_{\delta}.$ Moreover

$(b,\delta^{\ast})=(\gamma,\delta^{\ast})-2\pi(\gamma,\delta)\mid\delta
\mid^{-2}.$ Since $(\gamma,\delta^{\ast})\in2\pi Z,$ that is, $(\gamma
,\delta^{\ast})=2\pi n$ where $n\in Z$, we have $(\gamma,\delta)\mid\delta
\mid^{-2}=n-(2\pi)^{-1}(b,\delta^{\ast}).$ Therefore we obtain an orthogonal
decomposition
\begin{equation}
\gamma=b+\langle\gamma,\frac{\delta}{\mid\delta\mid}\rangle\frac{\delta}%
{\mid\delta\mid}=b+(n-(2\pi)^{-1}(b,\delta^{\ast}))\delta
\end{equation}
of $\gamma\in\Gamma$, where $b\in\Gamma_{\delta},$ and $n\in Z.$ Hence
$\Gamma\subset\Gamma^{^{\prime}}.$ On the other hand if $b\in\Gamma_{\delta},$
$h\in\Omega_{\delta}$ and $n,l\in Z$ then $(h+l\delta^{\ast},b+(n-(2\pi
)^{-1}(b,\delta^{\ast}))\delta)=(h,b)+2\pi nl\in2\pi Z.$ Thus, we have the
relations ( see definition of the sets $\Omega^{^{\prime}},\Gamma^{^{\prime}}$
)
\begin{equation}
\Omega\subset\Omega^{^{\prime}},\Gamma\subset\Gamma^{^{\prime}},(\omega
^{^{\prime}},\gamma^{^{\prime}})\in2\pi Z,\forall\omega^{^{\prime}}\in
\Omega^{^{\prime}},\forall\gamma^{^{\prime}}\in\Gamma^{^{\prime}}.
\end{equation}
Since $\Omega$ is the set of all vectors $\omega\in\mathbb{R}^{d}$ satisfying
$(\omega,\gamma)\in2\pi Z$ for all $\gamma\in\Gamma$ and $\Gamma$ is the set
of all vectors $\gamma\in\mathbb{R}^{d}$ satisfying $(\omega,\gamma)\in2\pi Z$
for all $\omega\in\Omega$ the relations in (49) imply $\Omega^{^{\prime}%
}\subset\Omega,$ $\Gamma^{^{\prime}}\subset\Gamma$ and hence $\Omega
=\Omega^{^{\prime}},\Gamma=\Gamma^{^{\prime}}$.

$(b)$ Since $\beta+\tau$ is orthogonal to $\delta,$ turning the coordinate
axis so that $\delta$ coincides with one of the coordinate axis and taking
into account that the Laplace operator is invariant under rotation, one can
easily verify that
\[
(-\Delta+q^{\delta}(x))\Phi_{j,\beta}(x)=\lambda_{j,\beta}\Phi_{j,\beta}(x)
\]
Now using the relation$(\delta,\omega)=2\pi l,$ where $\omega\in\Omega,$ $l\in
Z,$ and the definitions of $\Phi_{j,\beta}(x),\varphi_{j,v}(\delta,x)$ we
obtain
\[
\Phi_{j,\beta}(x+\omega)=e^{i(\beta+\tau,x+\omega)}\varphi_{j,v}%
(\delta,x+\omega)=\Phi_{j,\beta}(x)e^{i(\beta+\tau,\omega)+i2\pi lv(\beta
,t)}.
\]
Replacing $\tau$ and $\omega$ by $t-a-\mid\delta\mid^{-2}(t,\delta)\delta$ and
$h+l\delta^{\ast}$, where $h\in\Omega_{\delta},l\in Z,$ (see (47) and the
first equality of $(a)$) respectively, and then using

$(h,\delta)=0,$ $(\delta^{\ast},\delta)=2\pi$ one can easily verify that

$(\beta+\tau,\omega)=(t,\omega)+(\beta-a,h)-2\pi l[\mid\delta\mid
^{-2}(t,\delta)-(2\pi)^{-1}(\beta-a,\delta^{\ast})].$ From this using that
$(\beta-a,h)\in2\pi Z,$ ( since $\beta-a\in\Gamma_{\delta},$ $h\in
\Omega_{\delta}$), and $v(\beta,t)$ is a fractional part of the expression in
the last square bracket, we infer $\Phi_{j,\beta}(x+\omega)=e^{i(t,\omega
)}\Phi_{j,\beta}(x).$ Thus $\Phi_{j,\beta}(x)$ is an eigenfunction of
$L_{t}(q^{\delta}(x)).$

Now we prove that the system $\{\Phi_{j,\beta}(x):j\in Z,\beta\in
\Gamma_{\delta}\}$ contains all eigenfunctions of $L_{t}(q^{\delta}(x))$.
Assume the converse. Then there exists a nonzero function $f(x)\in L_{2}(F),$
which is orthogonal to all elements of this system. Using (47), (48) of and
the definition of $v(\beta,t)$ ( see Lemma 2(b)) we get%
\begin{equation}
\gamma+t=\beta+\tau+(j+v)\delta,
\end{equation}
where $\beta\in\Gamma_{\delta},\tau\in F_{\delta},$ $j\in Z,$ and
$v=v(\beta,t).$ Since $e^{i(j+v)\zeta}$ can be decomposed by basis
$\{\varphi_{j,v(\beta,t))}(\zeta):$ $j\in Z\}$ the function $e^{i(\gamma
+t,x)}=e^{i(\beta+\tau,x)}e^{i(j+v)\zeta}$ (see (50)) can be decomposed by
system $\{\Phi_{j,\beta}(x)=e^{i(\beta+\tau,x)}\varphi_{j,v(\beta,t))}%
(\zeta):$ $j\in Z\}.$ Then the above assumption $(\Phi_{j,\beta}(x),$
$f(x))=0$ for $j\in Z,\beta\in\Gamma_{\delta}$ implies that $(f(x),e^{i(\gamma
+t,x)})=0$ for all $\gamma\in$ $\Gamma.$ This is impossible, since the system
$\{e^{i(\gamma+t,x)}:\gamma\in\Gamma$ $\}$ is a basis of $L_{2}(F)$
\end{proof}

\begin{remark}
\bigskip Clearly every vectors $x$ of $\mathbb{R}^{d}$ has decompositions
$x=\gamma+t$ ,where $\gamma\in\Gamma,t\in F,$ and $x=\beta+\tau+(j+v)\delta$,
where $\beta\in\Gamma_{\delta},$ $\tau\in F_{\delta},$ $j\in Z,$ $v\in
\lbrack0,1).$ We say that the first and second decompositions are $\Gamma$ and
$\Gamma_{\delta}$ decompositions respectively. Writing $\gamma+t\equiv
\beta+\tau+(j+v(\beta,t))\delta$ (see (50)) we mean the $\Gamma_{\delta}$
decomposition of $\gamma+t.$ As it is noted in lemma 2 instead of $v(\beta,t)$
we write $v(\beta)$ (or $v)$ if $t$ (or $t$ and $\beta$), for which we
consider $v(\beta,t),$ is unambiguous.
\end{remark}

In section 2 we proved that if $\gamma+t\notin V_{\delta}(\rho^{\alpha_{1}})$
for all $\delta\in\Gamma(p\rho^{\alpha})$ then there is an eigenvalue
$\Lambda_{N}$ of $L_{t}(q(x)),$ which is close to the eigenvalue $\mid
\gamma+t\mid^{2}$ of $L_{t}(0),$ that is, the influence of the perturbation
$q(x)$ is not significant. If $\gamma+t\in V_{\delta}(\rho^{\alpha_{1}%
})\backslash E_{2}$, then using (50) and $\alpha_{1}=3\alpha,$ we get
\begin{equation}
\mid(j+v)\delta\mid<r_{1},\text{ }\mid j\delta\mid<r_{1},\text{ }r_{1}%
>2\rho^{\alpha},
\end{equation}
where $r_{1}=\frac{\rho^{\alpha_{1}}}{\mid2\delta\mid}+\mid2\delta\mid.$ To
the eigenvalue $\mid\gamma+t\mid^{2}=\mid\beta+\tau\mid^{2}+\mid
(j+v)\delta\mid^{2}$ (see (50)) of $L_{t}(0)$ assign the eigenvalue
$\lambda_{j,\beta}(v,\tau)=\mid\beta+\tau\mid^{2}+\mu_{j}(v)$ of
$L_{t}(q^{\delta}(x)),$ where $\mid(j+v)\delta\mid^{2}$ for $j\in Z$ is the
eigenvalue of $T_{v}(0)$ and $\mu_{j}(v)$ is the eigenvalue of $T_{v}%
(Q(\zeta))$ ( see lemma 2(b)) satisfying
\begin{equation}
\mid\mu_{j}(v)-\mid(j+v)\delta\mid^{2}\mid\leq\sup\mid Q(\zeta)\mid,\text{
}\forall j\in Z.
\end{equation}
The eigenvalue $\lambda_{j,\beta}(v,\tau)$ of $L_{t}(q^{\delta}(x))$ can be
considered as the perturbation of the eigenvalue \ $\mid\gamma+t\mid^{2}%
=\mid\beta+\tau\mid^{2}+\mid(j+v)\delta\mid^{2}$ of $L_{t}(0)$ by $q^{\delta
}(x).$ Lemma 2(b) shows that for this perturbation the influence of
$q^{\delta}(x)$ is significant for small value of $j.$ Now we prove that there
is an eigenvalue $\Lambda_{N}$ of $L_{t}(q(x))$ which is close to the
eigenvalue $\lambda_{j,\beta}(v,\tau)$ of $L_{t}(q^{\delta}(x)),$ that is, we
prove that the influence of $q(x)-q^{\delta}(x)$ is not significant. To prove
this we consider the operator $L_{t}(q(x))$ as perturbation of the operator
$L_{t}(q^{\delta}(x))$ with $q(x)-q^{\delta}(x)$ and use the formula
\begin{equation}
(\Lambda_{N}-\lambda_{j,\beta})(\Psi_{N,t}(x),\Phi_{j,\beta}(x))=(\Psi
_{N,t}(x),(q(x)-q^{\delta}(x))\Phi_{j,\beta}(x)),
\end{equation}
called binding formula for $L_{t}(q(x))$ and $L_{t}(q^{\delta}(x)),$ which can
be obtained from%
\[
(L_{t}(q^{\delta}(x))+(q(x)-(q^{\delta}(x)))\Psi_{N,t}(x)=\Lambda_{N}%
\Psi_{N,t}(x)
\]
by multiplying by $\Phi_{j,\beta}(x)$ and using $L_{t}(q^{\delta}%
(x))\Phi_{j,\beta}(x)=\lambda_{j,\beta}\Phi_{j,\beta}(x).$ Note that the
binding formula (53) can be obtained from the binding formula (16) by
replacing the perturbation $q(x)$ by the perturbation $q(x)-q^{\delta}(x)$,
the eigenvalues $\mid\gamma+t\mid^{2}$ and eigenfunctions $e^{i(\gamma+t,x)}$
of $L_{t}(0)$ by the eigenvalues and eigenfunctions of the operator
$L_{t}(q^{\delta}(x))$ respectively. Recall that we obtained the asymptotic
formulas for the perturbation of the non-resonance eigenvalue $\mid
\gamma+t\mid^{2}$ by iteration the binding formula (16) for the unperturbed
operator $L_{t}(0)$ and the perturbed operator $L_{t}(q(x))$ ( see section 2).
Similarly, now to obtain the asymptotic formulas for resonance eigenvalue we
iterate the binding formula (53) for the unperturbed operator$\ L_{t}%
(q^{\delta}(x))$ and perturbed operator$\ L_{t}(q(x))$. For this ( as in the
non-resonance case) we decompose $(q(x)-q^{\delta}(x))\Phi_{j,\beta}(x)$ by
the basis $\{\Phi_{j^{^{\prime}},\beta^{^{\prime}}}(x):j^{^{\prime}}\in
Z,\beta^{^{\prime}}\in\Gamma_{\delta}\}$ and put this decomposition into (53).
Let us find this decomposition. Using the decomposition (48) of $\gamma_{1}%
\in\Gamma(\rho^{\alpha})$ and (3), we get

$\gamma_{1}=\beta_{1}+(n_{1}-(2\pi)^{-1}(\beta_{1},\delta^{\ast}))\delta,$
$e^{i(\gamma_{1},x)}=e^{i(\beta_{1},x)}e^{i(n_{1}-(2\pi)^{-1}(\beta_{1}%
,\delta^{\ast}))\zeta},$%
\[
q(x)-Q(\zeta)=\sum\limits_{(n_{1},\beta_{1})\in\Gamma^{^{\prime}}(\rho
^{\alpha})}c(n_{1},\beta_{1})e^{i(\beta_{1},x)}e^{i(n_{1}-(2\pi)^{-1}%
(\beta_{1},\delta^{\ast}))\zeta}+O(\rho^{-p\alpha}),
\]%
\begin{align}
&  (q(x)-Q(\zeta))\Phi_{j,\beta}(x)\\
&  =\sum\limits_{(n_{1},\beta_{1})\in\Gamma^{^{\prime}}(\rho^{\alpha})}%
c(n_{1},\beta_{1})e^{i(\beta_{1}+\beta+\tau,x)}e^{i(n_{1}-(2\pi)^{-1}%
(\beta_{1},\delta^{\ast}))\zeta}\varphi_{j,v(\beta)}(\zeta)+O(\rho^{-p\alpha
}),\nonumber
\end{align}
where $c(n_{1},\beta_{1})=q_{\gamma_{1}},$%
\[
\Gamma^{^{\prime}}(\rho^{\alpha})=\{(n_{1},\beta_{1}):\beta_{1}\in
\Gamma_{\delta}\backslash\{0\},n_{1}\in Z,\beta_{1}+(n_{1}-(2\pi)^{-1}%
(\beta_{1},\delta^{\ast}))\delta\in\Gamma(\rho^{\alpha})\}.
\]
Note that if $(n_{1},\beta_{1})\in\Gamma^{^{\prime}}(\rho^{\alpha}),$ then
$\mid\beta_{1}+(n_{1}-(2\pi)^{-1}(\beta_{1},\delta^{\ast}))\delta\mid
<\rho^{\alpha}$ and
\begin{equation}
\mid\beta_{1}\mid<\rho^{\alpha},\text{ }\mid(n_{1}-(2\pi)^{-1}(\beta
_{1},\delta^{\ast}))\delta\mid<\rho^{\alpha}<\frac{1}{2}r_{1},
\end{equation}
since $\beta_{1}$ is orthogonal to $\delta$ and $r_{1}>2\rho^{\alpha}$ ( see
(51)). To decompose the right-hand side of (54) by basis $\{\Phi_{j^{^{\prime
}},\beta^{^{\prime}}}(x)\}$ we use the following lemma

\begin{lemma}
$(a)$ If integers $j,m$ satisfy the inequalities $\mid m\mid>2\mid j\mid,\mid
m\delta\mid\geq2r,$ then
\begin{align}
(\varphi_{j,v}(\zeta),e^{i(m+v)\zeta})  &  =O(\mid m\delta\mid^{-s-1}%
)=O(\rho^{-(s+1)\alpha}),\\
(\varphi_{m,v},e^{i(j+v)\zeta})  &  =O(\mid m\delta\mid^{-s-1}).
\end{align}
where $r\geq r_{1}$ and $r_{1}$ is defined in (51), $\varphi_{j,v}(\zeta)$ is
the eigenfunctions of the operator $T_{v}(Q(\zeta)),$ and $Q(\zeta)\in
W_{2}^{s}(0,2\pi).$

$(b)$ The set $W(\rho)\equiv\{v\in(0,1):\mid\mu_{j}(v)-\mu_{j^{^{\prime}}%
}(v)\mid>\frac{2}{\ln\rho},$ $\forall j^{^{\prime}},j\in Z,j^{^{\prime}}\neq
j\}$ contains a set $A(\varepsilon(\rho))\equiv(\varepsilon(\rho),\frac{1}%
{2}-\varepsilon(\rho))\cup(\frac{1}{2}+\varepsilon(\rho),1-\varepsilon
(\rho)),$ where $\varepsilon(\rho)>0$ and $\varepsilon(\rho)\rightarrow0$ as
$\rho\rightarrow\infty.$
\end{lemma}

\begin{proof}
$(a).$ To prove (56) we iterate the formula
\begin{equation}
(\mu_{j}(v)-\mid(m+v)\delta\mid^{2})(\varphi_{j,v}(\zeta),e^{i(m+v)\zeta
})=(\varphi_{j,v}(\zeta)Q(\zeta),e^{i(m+v)\zeta}),
\end{equation}
by using the decomposition
\begin{equation}
Q(\zeta)=\sum_{\mid l_{1}\mid<\frac{\mid m\mid}{2s}}q_{l_{1}\delta}%
e^{il_{1}\zeta}+O(\mid m\delta\mid^{-(s-1)})
\end{equation}
Note that (58), (59) is one dimensional case of (16), (3) and the iteration of
(58) is very simple. If $\mid j\mid<\frac{\mid m\mid}{2},$ and$\mid l_{i}%
\mid<\frac{\mid m\mid}{2s}$ for $i=1,2,...k\equiv\lbrack\frac{s}{2}],$ then
the inequalities: $\mid m+v-l_{1}-l_{2}-...-l_{q}\mid-\mid j\mid>\frac{1}%
{5}\mid m\mid,$

$\mid m\mid-\mid j+v-l_{1}-l_{2}-...-l_{q}\mid>\frac{1}{5}\mid m\mid$ hold for
$q=0,1,...,k$. Therefore by (52), we have
\begin{align}
(  &  \mid\mu_{j}-\mid(m-l_{1}-l_{2}-...-l_{q}+v)\delta\mid^{2}\mid
)^{-1}=O(\mid m\delta\mid^{-2}),\\
(  &  \mid\mu_{m}-\mid(j-l_{1}-l_{2}-...-l_{q}+v)\delta\mid^{2}\mid
)^{-1}=O(\mid m\delta\mid^{-2}),
\end{align}
for $q=0,1,...,k$. Iterating (58) $k$ times, by using (60), (61), we get
\begin{equation}
(\varphi_{j},e^{i(m+v)\zeta})=\sum_{\mid l_{1}\delta\mid,\mid l_{2}\delta
\mid,...,\mid l_{k+1}\delta\mid<\frac{\mid m\delta\mid}{2s}}q_{l_{1}\delta
}q_{l_{2}\delta}...q_{l_{k+1}\delta}\times
\end{equation}%
\[
\frac{(\varphi_{j},e^{i(m-l_{1}-l_{2}-...-l_{k+1}+v)\zeta})}{\sqcap_{q=0}%
^{k}(\mu_{j}-\mid(m-l_{1}-l_{2}-...-l_{q}+v)\delta\mid^{2})}+O(\mid
m\delta\mid^{-s-1}).
\]
Now (56) follows from (60), (62), and the first inequality in (4). Formula
(57) can be proved in the same way by using (61) instead of (60). Note that in
(56), and (57) instead of $O(\mid m\delta\mid^{-s-1})$ we can write
$O(\rho^{-(s+1)\alpha}),$ since

$\mid m\delta\mid\geq r\geq r_{1}>2\rho^{\alpha}$ ( see (51)).

$(b).$ During the proof of $(b)$ we numerate the eigenvalue of $T_{v}%
(Q(\zeta))$ in nondecreasing order, i.e., $\mu_{1}(v)\leq\mu_{2}(v)\leq...$.
It is well-known that the spectrum of Hill's operator $T(Q(\zeta))$ consists
of the intervals

$[\mu_{2j-1}(0),\mu_{2j-1}(\frac{1}{2})],[\mu_{2j}(\frac{1}{2}),\mu_{2j}(1)]$
for $j=1,2,....$ The length of the $j$th interval $\Delta_{j\text{ }}$of the
spectrum tends to infinity as $j$ tends to infinity. The distance between
neighbouring intervals, that is the length of gaps in spectrum, tends to zero.
The eigenvalues $\mu_{2j-1}(v)$ and $\mu_{2j}(v)$ are increasing continuous
functions in the intervals $(0,\frac{1}{2})$ and $(\frac{1}{2},1)$
respectively and $\mu_{j}(1+v)=\mu_{j}(v)=\mu_{j}(1-v).$ Since $(\ln\rho
)^{-1}\rightarrow0$ as $\rho\rightarrow\infty,$ the length of the interval
$\Delta_{j\text{ }}$ is sufficiently greater than $(\ln\rho)^{-1}$ for
$\rho\gg1$ and there are numbers $\varepsilon_{j}^{^{\prime}}(\rho
),\varepsilon_{j}^{^{\prime\prime}}(\rho)$ in $(0,\frac{1}{2})$ such that
\begin{align}
\mu_{2j-1}(\varepsilon_{2j-1}^{^{\prime}}(\rho))  &  =\mu_{2j-1}(0)+(\ln
\rho)^{-1},\\
\mu_{2j-1}(\frac{1}{2}-\varepsilon_{j}^{^{\prime\prime}}(\rho))  &
=\mu_{2j-1}(\frac{1}{2})-(\ln\rho)^{-1},\nonumber\\
\mu_{2j}(\frac{1}{2}+\varepsilon_{2j}^{^{\prime}}(\rho))  &  =\mu_{2j}%
(\frac{1}{2})+(\ln\rho)^{-1},\nonumber\\
\mu_{2j}(1-\varepsilon_{j}^{^{\prime\prime}}(\rho))  &  =\mu_{2j}(1)-(\ln
\rho)^{-1}.\nonumber
\end{align}
Denote $\varepsilon^{^{\prime}}(\rho)=\sup_{j}\varepsilon_{j}^{^{\prime}}%
(\rho),$ $\varepsilon^{^{\prime\prime}}(\rho)=\sup_{j}\varepsilon
_{j}^{^{\prime\prime}}(\rho),$ $\varepsilon(\rho)=\max\{\varepsilon^{^{\prime
}}(\rho),$ $\varepsilon^{^{\prime\prime}}(\rho)\}.$ To prove that
$\varepsilon(\rho)\rightarrow0$ as $\rho\rightarrow\infty$ we show that both
$\varepsilon^{^{\prime}}(\rho)$ and $\varepsilon^{^{\prime\prime}}(\rho)$
tends to zero as $\rho\rightarrow\infty.$ If $\rho_{1}<\rho_{2}$ then
$\varepsilon_{j}^{^{\prime}}(\rho_{2})<\varepsilon_{j}^{^{\prime}}(\rho_{1})$
and $\varepsilon^{^{\prime}}(\rho_{2})<\varepsilon^{^{\prime}}(\rho_{1}),$
since $\mu_{2j-1}(v)$ and $\mu_{2j}(v)$ are increasing functions in intervals
$(0,\frac{1}{2})$ and $(\frac{1}{2},1)$ respectively. Hence $\varepsilon
^{^{\prime}}(\rho)\rightarrow a\in\lbrack0,\frac{1}{2}]$ as $\rho
\rightarrow\infty.$ Suppose that $a>0.$ Then there is sequence $\rho
_{k}\rightarrow\infty$ as $k\rightarrow\infty$ such that $\varepsilon
^{^{\prime}}(\rho_{k})>\frac{a}{2}$ for all $k.$ This implies that there is a
sequence $\{i_{k}\}$ and without loss of generality it can be assumed that
there is a sequence $\{2j_{k}-1\}$ of odd numbers such that $\varepsilon
_{2j_{k}-1}^{^{\prime}}(\rho_{k})>\frac{a}{4}$ for all $k.$ Since $\mu
_{2j-1}(v)$ increases in $(0,\frac{1}{2})$ and $\mu_{2j_{k}-1}(\varepsilon
_{2j_{k}-1}^{^{\prime}}(\rho_{k}))-\mu_{2j_{k}-1}(0)=(\ln\rho_{k})^{-1}$ we have

$\mid\mu_{2j_{k}-1}(\frac{a}{4})-\mu_{2j_{k}-1}(0)\mid\leq(\ln\rho_{k}%
)^{-1}\rightarrow0$ as $k\rightarrow\infty,$ which contradicts the well-known
asymptotic formulas for eigenvalues $\mu_{j}(v),$ for $v=0$ and $v=\frac{a}%
{4},$ where $a\in(0,\frac{1}{2}].$ Thus we proved that $\varepsilon^{^{\prime
}}(\rho)\rightarrow0$ as $\rho\rightarrow\infty$. In the same way we prove
this for $\varepsilon^{^{\prime\prime}}(\rho)$, and hence for $\varepsilon
(\rho).$ Now suppose $v\in A(\varepsilon(\rho)).$ Using (63), the definition
of $\varepsilon(\rho),$ and taking into account that $\mu_{2j-1}(v)$ and
$\mu_{2j}(v)$ increase in $(0,\frac{1}{2})$ and $(\frac{1}{2},1)$
respectively, we obtain that the eigenvalues $\mu_{1}(v),\mu_{2}(v),...,$ are
contained in the intervals

$[\mu_{2j-1}(0)+(\ln\rho)^{-1},\mu_{2j-1}(\frac{1}{2})-(\ln\rho)^{-1}],$
$[\mu_{2j}(\frac{1}{2})+(\ln\rho)^{-1},\mu_{2j}(1)-(\ln\rho)^{-1}]$ for
$j=1,2,...,$ and in each interval there exists a unique eigenvalue of $T_{v}.$
Therefore the distance between eigenvalues of $T_{v}$ for $v\in A(\varepsilon
(\rho))$ is not less than the distance between these intervals, which is not
less than $2(\ln\rho)^{-1}.$ Hence the inequality in the definition of
$W(\rho)$ holds, that is, $A(\varepsilon(\rho))\subset W(\rho)$
\end{proof}

\begin{lemma}
If $\mid j\delta\mid<r$ and $(n_{1},\beta_{1})\in\Gamma^{^{\prime}}%
(\rho^{\alpha}),$ then
\begin{align}
&  \ e^{i(n_{1}-(2\pi)^{-1}(\beta_{1},\delta^{\ast}))\zeta}\varphi
_{j,v(\beta)}(\zeta)\nonumber\\
\  &  =\sum_{\mid j_{1}\delta\mid<9r}a(n_{1},\beta_{1},j,\beta,j+j_{1,}%
\beta+\beta_{1})\varphi_{j+j_{1},v(\beta+\beta_{1})}(\zeta)+O(\rho
^{-(s-1)\alpha}),
\end{align}

where $r,$ $\Gamma^{^{\prime}}(\rho^{\alpha})$ are defined in Lemma 3(a),
(54), and

$a(n_{1},\beta_{1},j,\beta,j+j_{1,}\beta+\beta_{1})=(e^{i(n_{1}-(2\pi
)^{-1}(\beta_{1},\delta^{\ast}))\zeta}\varphi_{j,v(\beta)}(\zeta
),\varphi_{j+j_{1},v(\beta+\beta_{1})}(\zeta)).$
\end{lemma}

\begin{proof}
Since $\ e^{i(n_{1}-(2\pi)^{-1}(\beta_{1},\delta^{\ast}))\zeta}\varphi
_{j,v(\beta)}(\zeta)$ is equal to its Fourier series with the orthonormal
basis $\{\varphi_{j+j_{1},v(\beta+\beta_{1})}(\zeta):j_{1}\in Z\}$ it suffices
to show that

$\sum_{j_{1}:\mid j_{1}\delta\mid\geq9r}\mid a(n_{1},\beta_{1},j,\beta
,j+j_{1,}\beta+\beta_{1})\mid=O(\rho^{-(s-1)\alpha}).$ For this we prove
\begin{equation}
\mid a(n_{1},\beta_{1},j,\beta,j+j_{1,}\beta+\beta_{1})\mid=O(\mid j_{1}%
\delta\mid^{-s})
\end{equation}
for all $j_{1}$ satisfying $\mid j_{1}\delta\mid\geq9r$ and take into account
that $r\geq r_{1}>\rho^{\alpha}$ ( see the last inequality in (51)).
Decomposing $\varphi_{j,v(\beta)}$ over $\{e^{i(m+v)\zeta}:m\in Z\}$ and using
the last inequality in (55), we have
\begin{equation}
e^{i(n_{1}-(2\pi)^{-1}(\beta_{1},\delta^{\ast}))\zeta}\varphi_{j}(\zeta
)=\sum_{m\in Z}(\varphi_{j},e^{i(m+v)\zeta})e^{i(m+n+v(\beta+\beta_{1}))\zeta
},
\end{equation}
where $n\in Z$ and $\mid n\delta\mid<r$. This and the decomposition

$\varphi_{j+j_{1}}(\zeta)=\sum_{m\in Z}(\varphi_{j+j_{1}},e^{i(m+v(\beta
+\beta_{1}))\zeta})e^{i(m+v(\beta+\beta_{1}))v}$ imply that
\begin{equation}
a(n_{1},\beta_{1},j,\beta,j+j_{1,}\beta+\beta_{1})=\sum_{m\in Z}(\varphi
_{j},e^{i(m-n+v)\zeta})(\varphi_{j+j_{1}},e^{i(m+v(\beta+\beta_{1}))\zeta})
\end{equation}
Consider two cases: 1. $\mid m\delta\mid>\frac{1}{3}\mid j_{1}\delta\mid
\geq3r$ and 2. $\mid m\mid\leq\frac{1}{3}\mid j_{1}\mid$ . In the first case
we have $\mid(m-n)\delta\mid>2r$ . Therefore (66) implies that
\begin{align*}
(\varphi_{j,v}(\zeta),e^{i(m-n+v)\zeta})  &  =O(\mid m\delta\mid^{-s-1}),\\
\sum_{\mid m\mid>\frac{1}{3}\mid j_{1}\mid}  &  \mid(\varphi_{j}%
,e^{i(m-n+v)\zeta})\mid=O(\mid j_{1}\delta\mid^{-s}).
\end{align*}
In the second case using $\mid j_{1}\delta\mid\geq9r$, $\mid j\delta\mid<r,$
(57), we get$\mid j_{1}+j\mid>2\mid m\mid$,
\begin{align*}
(\varphi_{j+j_{1}},e^{i(m+v(\beta+\beta_{1}))\zeta})  &  =O(\mid
(j_{1}+j)\delta\mid^{-(s-1)})=O(\mid j_{1}\delta\mid^{-s-1}),\\
\sum_{\mid m\mid\leq\frac{1}{3}\mid j_{1}\mid}  &  \mid(\varphi_{j+j_{1}%
}(\zeta),e^{i(m+v(\beta+\beta_{1}))\zeta})\mid=O(\mid j_{1}\delta\mid^{-s}).
\end{align*}
These estimations for these two cases together with (67) yield (65)
\end{proof}

Now it follows from (54) and (64) that
\[
(q(x)-Q(\zeta))\Phi_{j^{^{\prime}},\beta^{^{\prime}}}(x)=O(\rho^{-p\alpha
})+\sum\limits_{(n_{1,}j_{_{1}},\beta_{1})\in G(\rho^{\alpha},9r)}%
c(n_{1},\beta_{1})\times
\]%
\begin{equation}
a(n_{1},\beta_{1},j,\beta^{^{\prime}},j^{^{\prime}}+j_{1,}\beta^{^{\prime}%
}+\beta_{1})e^{i(\beta_{1}+\beta^{^{\prime}}+\tau,x)}\varphi_{j^{^{\prime}%
}+j_{1},v(\beta^{^{\prime}}+\beta_{1})}(\zeta)
\end{equation}
for all $j^{^{\prime}}$ satisfying $\mid j^{^{\prime}}\delta\mid<r,$ where

$G(\rho^{\alpha},9r)=\{(n,j,\beta):\mid j\delta\mid<9r,(n,\beta)\in
\Gamma^{^{\prime}}(\rho^{\alpha}),\beta\neq0\}.$ In (68) the multiplicand
$e^{i(\beta_{1}+\beta^{^{\prime}}+\tau,x)}\varphi_{j^{^{\prime}}+j_{1}%
,v(\beta+\beta_{1})}(\zeta)=\Phi_{j^{^{\prime}}+j_{1,}\beta^{^{\prime}}%
+\beta_{1}}(x)$ does not depend on $n_{1}.$ Its coefficient is
\begin{equation}
\overline{A(j^{^{\prime}},\beta^{^{\prime}},j^{^{\prime}}+j_{1,}%
\beta^{^{\prime}}+\beta_{1}}=\sum\limits_{n_{1}:(n_{1},\beta_{1})\in
\Gamma^{^{\prime}}(\rho^{\alpha})}c(n_{1},\beta_{1})a(n_{1},\beta
_{1},j^{^{\prime}},\beta^{^{\prime}},j^{^{\prime}}+j_{1,}\beta^{^{\prime}%
}+\beta_{1}).
\end{equation}

\begin{lemma}
If $\mid\beta^{^{\prime}}\mid\sim\rho$ and $\mid j^{^{\prime}}\delta\mid<r,$
where $r$ is defined in lemma 3(a), then
\begin{align}
&  (q(x)-Q(\zeta))\Phi_{j^{^{\prime}},\beta^{^{\prime}}}(x)\\
&  =\sum\limits_{(j_{_{1}},\beta_{1})\in Q(\rho^{\alpha},9r)}\overline
{A(j^{^{\prime}},\beta^{^{\prime}},j^{^{\prime}}+j_{1,}\beta^{^{\prime}}%
+\beta_{1}})\Phi_{j^{^{\prime}}+j_{1,}\beta^{^{\prime}}+\beta_{1}}%
(x)+O(\rho^{-p\alpha}),\nonumber
\end{align}
where $Q(\rho^{\alpha},9r)=\{(j,\beta):\mid j\delta\mid<9r,$ $0<\mid\beta
\mid<\rho^{\alpha}\}.$ Moreover,
\begin{equation}
\sum\limits_{(j_{1},\beta_{1})\in Q(\rho^{\alpha},9r)}\mid A(j^{^{\prime}%
},\beta^{^{\prime}},j^{^{\prime}}+j_{1,}\beta^{^{\prime}}+\beta_{1})\mid
<c_{9},
\end{equation}
where $c_{9}$ does not depend on $(j^{^{\prime}},\beta^{^{\prime}}).$
\end{lemma}

\begin{proof}
The formula (70) follows from (68), (69). Now we prove (71). Since

$\sum_{(n_{1},\beta_{1})\in\Gamma^{^{\prime}}(\rho^{\alpha})}\mid
c(n_{1},\beta_{1})\mid\leq\sum_{\gamma}\mid q_{\gamma}\mid<c_{3}$ (see
definition of $c(n_{1},\beta_{1})$ and (4)), it follows from (69) that we need
to prove
\begin{equation}
\sum\limits_{j_{_{1}}}\mid a(n_{1},\beta_{1},j^{^{\prime}},\beta^{^{\prime}%
},j^{^{\prime}}+j_{1,}\beta^{^{\prime}}+\beta_{1})\mid<c_{9}(c_{3})^{-1}.
\end{equation}
For this we use (67) and prove the inequalities:
\begin{align}
\sum_{m\in Z}  &  \mid(\varphi_{j^{^{\prime}}},e^{i(m-n+v(\beta^{^{\prime}%
})\zeta})\mid<c_{10},\\
\sum_{j_{1}\in Z}  &  \mid(\varphi_{j^{^{\prime}}+j_{1}},e^{i(m+v(\beta
_{1}+\beta^{^{\prime}}))\zeta})\mid<c_{11}.
\end{align}
Since the distance between numbers $\mid v\delta\mid^{2},\mid(1+v)\delta
\mid^{2},...,$ for $v\in\lbrack0,1]$ and similarly the distance between
numbers $\mid(-1+v)\delta\mid^{2},\mid(-2+v)\delta\mid^{2},...,$ is not less
than $c_{12}$, it follows from (52) that the number of elements of the sets

$A=\{m:\mid(m-n+v(\beta^{^{\prime}}))\delta\mid^{2}\in\lbrack\mu_{j^{^{\prime
}}}(v(\beta^{^{\prime}}))-1,\mu_{j^{^{\prime}}}(v(\beta^{^{\prime}}))+1]\}$ and

$B=\{j_{1}:\mu_{j^{^{\prime}}+j_{1}}(v(\beta_{1}+\beta^{^{\prime}}))\in
\lbrack\mid(m+v(\beta_{1}+\beta^{^{\prime}}))\delta\mid^{2}-1,\mid
(m+v)\delta\mid^{2}+1]\}$ is less than $c_{13}.$ Now in (73) and (74)
isolating the term with $m\in A$ and $j_{1}\in B$ respectively, applying (58)
to other terms and then using
\begin{align*}
\sum_{m\notin A}\frac{1}{\mid\mu_{j^{^{\prime}}}(v^{^{\prime}})-\mid
(m-n+v^{^{\prime}})\delta\mid^{2}\mid}  &  <c_{14},\\
\sum_{j_{1}\notin B}\frac{1}{\mid\mu_{j^{^{\prime}}+j_{1}}(v_{1}^{^{\prime}%
})-\mid(m+v_{1}^{^{\prime}})\delta\mid^{2}\mid}  &  <c_{14}%
\end{align*}
we get (73), (74), and hence (64). Clearly the constants $c_{14},c_{13}%
,c_{12},c_{11},c_{10}$ can be chosen independently on $(j^{^{\prime}}%
,\beta^{^{\prime}}).$ Therefore $c_{9}$ does not depend on $(j^{^{\prime}%
},\beta^{^{\prime}})$
\end{proof}

Replacing $(j,\beta)$ by $(j^{^{\prime}},\beta^{^{\prime}})$ in (53) and using
(70), we get%

\[
(\Lambda_{N}-\lambda_{j^{^{\prime}},\beta^{^{\prime}}})b(N,j^{^{\prime}}%
,\beta^{^{\prime}})=(\Psi_{N}(x),(q(x)-Q(\zeta))\Phi_{j^{^{\prime}}%
,\beta^{^{\prime}}}(x))=O(\rho^{-p\alpha})
\]

\begin{equation}
+\sum\limits_{(j_{1},\beta_{1})\in Q(\rho^{\alpha},9r)}A(j^{^{\prime}}%
,\beta^{^{\prime}},j^{^{\prime}}+j_{1,}\beta^{^{\prime}}+\beta_{1}%
)b(N,j^{^{\prime}}+j_{1},\beta^{^{\prime}}+\beta_{1})
\end{equation}
for $\mid\beta^{^{\prime}}\mid\sim\rho$ and $\mid j^{^{\prime}}\delta\mid<r,$
where $b(N,j,\beta)=(\Psi_{N}(x),\Phi_{j,\beta}(x)).$ We would like to
emphasize that if $\mid j^{^{\prime}}\delta\mid<r,$ then the summation in (75)
is taken over $Q(\rho^{\alpha},9r).$ Therefore if $\mid j\delta\mid<r_{1}$ (
see (51)), then we have the formula
\[
(\Lambda_{N}-\lambda_{j,\beta})b(N,j,\beta)=O(\rho^{-p\alpha})
\]

\begin{equation}
+\sum\limits_{(j_{1},\beta_{1})\in Q(\rho^{\alpha},9r_{1})}A(j,\beta
,j+j_{1,}\beta+\beta_{1})b(N,j+j_{1},\beta+\beta_{1}).
\end{equation}
So (76) is obtained from (75) by interchanging $j^{^{\prime}},\beta^{^{\prime
}},r,$ and $j,\beta,r_{1}$. Besides (76) is obtained from (53) by applying
lemma 5. Now to find the eigenvalue $\Lambda_{N},$ which is close to
$\lambda_{j,\beta}$ , where $\mid j\delta\mid<r_{1}$ (see (51)) we are going
to iterate (76) as follows. Since $\mid j\delta\mid<r_{1}$ and $(j_{1}%
,\beta_{1})\in Q(\rho^{\alpha},9r_{1}),$ we have $\mid(j+j_{1})\delta
\mid<10r_{1}.$ Therefore in (75) interchanging $j^{^{\prime}},\beta^{^{\prime
}},r,$ and $j+j_{1,}\beta+\beta_{1},10r_{1}$ and introducing the notations
$r_{2}=10r_{1},$ $j^{2}=j+j_{1}+j_{2},$ $\beta^{2}=\beta+\beta_{1}+\beta_{2},$
we obtain%
\[
(\Lambda_{N}-\lambda_{j+j_{1},\beta_{1}+\beta})b(N,j+j_{1},\beta+\beta
_{1})=O(\rho^{-p\alpha})+
\]%
\begin{equation}%
{\displaystyle\sum_{(j_{2},\beta_{2})\in Q(\rho^{\alpha},9r_{2})}}
b(N,j^{2},\beta^{2})A(j+j_{1},\beta+\beta_{1},j^{2},\beta^{2}).
\end{equation}
Clearly, there exist an eigenvalue $\Lambda_{N}(t)$ satisfying $\mid
\lambda_{j,\beta}-\Lambda_{N}\mid\leq2M,$ where $M=\sup\mid q(x)\mid.$
Moreover, we will prove that \ if $\mid\beta\mid\sim\rho,$ and

$(j_{1},\beta_{1})\in Q(\rho^{\alpha},9r_{1})$ ( see Lemma 6 ), then%
\begin{equation}
\mid\lambda_{j,\beta}-\lambda_{j+j_{1},\beta+\beta_{1}}\mid>\frac{5}{9}%
\rho^{\alpha_{2}},\text{ }\mid\Lambda_{N}-\lambda_{j+j_{1},\beta+\beta_{1}%
}\mid>\frac{1}{2}\rho^{\alpha_{2}}.
\end{equation}
Then dividing both side of (77) by $\Lambda_{N}-\lambda_{j+j_{1},\beta
+\beta_{1}}$ and using (78), we get
\[
b(N,j+j_{1},\beta_{1}+\beta)=O(\rho^{-p\alpha-\alpha_{2}})+
\]%
\begin{equation}%
{\displaystyle\sum_{(j_{2},\beta_{2})\in Q(\rho^{\alpha},9r_{2})}}
\dfrac{A(j+j_{1},\beta+\beta_{1},j^{2},\beta^{2})b(N,j^{2},\beta^{2})}%
{\Lambda_{N}-\lambda_{j+j_{1},\beta_{1}+\beta}}.
\end{equation}
Putting the obtained formula for $b(N,j+j_{1},\beta_{1}+\beta)$ into (76), we
obtain
\[
(\Lambda_{N}-\lambda_{j,\beta})b(N,j,\beta)=O(\rho^{-p\alpha})+
\]%
\begin{equation}%
{\displaystyle\sum_{\substack{(j_{1},\beta_{1})\in Q(\rho^{\alpha}%
,9r_{1})\\(j_{2},\beta_{2})\in Q(\rho^{\alpha},9r_{2})}}}
\dfrac{A(j,\beta,j+j_{1,}\beta+\beta_{1})A(j+j_{1},\beta+\beta_{1},j^{2}%
,\beta^{2})b(N,j^{2},\beta^{2})}{\Lambda_{N}-\lambda_{j+j_{1},\beta+\beta_{1}%
}},
\end{equation}
Thus we got the one time iteration of (76). It will give the first term of
asymptotic formula for $\Lambda_{N}.$ For this we find the index $N$ such that
$b(N,j,\beta)$ is not very small (see Lemma 7) and (78) is satisfied, i.e.,
the denominator of the fraction in (80) is a big number. Then dividing both
sides of (80) by $b(N,j,\beta),$ we get the asymptotic formula for
$\Lambda_{N}$ (see Theorem 3).

\begin{lemma}
Let $\gamma+t\equiv\beta+\tau+(j+v)\delta\in V_{\delta}^{^{\prime}}%
(\rho^{\alpha_{1}})$ (see (50) and Remark 3), and $(j_{1},\beta_{1})\in
Q(\rho^{\alpha},9r_{1}),$ $(j_{k},\beta_{k})\in Q(\rho^{\alpha},9r_{k}),$
where $r_{1}$ is defined in (51) and $r_{k}=10r_{k-1}$ for $k=2,3,...,p-1$.
Then
\begin{align}
&  \mid j\delta\mid=O(\rho^{\alpha_{1}}),\text{ }\mid j_{k}\delta\mid
=O(\rho^{\alpha_{1}}),\text{ }\mid\beta_{k}\mid<\rho^{\alpha},\forall
k=1,2,...,p-1,\\
&  \mid\lambda_{j,\beta}(v,\tau)-\lambda_{j^{^{\prime}},\beta}\mid>2(\ln
\rho)^{-1},\text{ }\forall j^{^{\prime}}\neq j\text{, }\forall v(\beta)\in
W(\rho).
\end{align}

Moreover if $\mid j^{^{\prime}}\delta\mid<\frac{1}{2}\rho^{\frac{1}{2}%
\alpha_{2}},\mid\beta^{^{\prime}}-\beta\mid<(p-1)\rho^{\alpha},$
$\beta^{^{\prime}}\in\Gamma_{\delta}$, $j^{k}=j+j_{1}+...+j_{k},$ $\beta
^{k}=\beta+\beta_{1}+...+\beta_{k},$ where $k=1,2,...,p-1,$ then
\begin{align}
&  \mid\lambda_{j,\beta}-\lambda_{j^{^{\prime}},\beta^{^{\prime}}}\mid
>\frac{5}{9}\rho^{\alpha_{2}},\text{ }\forall\beta^{^{\prime}}\neq\beta,\\
&  \mid\lambda_{j,\beta}(v,\tau)-\lambda_{j^{k},\beta^{k}}\mid>\frac{5}{9}%
\rho^{\alpha_{2}},\text{ }\forall\beta^{k}\neq\beta.
\end{align}

\end{lemma}

\begin{proof}
The relations in (81) follows from (51) and the definitions of $r_{1},r_{k},$
$Q(\rho^{\alpha},9r_{k})$ (see Lemma 5). Inequality (82) is a consequence of
the definition of $W(\rho)$ (see Lemma 3(b)). Inequality (84) follows from
(83) and (81). It remains to prove (83). Since
\begin{equation}
\mid\lambda_{j,\beta}-\lambda_{j^{^{\prime}},\beta^{^{\prime}}}\mid\geq
\mid\mid\beta^{^{\prime}}+\tau\mid^{2}-\mid\beta+\tau\mid^{2}\mid-\mid\mu
_{j}-\mu_{j^{^{\prime}}}\mid,
\end{equation}
it is enough to prove the following two inequalities $\mid\mu_{j}%
-\mu_{j^{^{\prime}}}\mid<\frac{1}{3}\rho^{\alpha_{2}},$
\begin{equation}
\mid\mid\beta+\tau\mid^{2}-\mid\beta^{^{\prime}}+\tau\mid^{2}\mid>\frac{8}%
{9}\rho^{\alpha_{2}}.
\end{equation}
The first inequality follows from $\mid j^{^{\prime}}\delta\mid<\frac{1}%
{2}\rho^{\frac{1}{2}\alpha_{2}},\mid j\delta\mid=O(\rho^{\alpha_{1}})$ ( see
the conditions of this lemma and (81)) and (52), since $\alpha_{2}=3\alpha
_{1}.$ Now we prove (86). The conditions $\mid\beta^{^{\prime}}-\beta
\mid<(p-1)\rho^{\alpha},\mid\delta\mid<\rho^{\alpha}$ imply that there exist
$n\in Z$ and $\gamma^{^{\prime}}\in\Gamma$ such that
\begin{equation}
\gamma^{^{\prime}}=\beta^{^{\prime}}-\beta+(n+(2\pi)^{-1}(\beta^{^{\prime}%
}-\beta,\delta^{\ast}))\delta\in\Gamma(p\rho^{\alpha}).
\end{equation}
Since $\beta^{^{\prime}}-\beta$ is nonzero element of $\Gamma_{\delta}$ we
have $\gamma^{^{\prime}}\in\Gamma(p\rho^{\alpha})\backslash\delta R.$ This
together with the inclusion $\gamma+t=\beta+\tau+(j+v)\delta\in V_{\delta
}^{^{\prime}}(\rho^{\alpha_{1}})=V_{\delta}(\rho^{\alpha_{1}})\backslash
E_{2}$ ( see assumption of the lemma and definition of $E_{2}$) implies that
$\gamma+t\notin V_{\gamma^{^{\prime}}}(\rho^{\alpha_{2}}),$ that is, $\mid
\mid\gamma+t\mid^{2}-\mid\gamma+t+\gamma^{^{\prime}}\mid^{2}\mid\geq
\rho^{\alpha_{2}}.$ From this using the orthogonal decompositions (50), (87)
of $\gamma+t,\gamma^{^{\prime}}$ and taking into account that $\beta
,\tau,\beta^{^{\prime}}$ are orthogonal to $\delta$, $\mid j\delta\mid
=O(\rho^{\alpha_{1}})$ (see (81)), $\mid n+(2\pi)^{-1}(\beta^{^{\prime}}%
-\beta,\delta^{\ast}))\delta\mid=O(\rho^{\alpha}),$ $\alpha_{2}>2\alpha,$ we
obtain (86)
\end{proof}

\begin{lemma}
Suppose $h_{1}(x),h_{2}(x),...,h_{m}(x)\in L_{2}(F),$ where $m=p_{1}-1,$

$p_{1}=[\frac{p}{3}]+1$. Then for every eigenvalue $\lambda_{j,\beta}\sim
\rho^{2}$ of the operator $L_{t}(q^{\delta}(x))$ there exists an eigenvalue
$\Lambda_{N}$ and a corresponding normalized eigenfunction $\Psi_{N}$ of the
operator $L_{t}(q(x))$ such that:

$(i)$ $\mid\lambda_{j,\beta}-\Lambda_{N}\mid\leq2M,$ where $M=\sup\mid
q(x)\mid,$

$(ii)\mid b(N,j,\beta)\mid>\frac{1}{2}(c_{5})^{-\frac{1}{2}}\rho^{-\frac{1}%
{2}(d-1)q\alpha},$

$(iii)$ $\mid b(N,j,\beta)\mid^{2}>\frac{1}{2m}%
{\displaystyle\sum_{i=1}^{m}}
(\Psi_{N},\frac{h_{i}}{\mid\mid h_{i}\mid\mid})\mid^{2}>\frac{1}{2m}\mid
(\Psi_{N},\frac{h_{i}}{\mid\mid h_{i}\mid\mid})\mid^{2},$ $\forall i.$
\end{lemma}

\begin{proof}
Let $A,B,C$ be the set of indexes $N$ satisfying $(i),(ii),(iii)$
respectively. Using (53) and the Bessel inequality we obtain
\begin{align*}
\sum_{N\notin A}  &  \mid b(N,j,\beta)\mid^{2}=\sum_{N\notin A}\mid
\dfrac{(\Psi_{N}(x),(q(x)-Q(\zeta))\Phi_{j,\beta}(x))}{\Lambda_{N}%
-\lambda_{j,\beta}}\mid^{2}\\
&  <(2M)^{-2}\mid\mid q(x)-Q(\zeta))\Phi_{j,\beta}(x))\mid\mid^{2}\leq\frac
{1}{4}.
\end{align*}
This and the inequality $\mid A\mid<c_{5}\rho^{(d-1)q\alpha}$ ( see the end of
the introduction) imply that $\sum_{N\in A\backslash B}\mid b(N,j,\beta
)\mid^{2}<\frac{1}{4}.$ Therefore using the Parseval equality, we obtain
\[
\sum_{N\in A}\mid b(N,j,\beta)\mid^{2}\geq\frac{3}{4},\text{ }\sum_{N\in A\cap
B}\mid b(N,j,\beta)\mid^{2}>\frac{1}{2}.
\]
Now to prove the lemma we show that there exists $N\in A\cap B$ satisfying
$(iii)$. Assume the converse, that is, assume that the condition $(iii)$ does
not hold for all $N\in A\cap B$ . Then using the Bessel inequality, and the
last inequalities we get the contradiction $\frac{1}{2}<\sum_{N\in A\cap
B}\mid b(N,j,\beta)\mid^{2}<\frac{1}{2m}\sum_{i=1}^{m}\sum_{N\in A}\mid
(\Psi_{N},\frac{h_{i}}{\mid\mid h_{i}\mid\mid})\mid^{2}$

$\leq\frac{1}{2m}\sum_{i=1}^{m}\mid\mid\frac{h_{i}}{\mid\mid h_{i}\mid\mid
}\mid\mid^{2}=\frac{1}{2}.$
\end{proof}

\begin{theorem}
For every eigenvalue $\lambda_{j,\beta}(v,\tau)$ of $L_{t}(q^{\delta}(x))$
such that

$\beta+\tau+(j+v)\delta\in V_{\delta}^{^{\prime}}(\rho^{\alpha_{1}})$ there
exists an eigenvalue $\Lambda_{N}$ , denoted by $\Lambda_{N}(\lambda_{j,\beta
}(v,\tau)),$ of $L_{t}(q(x))$ satisfying
\begin{equation}
\Lambda_{N}(\lambda_{j,\beta}(v,\tau))=\lambda_{j,\beta}(v,\tau)+O(\rho
^{-\alpha_{2}}).
\end{equation}

\end{theorem}

\begin{proof}
By Lemma 7 there is an eigenvalue $\Lambda_{N}$ satisfying $(i)$-$(iii)$ for
\[
h_{i}(x)=%
{\displaystyle\sum_{\substack{(j_{1},\beta_{1})\in Q(\rho^{\alpha}%
,9r_{1}),\\(j_{2},\beta_{2})\in Q(\rho^{\alpha},9r_{2})}}}
\dfrac{\overline{A(j,\beta,j^{1},\beta^{1}})\overline{A(j^{1},\beta^{1}%
,j^{2},\beta^{2}})\Phi_{j^{2},\beta^{2}}(x)}{(\lambda_{j,\beta}-\lambda
_{j+j_{1},\beta+\beta_{1}})^{i}},
\]
where $i=1,2,...,m;$ $m=p_{1}-1.$ Since $\beta_{1}\neq0$, the inequality (83)
and condition $(i)$ of lemma 7 yield (78). Hence, in brief notations
$a=\lambda_{j,\beta},$ $z=\lambda_{j+j_{1},\beta+\beta_{1}},$ we have
$\mid\Lambda_{N}-a\mid<2M,$ $\mid z-a\mid>\frac{1}{2}\rho^{\alpha_{2}}.$ Using
the relations
\[
\frac{1}{\Lambda_{N}-z}=-\sum_{i=1}^{\infty}\frac{(\Lambda_{N}-a)^{i-1}%
}{(z-a)^{i}}=-\sum_{i=1}^{m}\frac{(\Lambda_{N}-a)^{i-1}}{(z-a)^{i}}%
+O(\rho^{-p_{1}\alpha_{2}})
\]
and $p_{1}\alpha_{2}>p\alpha$ we see that formula (80) can be written as
\[
(\Lambda_{N}-\lambda_{j,\beta})b(N,j,\beta,)=\sum_{i=1}^{m}(\Lambda
_{N}-a)^{i-1}(\Psi_{N},\frac{h_{i}}{\mid\mid h_{i}\mid\mid})\parallel
h_{i}\parallel+O(\rho^{-p\alpha}).
\]
Dividing both sides by $b(N,j,\beta)$, using $(ii),(iii)$ of lemma 7, and the
obvious inequality $p\alpha-\frac{1}{2}(d-1)q\alpha>\alpha_{2}$ ( see the
third inequality in (15) and the definitions of $k_{1},\alpha,\alpha_{2}$ in
the end of introduction), we get
\begin{equation}
\mid(\Lambda_{N}-\lambda_{j,\beta})\mid<(2m)^{\frac{1}{2}}\sum_{i=1}^{m}%
\mid\Lambda_{N}-a\mid^{i-1}\parallel h_{i}\parallel+O(\rho^{-\alpha_{2}})
\end{equation}
On the other hand the inequalities (71) and (84) imply that $\parallel
h_{i}\parallel=O(\rho^{-\alpha_{2}}).$ Therefore (89) and $\mid\Lambda
_{N}-a\mid<2M,$ yield the proof of the theorem
\end{proof}

It follow from formulas (82), (84), and (88) that
\begin{equation}
\mid\Lambda_{N}(\lambda_{j,\beta})-\lambda_{j^{k},\beta^{k}}(v,\tau
)\mid>c(\beta^{k},\rho),\forall\text{ }v(\beta)\in W(\rho),
\end{equation}
where $(j_{k},\beta_{k})\in Q(\rho^{\alpha},9r_{k}),$ $k=1,2,...,p-1$;
$c(\beta^{k},\rho)=(\ln\rho)^{-1}$ when $\beta^{k}=\beta,$ $j^{k}\neq j$ and
$c(\beta^{k},\rho)=\frac{1}{2}\rho^{\alpha_{2}}$ when $\beta^{k}\neq\beta.$ We
iterated (76) one time and got (80) from which the formula (88) is obtained.
Now to obtain the asymptotic formulas of the arbitrary order for $\Lambda_{N}$
we repeat this iteration $2p_{1}$ times as follows. Since $\mid j\delta
\mid<r_{1}$ ( see (51)), $(j_{1},\beta_{1})\in Q(\rho^{\alpha},9r_{1}%
),(j_{2},\beta_{2})\in Q(\rho^{\alpha},9r_{2})$ (see (80)) and $j^{2}%
=j+j_{1}+j_{2}$ (see definition of $j^{2}$ in (77)), we have $\mid j^{2}%
\delta\mid<10r_{2}$. Therefore in (75) interchanging $j^{^{\prime}}%
,\beta^{^{\prime}},r,$ and $j^{2},\beta^{2},10r_{2}$ and using the notations
$r_{3}=10r_{2},$ $j^{3}=j^{2}+j_{3},$ $\beta^{3}=\beta^{2}+\beta_{3}$ ( see
Lemma 6), we obtain%
\[
(\Lambda_{N}-\lambda_{j^{2},\beta^{2}})b(N,j^{2},\beta^{2})=O(\rho^{-p\alpha
})+
\]%
\begin{equation}%
{\displaystyle\sum_{(j_{3},\beta_{3})\in Q(\rho^{\alpha},9r_{3})}}
b(N,j^{3},\beta^{3})A(j^{2},\beta^{2},j^{3},\beta^{3}).
\end{equation}
Dividing both side of (91) by $\Lambda_{N}-\lambda_{j^{2},\beta^{2}}$ and
using (90), we get
\[
b(N,j^{2},\beta^{2})=O(\rho^{-p\alpha}(c(\beta^{2},\rho))^{-1})+
\]%
\begin{equation}%
{\displaystyle\sum_{(j_{3},\beta_{3})\in Q(\rho^{\alpha},9r_{3})}}
\dfrac{b(N,j^{3},\beta^{3})A(j^{2},\beta^{2},j^{3},\beta^{3})}{\Lambda
_{N}-\lambda_{j^{2},\beta^{2}}}.
\end{equation}
for $(j^{2},\beta^{2})\neq(j,\beta).$ In the same way we obtain
\[
b(N,j^{k},\beta^{k})=O(\rho^{-p\alpha}(c(\beta^{k},\rho))^{-1})+
\]%
\begin{equation}%
{\displaystyle\sum_{(j_{k+1},\beta_{k+1})\in Q(\rho^{\alpha},9r_{k+1})}}
\dfrac{b(N,j^{k+1},\beta^{k+1})A(j^{k},\beta^{k},j^{k+1},\beta^{k+1})}%
{\Lambda_{N}-\lambda_{j^{k},\beta^{k}}}.
\end{equation}
for $(j^{k},\beta^{k})\neq(j,\beta),$ $k=3,4,....$ Now we isolate the terms in
the right-hand side of (80) with multiplicand $b(N,j,\beta)$ , i.e.,the case
$(j^{2},\beta^{2})=(j,\beta)$, and replace $b(N,j^{2},\beta^{2})$ in (80) by
the right-hand side of (92) when $(j^{2},\beta^{2})\neq(j,\beta)$ and use
(78), (90) to get
\begin{equation}
(\Lambda_{N}-\lambda_{j,\beta})b(N,j,\beta)=S_{1}^{^{\prime}}(\Lambda
_{N},\lambda_{j,\beta}))b(N,j,\beta)+O(\rho^{-p\alpha})+
\end{equation}%
\[%
{\displaystyle\sum_{\substack{(j_{1},\beta_{1})\in Q(\rho^{\alpha}%
,9r_{1}),\\(j_{2},\beta_{2})\in Q(\rho^{\alpha},9r_{2}),(j^{2},\beta^{2}%
)\neq(j,\beta)}}}
\dfrac{A(j_{,}\beta,j^{1},\beta^{1})A(j_{,}^{1}\beta^{1},j^{2},\beta
^{2})b(N,j^{3},\beta^{3})}{(\Lambda_{N}-\lambda_{j+j_{1},\beta+\beta_{1}%
})(\Lambda_{N}-\lambda_{j^{2},\beta^{2}})},
\]

where
\begin{equation}
S_{1}^{\prime}(\Lambda_{N},\lambda_{j,\beta})=%
{\displaystyle\sum_{(j_{1},\beta_{1})\in Q(\rho^{\alpha},9r_{1})}}
\dfrac{A(j,\beta,j+j_{1,}\beta+\beta_{1})A(j+j_{1},\beta+\beta_{1},j,\beta
)}{\Lambda_{N}-\lambda_{j+j_{1},\beta+\beta_{1}}}.
\end{equation}
The formula (94) is the two times iteration of (76). Repeating these process
$2p_{1}$ times, i.e., in (94) isolating the terms with multiplicand
$b(N,j,\beta)$ (i.e., the case $(j^{3},\beta^{3})=(j,\beta))$ and replacing
$b(N,j^{3},\beta^{3})$ by the right-hand side of (93) (for $k=3)$ when
$(j^{3},\beta^{3})\neq(j,\beta)$ etc., we obtain
\begin{equation}
(\Lambda_{N}-\lambda_{j,\beta})b(N,j,\beta)=(\sum_{k=1}^{2p_{1}}%
S_{k}^{^{\prime}}(\Lambda_{N},\lambda_{j,\beta}))b(N,j,\beta)+C_{2p_{1}%
}^{^{\prime}}+O(\rho^{-p\alpha}),
\end{equation}

where%
\[
S_{k}^{\prime}(\Lambda_{N},\lambda_{j,\beta})=%
{\displaystyle\sum}
(\prod_{i=1}^{k}\dfrac{A(j_{,}^{i-1}\beta^{i-1},j^{i},\beta^{i})}{(\Lambda
_{N}-\lambda_{j^{i},\beta^{i}})})A(j^{k},\beta^{k},j,\beta),
\]%
\[
C_{k}^{\prime}=%
{\displaystyle\sum}
(\prod_{i=1}^{k}\dfrac{A(j_{,}^{i-1}\beta^{i-1},j^{i},\beta^{i})}{(\Lambda
_{N}-\lambda_{j^{i},\beta^{i}})})A(j_{,}^{k}\beta^{k},j^{k+1},\beta
^{k+1})b(N,j^{k+1},\beta^{k+1})
\]
Here $j^{0}=j,$ $\beta^{0}=\beta$ and the summation for $S_{k}^{\prime},$ and
$C_{k}^{^{\prime}}$ are taken under the conditions $(j_{i},\beta_{i})\in
Q(\rho^{\alpha},9r_{i}),(j^{i},\beta^{i})\neq(j,\beta),$ for $i=2,3,...,k$ and
for $i=2,3,...,k+1$ respectively. Since $\beta_{k}\neq0$ for every integer
$k,$ the relation $\beta_{1}+...+\beta_{i}=0$ implies that $\beta
_{1}+...+\beta_{i\pm1}\neq0.$ Hence $\beta^{1}\neq\beta.$ Moreover if
$\beta^{i}=\beta,$ then $\beta^{i\pm1}\neq\beta.$ Therefore the number of the
multiplicands $\Lambda_{N}-\lambda_{j^{i},\beta^{i}}$ in the denominators of
$S_{k}^{^{\prime}}$ and $C_{2p_{1}}^{^{\prime}}$ satisfying $\mid\Lambda
_{N}(\lambda_{j,\beta})-\lambda_{j^{i},\beta^{i}}\mid>\frac{1}{2}\rho
^{\alpha_{2}}$ ( see 90) is not less than $\frac{k}{2}$ and $p_{1}$
respectively. Hence using (71) and $p_{1}\alpha_{2}>p\alpha,$ we obtain
\begin{align}
C_{2p_{1}}^{^{\prime}}  &  =O((\rho^{-\alpha_{2}}\ln\rho)^{p_{1}}%
)=O(\rho^{-p\alpha}),\text{ }S_{1}^{^{\prime}}(\Lambda_{N},\lambda_{j,\beta
})=O(\rho^{-\alpha_{2}}),\\
S_{k}^{^{\prime}}(\Lambda_{N},\lambda_{j,\beta})  &  =O((\rho^{-\alpha_{2}}%
\ln\rho)^{\frac{k}{2}}),\forall k=2,3,...,2p_{1}.\nonumber
\end{align}
To prove this estimation we used (90). Moreover, if a real number $a$
satisfies $\mid a-\lambda_{j,\beta}\mid<(\ln\rho)^{-1}$ then, by (82), (84) we
have $\mid a-\lambda_{j^{k},\beta^{k}}(v,\tau)\mid>c(\beta^{k},\rho).$
Therefore using this instead of (90) and repeating the proof of (97) we
obtain
\begin{equation}
S_{1}^{^{\prime}}(a,\lambda_{j,\beta})=O(\rho^{-\alpha_{2}}),S_{k}^{^{\prime}%
}(a,\lambda_{j,\beta})=O(\rho^{-\alpha_{2}}\ln\rho)^{\frac{k}{2}}),\forall
k=2,3,...,2p_{1}.
\end{equation}

\begin{theorem}
For every eigenvalue $\lambda_{j,\beta}(v,\tau)$ of the operator
$L_{t}(q^{\delta}(x))$ such that $\beta+\tau+(j+v)\delta\in V_{\delta
}^{^{\prime}}(\rho^{\alpha_{1}})$, $v(\beta)\in W(\rho)$ there exists an
eigenvalue $\Lambda_{N},$ denoted by $\Lambda_{N}(\lambda_{j,\beta}(v,\tau)),$
of $L_{t}(q(x))$ satisfying the formulas
\begin{equation}
\Lambda_{N}(\lambda_{j,\beta}(v,\tau))=\lambda_{j,\beta}(v,\tau)+E_{k-1}%
(\lambda_{j,\beta})+O(\rho^{-k\alpha_{2}}(\ln\rho)^{2k}),
\end{equation}
where $E_{0}=0,$ $E_{s}=\sum_{k=1}^{2p_{2}}S_{k}^{^{\prime}}(\lambda_{j,\beta
}+E_{s-1},\lambda_{j,\beta}),$ for $s=1,2,...$ , and
\begin{equation}
E_{k-1}(\lambda_{j,\beta})=O(\rho^{-\alpha_{2}}(\ln\rho)),
\end{equation}
for $k=1,2,...,[\frac{1}{9}(p-\frac{1}{2}q(d-1)]$
\end{theorem}

\begin{proof}
The proof of this Theorem is similar to the proof of Theorem 1(a). By Theorem
3 formula (99) for the case $k=1$ is proved and $E_{0}=0$. Hence (100) for
$k=1$ is also proved. The proof of (100), for arbitrary $k,$ follows from (98)
and the definition of $E_{s}$ by induction. Now we prove (99) by induction.
Assume that (99) is true for $k=s<[\frac{1}{9}(p-\frac{1}{2}q(d-1)]$ ,i.e.,

$\Lambda_{N}=\lambda_{j,\beta}+E_{s-1}+O(\rho^{-s\alpha_{2}}(\ln\rho)^{2s})).$
Putting this expression for $\Lambda_{N\text{ }}$ into

$\sum_{k=1}^{2p_{1}}S_{k}^{^{\prime}}(\Lambda_{N},\lambda_{j,\beta})$,
dividing both sides of (96) by $b(N,j,\beta),$ using (97), (98), condition
$(ii)$ of Lemma 7 and the equality $\alpha_{2}=9\alpha,$ we get

$\Lambda_{N}=\lambda_{j,\beta}+\sum_{k=1}^{2p_{1}}S_{k}^{^{\prime}}%
(\lambda_{j,\beta}+E_{s-1}+O(\frac{(\ln\rho)^{2s}}{\rho^{s\alpha_{2}}%
}),\lambda_{j,\beta})+O(\rho^{-\frac{1}{9}(p-\frac{1}{2}q(d-1))\alpha_{2}})$%
\begin{align*}
&  =O(\rho^{-\frac{1}{9}(p-\frac{1}{2}q(d-1))\alpha_{2}})+\lambda_{j,\beta
}+\sum_{k=1}^{2p_{1}}S_{k}^{^{\prime}}(\lambda_{j,\beta}+E_{s-1}%
,\lambda_{j,\beta})+\\
&  \{\sum_{k=1}^{2p_{1}}S_{k}^{^{\prime}}(\lambda_{j,\beta}+E_{s-1}%
+O(\rho^{-s\alpha_{2}}(\ln\rho)^{2s}),\lambda_{j,\beta})-\sum_{k=1}^{2p_{1}%
}S_{k}^{^{\prime}}(\lambda_{j,\beta}+E_{s-1},\lambda_{j,\beta})\}.
\end{align*}
To prove (99) for $k=s+1$ we need to show that the expression in the curly
brackets is equal to $O((\rho^{-(s+1)\alpha_{2}}(\ln\rho)^{2s+1}).$ This can
be checked by using the estimations (71), (100), (82), (84) and the obvious
relation
\begin{align*}
&  \dfrac{1}{\prod_{i=1}^{n}(\lambda_{j,\beta}+E_{s-1}+O(\rho^{-s\alpha_{2}%
}(\ln\rho)^{2s})-\lambda_{j^{i},\beta^{i}})}-\dfrac{1}{\prod_{i=1}^{n}%
(\lambda_{j,\beta}+E_{s-1}-\lambda_{j^{i},\beta^{i}})}\\
&  =\dfrac{1}{\prod_{i=1}^{n}(\lambda_{j,\beta}+E_{s-1}-\lambda_{j^{i}%
,\beta^{i}})}(\frac{1}{1+O(\rho^{-s\alpha_{2}}(\ln\rho)^{2s}\ln\rho)}-1)
\end{align*}

$=O(\rho^{-(s+1)\alpha_{2}}(\ln\rho)^{2(s+1)})$ for all $n=1,2,...,2p_{1}$.
The theorem is proved
\end{proof}

\begin{remark}
Here we note some properties of the known parts $\lambda_{j,\beta}+E_{k}$ (see
(99)), where $\lambda_{j,\beta}=\mu_{j}(v)+\mid\beta+\tau\mid^{2}$ ( see Lemma
2(b)), of the eigenvalues of $L_{t}(q(x))$. We prove the equality
\begin{equation}
\frac{\partial(E_{k}(\mu_{j}(v)+\mid\beta+\tau\mid^{2}))}{\partial\tau_{i}%
}=O(\rho^{-2\alpha_{2}+\alpha}\ln\rho),
\end{equation}
for $i=1,2,...,d-1,$ where $\tau=(\tau_{1},\tau_{2},...,\tau_{d-1}%
),k=1,2,...,[\frac{1}{9}(p-\frac{1}{2}q(d-1)]$, and $v(\beta)\in W(\rho).$ To
prove (101) for $k=1$ we calculate the derivatives of the expression
$H(\beta^{k},\tau)\equiv(\mu_{j}+\mid\beta+\tau\mid^{2}-\mu_{j^{k}}-\mid
\beta^{k}+\tau\mid^{2})^{-1}.$ Since $\mu_{j}$,and $\mu_{j^{^{\prime}}}$ do
not depend on $\tau_{i},$ the function $H(\beta^{k},\tau)$ for $\beta
^{k}=\beta$ do not depend on $\tau_{i}$ and it follows from Lemma 3(b) that
$H(\beta,\tau)=O(\ln\rho).$ For $\beta^{k}\neq\beta$ using (84), and equality
$\mid\beta^{k}-\beta\mid=\mid\beta_{1}+\beta_{2}...+\beta_{i}\mid
=O(\rho^{\alpha})$ (see last inequality in (81)) we obtain that the
derivatives of $H(\beta^{k},\tau)$ is equal to $O(\rho^{-2\alpha_{2}+\alpha
}).$ Therefore using (71) and the definition of $E_{1}(\lambda_{j,\beta})$ (
see (99) and (96)), by direct calculation, we get (101) for $k=1.$ Now suppose
that (101) holds for $k=s-1.$ Using this, replacing $\mu_{j}+\mid\beta
+\tau\mid^{2}$ by $\mu_{j}+\mid\beta+\tau\mid^{2}+E_{s-1}$ in $H(\beta
^{k},\tau)$, arguing as above we get (101) for $k=s$.
\end{remark}

\section{Asymptotic Formulas for the Bloch Functions}

In this section using the asymptotic formulas for the eigenvalues and the
simplicity conditions (12), (13), we prove the asymptotic formulas for the
Bloch functions with a quasimomentum of the simple set $B$.

\begin{theorem}
If $\gamma+t\in B,$ then there exists a unique eigenvalue $\Lambda_{N}(t)$
satisfying (5) for $k=1,2,...,[\frac{p}{3}],$ where $p$ is defined in (3).
This is a simple eigenvalue and the corresponding eigenfunction $\Psi
_{N,t}(x)$ of $L(q(x))$ satisfies (10) if

$q(x)\in W_{2}^{s_{0}}(F),$ where $s_{0}=\frac{3d-1}{2}(3^{d}+d+2)+\frac{1}%
{4}d3^{d}+d+6.$
\end{theorem}

\begin{proof}
By Theorem 1(b) if $\gamma+t\in B\subset U(\rho^{\alpha_{1}},p),$ then there
exists an eigenvalue $\Lambda_{N}(t)$ satisfying (5) for $k=1,2,...,[\frac
{1}{3}(p-\frac{1}{2}q(d-1))].$ Since

$k_{1}=[\frac{d}{3\alpha}]+2\leq\frac{1}{3}(p-\frac{1}{2}q(d-1))$ (see the
third inequality in (15)) formula (5) holds for $k=k_{1}.$ Therefore using
(5), the relation $3k_{1}\alpha>d+2\alpha$ ( see the fifth inequality in
(15)), and notations $F(\gamma+t)=\mid\gamma+t\mid^{2}+F_{k_{1}-1}(\gamma+t)$,
$\varepsilon_{1}=\rho^{-d-2\alpha}$ ( see Step 1 in introduction), we obtain
\begin{equation}
\Lambda_{N}(t)=F(\gamma+t)+o(\varepsilon_{1}).
\end{equation}
Let $\Psi_{N}$ be any normalized eigenfunction corresponding to $\Lambda_{N}$.
Since the normalized eigenfunction is defined up to constant of modulus $1,$
without loss of generality it can assumed that $\arg b(N,\gamma)=0,$ where
$b(N,\gamma)=(\Psi_{N},e^{i(\gamma+t,x)}).$ Therefore to prove (10) it
suffices to show that (14) holds. To prove (14) first \ we estimate
\ $\sum_{\gamma^{^{\prime}}\notin K}\mid b(N,\gamma^{^{\prime}})\mid^{2}$and
then $\sum_{\gamma^{^{\prime}}\in K\backslash\{\gamma\}}\mid b(N,\gamma
^{^{\prime}})\mid^{2},$ where $K$ is defined in (12), (13). Using (102), the
definition of $K$, and (16), we get
\begin{align}
&  \mid\Lambda_{N}-\mid\gamma^{^{\prime}}+t\mid^{2}\mid>\frac{1}{4}%
\rho^{\alpha_{1}},\text{ }\forall\gamma^{^{\prime}}\notin K,\\
\sum_{\gamma^{^{\prime}}\notin K}  &  \mid b(N,\gamma^{^{\prime}})\mid
^{2}=\parallel q(x)\Psi_{N}\parallel^{2}O(\rho^{-2\alpha_{1}})=O(\rho
^{-2\alpha_{1}}).\nonumber
\end{align}
If $\gamma^{^{\prime}}\in K$ , then by (102) and by definition of $K,$ it
follows that
\begin{equation}
\mid\Lambda_{N}-\mid\gamma^{^{\prime}}+t\mid^{2}\mid<\frac{1}{2}\rho
^{\alpha_{1}}%
\end{equation}
Now we prove that the simplicity conditions (12), (13) imply
\begin{equation}
\mid b(N,\gamma^{^{\prime}})\mid\leq c_{4}\rho^{-c\alpha},\text{ }%
\forall\gamma^{^{\prime}}\in K\backslash\{\gamma\},
\end{equation}
where $c=p-dq-\frac{1}{4}d3^{d}-3.$ The conditions $\gamma^{^{\prime}}\in K,$
$\gamma+t\in B$ and (24) imply the inclusion $\gamma^{^{\prime}}+t\in
R(\frac{3}{2}\rho)\backslash R(\frac{1}{2}\rho).$ If for $\gamma^{^{\prime}%
}+t\in U(\rho^{\alpha_{1}},p)$ and $\gamma^{^{\prime}}\in K\backslash
\{\gamma\}$ the inequality in (105) is not true, then by (104) and Theorem
1(a), we have
\begin{equation}
\Lambda_{N}=\mid\gamma^{^{\prime}}+t\mid^{2}+F_{k-1}(\gamma^{^{\prime}%
}+t)+O(\rho^{-3k\alpha})
\end{equation}
for $k=1,2,...,[\frac{1}{3}(p-c)]=[\frac{1}{3}(dq+\frac{1}{4}d3^{d}+3)].$
Since $\alpha=\frac{1}{q}$ and

$k_{1}\equiv\lbrack\frac{d}{3\alpha}]+2<\frac{1}{3}(dq+\frac{1}{4}d3^{d}+3)$,
\ the formula (106) holds for $k=k_{1}.$ Therefore arguing as in the prove of
(102), we get $\Lambda_{N}-F(\gamma^{^{\prime}}+t)=o(\varepsilon_{1})$. \ This
with (102) contradicts (12). Similarly, if the inequality in (105) does not
hold for $\gamma^{^{\prime}}+t\in(E_{k}\backslash E_{k+1})$ and $\gamma
^{^{\prime}}\in K,$ then by Theorem 2(a)
\begin{equation}
\Lambda_{N}=\lambda_{j}(\gamma^{^{\prime}}+t)+O(\rho^{-(p-c-\frac{1}{4}%
d3^{d})\alpha}),
\end{equation}
where $(p-c-\frac{1}{4}d3^{d})\alpha=(dq+3)\alpha>d+2\alpha$ . Hence we have

$\Lambda_{N}-\lambda_{j}(\gamma^{^{\prime}}+t)=o(\varepsilon_{1}).$ This with
(102) contradicts (13). So the inequality in (105) holds. Therefore, using
$\mid K\mid=O(\rho^{d-1}),$ $q\alpha=1,$ we get
\begin{equation}
\sum_{\gamma^{^{\prime}}\in K\backslash\{\gamma\}}\mid b(N,\gamma^{^{\prime}%
})\mid^{2}=O(\rho^{-(2c-q(d-1))\alpha})=O(\rho^{-(2p-(3d-1)q-\frac{1}{2}%
d3^{d}-6)\alpha}).
\end{equation}
If $s=s_{0},$ that is, $p=s_{0}-d,$ then $2p-(3d-1)q-\frac{1}{2}d3^{d}-6=6.$
Since $\alpha_{1}=3\alpha,$ the equality (108) and the equality in (103) imply
(14). Thus we proved that the equality (10) holds for any normalized
eigenfunction $\Psi_{N}$ corresponding to any eigenvalue $\Lambda_{N}$
\ satisfying (5). If there exist two different eigenvalues or multiple
eigenvalue satisfying (5), then there exist two orthogonal normalized
eigenfunction satisfying (10), which is impossible. Therefore $\Lambda_{N}%
$\ is a simple eigenvalue. It follows from Theorem 1(a) that $\Lambda_{N}$
satisfies (5) for $k=1,2,...,[\frac{p}{3}],$ since the inequality (7) holds
for $c=0$ ( see (10)).
\end{proof}

\begin{remark}
Since for $\gamma+t\in B$ \ there exists a unique eigenvalue satisfying (5),
(102) we denote this eigenvalue by $\Lambda(\gamma+t).$ Since this eigenvalue
is simple, we denote the \ corresponding eigenfunction by $\Psi_{\gamma
+t}(x).$ By Theorem 5 this eigenfunction satisfies (10). Clearly, for
$\gamma+t\in B$ \ there exists a unique index $N\equiv N(\gamma+t)$ such that
$\Lambda(\gamma+t)=\Lambda_{N(\gamma+t)}$) and $\Psi_{\gamma+t}(x)=\Psi
_{N(\gamma+t)}(x)).$
\end{remark}

Now we prove the asymptotic formulas of arbitrary order for $\Psi_{\gamma
+t}(x).$

\begin{theorem}
If $\gamma+t\in B,$ then the eigenfunction $\Psi_{\gamma+t}(x)\equiv
\Psi_{N(\gamma+t)}(x)$ corresponding to the eigenvalue $\Lambda_{N}%
\equiv\Lambda(\gamma+t)$ satisfies formulas (11), for

$k=1,2,...,n$, where $n=[\frac{1}{6}(2p-(3d-1)q-\frac{1}{2}d3^{d}-6)],$

$\Phi_{0}(x)=0,$ $\Phi_{1}(x)=%
{\displaystyle\sum_{\gamma_{1}\in\Gamma(\rho^{\alpha})}}
\dfrac{q_{\gamma_{1}}e^{i(\gamma+t+\gamma_{1},x)}}{(\mid\gamma+t\mid^{2}%
-\mid\gamma+\gamma_{1}+t\mid^{2})},$

and $\Phi_{k-1}(x)$ for $k>2$ is a linear combination of $e^{i(\gamma
+t+\gamma^{^{\prime}},x)}$ for

$\gamma^{^{\prime}}\in\Gamma((k-1)\rho^{\alpha})\cup\{0\}$ with coefficients
(114), (115).
\end{theorem}

\begin{proof}
By Theorem 5, formula (11) for $k=1$ is proved. To prove formula (11) for
arbitrary $k\leq n$ we prove the following equivalent relations
\begin{equation}
\sum_{\gamma^{^{\prime}}\in\Gamma^{c}(k-1)}\mid b(N,\gamma+\gamma^{^{\prime}%
})\mid^{2}=O(\rho^{-2k\alpha_{1}}),
\end{equation}%
\begin{equation}
\Psi_{N}=b(N,\gamma)e^{i(\gamma+t,x)}+\sum_{\gamma^{^{\prime}}\in
\Gamma((k-1)\rho^{\alpha})}b(N,\gamma+\gamma^{^{\prime}})e^{i(\gamma
+t+\gamma^{^{\prime}},x)}+H_{k}(x),
\end{equation}
where $\Gamma^{c}(m)\equiv\Gamma\backslash(\Gamma(m\rho^{\alpha})\cup\{0\})$
and $\parallel H_{k}(x)\parallel=O(\rho^{-k\alpha_{1}}).$ The case $k=1$ is
proved due to (14). Assume that (109) is true for $k=m$ . Then using (110) for
$\ \ \ k=m,$ and (3), we have $\Psi_{N}(x)(q(x))=H(x)+O(\rho^{-m\alpha_{1}}),$
where $H(x)$ is a linear combination of $e^{i(\gamma+t+\gamma^{^{\prime}},x)}$
for $\gamma^{^{\prime}}\in\Gamma(m\rho^{\alpha})\cup\{0\}.$ Hence
$(H(x),e^{i(\gamma+t+\gamma^{^{\prime}},x)})=0$ for $\gamma^{^{\prime}}%
\in\Gamma^{c}(m).$ So using (16) and (103), we get%
\begin{equation}
\sum_{\gamma^{^{\prime}}}\mid b(N,\gamma+\gamma^{^{\prime}})\mid^{2}%
=\sum_{\gamma^{^{\prime}}}\mid\dfrac{(O(\rho^{-m\alpha_{1}}),e^{i(\gamma
+t+\gamma^{^{\prime}},x)})}{\Lambda_{N}-\mid\gamma+\gamma^{^{\prime}}%
+t\mid^{2}}\mid^{2}=O(\rho^{-2(m+1)\alpha_{1}}),
\end{equation}
where the summation is taken under conditions $\gamma^{^{\prime}}\in\Gamma
^{c}(m)$, $\gamma+\gamma^{^{\prime}}\notin K.$ On the other hand, using
$\alpha_{1}=3\alpha,$ (108), and the definition of $n$, we obtain
\[
\sum_{\gamma^{^{\prime}}\in K\backslash\{\gamma\}}\mid b(N,\gamma^{^{\prime}%
})\mid^{2}=O(\rho^{-2n\alpha_{1}}).
\]
This with (111) implies (109) for $k=m+1.$ Thus (110) is also proved. Here
$b(N,\gamma)$ and $b(N,\gamma+\gamma^{^{\prime}})$ for $\gamma^{^{\prime}}%
\in\Gamma((n-1)\rho^{\alpha})$ can be calculated as follows. First we express
$b(N,\gamma+\gamma^{^{\prime}})$ by $b(N,\gamma)$. For this we apply (18) for
$b(N,\gamma+\gamma^{^{\prime}}),$ where $\gamma^{^{\prime}}\in\Gamma
((n-1)\rho^{\alpha}),$ that is, in (18) replace $\gamma^{^{\prime}}$ by
$\gamma+\gamma^{^{\prime}}$. Iterate it $n$ times and every times isolate the
terms with multiplicand $b(N,\gamma).$ In other word apply (18) for
$b(N,\gamma+\gamma^{^{\prime}})$ and isolate the terms with multiplicand
$b(N,\gamma).$ Then apply (18) for $b(N,\gamma+\gamma^{^{\prime}}-\gamma_{1})$
when $\gamma^{^{\prime}}-\gamma_{1}\neq0.$ Then apply (18) for

$b(N,\gamma+\gamma^{^{\prime}}-\sum_{i=1}^{2}\gamma_{i})$ when $\gamma
^{^{\prime}}-\sum_{i=1}^{2}\gamma_{i}\neq0,$ etc. Apply (18) for

$b(N,\gamma+\gamma^{^{\prime}}-\sum_{i=1}^{j}\gamma_{i})$ when $\gamma
^{^{\prime}}-\sum_{i=1}^{j}\gamma_{i}\neq0,$ where $\gamma_{i}\in\Gamma
(\rho^{\alpha}),$

$j=3,4,...,n-1.$ Then using (4) and the relations

$\mid\Lambda_{N}-\mid\gamma+t+\gamma^{^{\prime}}-\sum_{i=1}^{j}\gamma_{i}%
\mid^{2}\mid>\frac{1}{2}\rho^{\alpha_{1}}$ ( see (20) and take into account that

$\gamma^{^{\prime}}-\sum_{i=1}^{j}\gamma_{i}\in\Gamma(p\rho^{\alpha}),$ since
$p>2n$), $\Lambda_{N}=P(\gamma+t)+O(\rho^{-n\alpha_{1}}),$ where
$P(\gamma+t)=\mid\gamma+t\mid^{2}+F_{[\frac{p}{3}]}(\gamma+t)$ ( see Theorem
5), we obtain%
\begin{equation}
b(N,\gamma+\gamma^{^{\prime}})=\sum_{k=1}^{n-1}A_{k}(\gamma^{^{\prime}%
})b(N,\gamma)+O(\rho^{-n\alpha_{1}}),
\end{equation}
where

$A_{1}(\gamma^{^{\prime}})=\dfrac{q_{\gamma^{^{\prime}}}}{P(\gamma
+t)-\mid\gamma+\gamma^{^{\prime}}+t\mid^{2}}=\dfrac{q_{\gamma^{^{\prime}}}%
}{\mid\gamma+t\mid^{2}-\mid\gamma+\gamma^{^{\prime}}+t\mid^{2}}+O(\frac
{1}{\rho^{3\alpha_{1}}}),$%

\[
A_{k}(\gamma^{^{\prime}})=%
{\displaystyle\sum_{\gamma_{1},...,\gamma_{k-1}}}
\dfrac{q_{\gamma_{1}}q_{\gamma_{2}}...q_{\gamma_{k-1}}q_{\gamma^{^{\prime}%
}-\gamma_{1}-\gamma_{2}-...-\gamma_{k-1}}}{\prod_{j=0}^{k-1}(P(\gamma
+t)-\mid\gamma+t+\gamma^{^{\prime}}-\sum_{i=1}^{j}\gamma_{i}\mid^{2})}%
=O(\rho^{-k\alpha_{1}}),
\]%
\begin{equation}
\sum_{\gamma^{\ast}\in\Gamma((n-1)\rho^{\alpha})}\mid A_{1}(\gamma^{\ast}%
)\mid^{2}=O(\rho^{-2\alpha_{1}}),\sum_{\gamma^{\ast}\in\Gamma((n-1)\rho
^{\alpha})}\mid A_{k}(\gamma^{\ast})\mid=O(\rho^{-k\alpha_{1}})
\end{equation}
for $k>1.$ Now from (110) for $k=n$ and (112), we obtain
\begin{align*}
\Psi_{N}(x)  &  =b(N,\gamma)e^{i(\gamma+t,x)}+\\
&  \sum_{\gamma^{\ast}\in\Gamma((n-1)\rho^{\alpha})}\sum_{k=1}^{n-1}%
(A_{k}(\gamma^{\ast})b(N,\gamma)+O(\rho^{-n\alpha_{1}}))e^{i(\gamma
+t+\gamma^{\ast},x)})+H_{n}(x).
\end{align*}
Using the equalities $\parallel\Psi_{N}\parallel=1,$ $\arg b(N,\gamma)=0,$
$\parallel H_{n}\parallel=O(\rho^{-n\alpha_{1}})$ and taking into account that
the functions $e^{i(\gamma+t,x)},$ $H_{n}(x),$ $e^{i(\gamma+t+\gamma^{\ast
},x)},$ $(\gamma^{\ast}\in\Gamma((n-1)\rho^{\alpha}))$ are orthogonal, we get

$1=\mid b(N,\gamma)\mid^{2}+\sum_{k=1}^{n-1}(\sum_{\gamma^{\ast}\in
\Gamma((n-1)\rho^{\alpha})}\mid A_{k}(\gamma^{\ast})b(N,\gamma)\mid^{2}%
+O(\rho^{-n\alpha_{1}})),$
\begin{equation}
b(N,\gamma)=(1+\sum_{k=1}^{n-1}(\sum_{\gamma^{\ast}\in\Gamma((n-1)\rho
^{\alpha})}\mid A_{k}(\gamma^{\ast})\mid^{2}))^{-\frac{1}{2}}+O(\rho
^{-n\alpha_{1}}))
\end{equation}
(see the second equality in (113)). Thus from (112), we obtain
\begin{equation}
b(N,\gamma+\gamma^{^{\prime}})=(\sum_{k=1}^{n-1}A_{k}(\gamma^{^{\prime}%
}))(1+\sum_{k=1}^{n-1}\sum_{\gamma^{\ast}}\mid A_{k}(\gamma^{\ast})\mid
^{2})^{-\frac{1}{2}}+O(\rho^{-n\alpha_{1}}).
\end{equation}
Consider the case $n=2.$ By (114), (113), (115) we have $b(N,\gamma
)=1+O(\rho^{-2\alpha_{1}}),$

$b(N,\gamma+\gamma^{^{\prime}})=A_{1}(\gamma^{^{\prime}})+O(\rho^{-2\alpha
_{1}})=\dfrac{q_{\gamma^{^{\prime}}}}{\mid\gamma+t\mid^{2}-\mid\gamma
+\gamma^{^{\prime}}+t\mid^{2}}+O(\rho^{-2\alpha_{1}})$ for all $\gamma
^{^{\prime}}\in\Gamma(\rho^{\alpha}).$ These and (110) for $k=2$ imply the
formula for $\Phi_{1}$
\end{proof}

\section{Simple Sets and Isoenergetic Surfaces}

In this section we consider the simple sets $B$ and construct a big part of
the isoenergetic surfaces corresponding to $\rho^{2}$ for big $\rho.$ The
isoenergetic surfaces of $L(q)$ corresponding to $\rho^{2}$ is the set
$I_{\rho}(q(x))=\{t\in F^{\ast}:\exists N,\Lambda_{N}(t)=\rho^{2}\}.$ In the
case $q(x)=0$ \ the isoenergetic surface $I_{\rho}(0)=$ $\{t\in F^{\ast
}:\exists\gamma\in\Gamma,\mid\gamma+t\mid^{2}=\rho^{2}\}$ is the translation
of the sphere

$B(\rho)=\{\gamma+t:t\in F^{\ast},\gamma\in\Gamma,\mid\gamma+t\mid^{2}%
=\rho^{2}\}$ by the vectors $\gamma\in\Gamma.$ We call $B(\rho)$ the
translated isoenergetic surfaces of $L(0)$ corresponding to $\rho^{2}.$
Similarly, we call the sets $P_{\rho}^{^{\prime}}=\{\gamma+t:\Lambda
(\gamma+t)=\rho^{2}\}$ and

$P_{\rho}^{^{\prime\prime}}=\{t\in F^{\ast}:\exists\gamma\in\Gamma
,\Lambda(\gamma+t)=\rho^{2}\},$ where $\Lambda(\gamma+t)=\Lambda_{N(\gamma
+t)}(t)$ is defined in Remark 5, the parts of translated isoenergetic surfaces
and isoenergetic surfaces of $L(q).$ In this section we construct the subsets
$I_{\rho}^{^{\prime}}$ and $I_{\rho}^{^{\prime\prime}}$ of $P_{\rho}%
^{^{\prime}}$ and $P_{\rho}^{^{\prime\prime}}$ respectively and prove that the
measures of these subsets are asymptotically equal to the measure of the
isoenergetic surfaces $I_{\rho}(0)$ of $L(0)$. In other word we construct a
big part (in some sense) of isoenergetic surfaces $I_{\rho}(q(x))$ of $L(q)$.
As we see below the set $I_{\rho}^{^{\prime\prime}}$ is a translation of
$I_{\rho}^{^{\prime}}$ by vectors $\gamma\in\Gamma$ to $F^{\ast}$ and the set
$I_{\rho}^{^{\prime}}$ lies in $\varepsilon$ nieghborhood of the surface
$S_{\rho}=\{x\in U(2\rho^{\alpha_{1}},p):F(x)=\rho^{2}\},$ where $F(x)$ is
defined in Step 1 of introduction. Due to (102) it is natural to call
$S_{\rho}$ the approximated isoenergetic surfaces in the non-resonance domain.
Here we construct a part of the simple set $B$ in nieghborhood of $S_{\rho}$
that contains $I_{\rho}^{^{\prime}}$. For this we consider the surface
$S_{\rho}$. As we noted in introduction \ ( see Step 2 and (12)) the
non-resonance eigenvalue $\Lambda(\gamma+t)$ does not coincide with other
non-resonance eigenvalue $\Lambda(\gamma+t+b)$ if $\mid F(\gamma
+t)-F(\gamma+t+b)\mid>2\varepsilon_{1}$ for $\gamma+t+b\in U(\rho^{\alpha_{1}%
},p)$ and $b\in\Gamma\backslash\{0\}$. Therefore we eliminate
\begin{equation}
P_{b}=\{x:x,x+b\in U(\rho^{\alpha_{1}},p),\mid F(x)-F(x+b)\mid<3\varepsilon
_{1}\}
\end{equation}
for $b\in\Gamma\backslash\{0\}$ from $S_{\rho}$. Denote the remaining part of
$S_{\rho}$ by $S_{\rho}^{^{\prime}}.$ Then we consider the $\varepsilon$
neighbourhood $U_{\varepsilon}(S_{\rho}^{^{\prime}})=\cup_{a\in S_{\rho
}^{^{\prime}}}U_{\varepsilon}(a)\}$ of $S_{\rho}^{^{\prime}}$ ,

where $\varepsilon=\frac{\varepsilon_{1}}{7\rho},$ $U_{\varepsilon}(a)=\{x\in
R^{d}:\mid x-a\mid<\varepsilon\}.$ In this set the first simplicity condition
(12) holds (see Lemma 8(a)). Denote by

$Tr(E)=\{\gamma+x\in U_{\varepsilon}(S_{\rho}^{^{\prime}}):\gamma\in
\Gamma,x\in E\}$ and

$Tr_{F^{\star}}(E)\equiv\{\gamma+x\in F^{\star}:\gamma\in\Gamma,x\in E\}$ the
translations of $E\subset R^{d}$ into $U_{\varepsilon}(S_{\rho}^{^{\prime}})$
and $F^{\star}$ respectively. In order that the second simplicity condition
(13) holds, we discard from $U_{\varepsilon}(S_{\rho}^{^{\prime}})$ the
translation $Tr(A(\rho))$ of
\begin{equation}
A(\rho)\equiv\cup_{k=1}^{d-1}(\cup_{\gamma_{1},\gamma_{2},...,\gamma_{k}%
\in\Gamma(p\rho^{\alpha})}(\cup_{i=1}^{b_{k}}A_{k,i}(\gamma_{1},\gamma
_{2},...,\gamma_{k}))),
\end{equation}

where $A_{k,i}(\gamma_{1},...,\gamma_{k})=$

$\{x\in(\cap_{i=1}^{k}V_{\gamma_{i}}(\rho^{\alpha_{k}})\backslash E_{k+1})\cap
K_{\rho}:\lambda_{i}(x)\in(\rho^{2}-3\varepsilon_{1},\rho^{2}+3\varepsilon
_{1})\},$

$\lambda_{i}(x),$ $b_{k}$ is defined in Theorem 2 and
\begin{equation}
K_{\rho}=\{x\in\mathbb{R}^{d}:\mid\mid x\mid^{2}-\rho^{2}\mid<\rho^{\alpha
_{1}}\}.
\end{equation}
As a result we construct the part $U_{\varepsilon}(S_{\rho}^{^{\prime}%
})\backslash Tr(A(\rho))$ of the simple set $B$ (see Theorem 7(a)) which
contains the set $I_{\rho}^{^{\prime}}$ (see Theorem 7(c)). For this we need
the following lemma.

\begin{lemma}
$(a)$ If $x\in U_{\varepsilon}(S_{\rho}^{^{\prime}})$ and $x+b\in
U(\rho^{\alpha_{1}},p),$ where $b\in\Gamma,$ then
\begin{equation}
\mid F(x)-F(x+b)\mid>2\varepsilon_{1},
\end{equation}
where $\varepsilon=\frac{\varepsilon_{1}}{7\rho},\varepsilon_{1}%
=\rho^{-d-2\alpha},$ $F(x)=\mid x\mid^{2}+F_{k_{1}-1}(x),$ $k_{1}=[\frac
{d}{3\alpha}]+2,$ hence for $\gamma+t\in U_{\varepsilon}(S_{\rho}^{^{\prime}%
})$ the simplicity condition (12) holds.

$(b)$ If $x\in U_{\varepsilon}(S_{\rho}^{^{\prime}}),$ then $x+b\notin
U_{\varepsilon}(S_{\rho}^{^{\prime}})$ for all $b\in\Gamma$ $.$

$(c)$ If $E$ is a bounded subset of $\mathbb{R}^{d}$, then $\mu(Tr(E))\leq
\mu(E)$.

$(d)$ If $E\subset U_{\varepsilon}(S_{\rho}^{^{\prime}}),$ then $\mu
(Tr_{F^{\star}}(E))=\mu(E).$
\end{lemma}

\begin{proof}
$(a)$ If $x\in U_{\varepsilon}(S_{\rho}^{^{\prime}}),$ then there exists a
point $a$ in $S_{\rho}^{^{\prime}}$ such that $x\in U_{\varepsilon}(a)$. Since
$S_{\rho}^{^{\prime}}\cap P_{b}=\emptyset$ ( see (116) and the definition of
$S_{\rho}^{^{\prime}}$), we have
\begin{equation}
\mid F(a)-F(a+b)\mid\geq3\varepsilon_{1}%
\end{equation}
On the other hand, using (44) and the obvious relations

$\mid x\mid<\rho+1,$ $\mid x-a\mid<\varepsilon,$ $\mid x+b-a-b\mid
<\varepsilon,$ we obtain
\begin{equation}
\mid F(x)-F(a)\mid<3\rho\varepsilon,\mid F(x+b)-F(a+b)\mid<3\rho\varepsilon
\end{equation}
These inequalities together with (120) give (119), since $6\rho\varepsilon
<\varepsilon_{1}.$

$(b)$ If $x$ and $x+b$ lie in $U_{\varepsilon}(S_{\rho}^{^{\prime}}),$ then
there exist points $a$ and $c$ in $S_{\rho}^{^{\prime}}$ such that $x\in
U_{\varepsilon}(a)$ and $x+b\in U_{\varepsilon}(c).$ Repeating the proof of
(121), we get

$\mid F(c)-F(x+b)\mid<3\rho\varepsilon.$ This, the first inequality in (121),
and the relations $F(a)=\rho^{2},$ $F(c)=\rho^{2}$ (see the definition of
$S_{\rho})$ give

$\mid F(x)-F(x+b)\mid<\varepsilon_{1},$ which contradicts (119).

$(c)$ Clearly, for any bounded set $E$ there exist only a finite number of
vectors $\gamma_{1},\gamma_{2},...,\gamma_{s}$ such that $E(k)\equiv
(E+\gamma_{k})\cap U_{\varepsilon}(S_{\rho}^{^{\prime}})\neq\emptyset$ for
$k=1,2,...,s$ and $Tr(E)$ is the union of the sets $E(k)$. For $E(k)-\gamma
_{k}$ we have the relations $\mu(E(k)-\gamma_{k})=\mu(E(k)),$ $E(k)-\gamma
_{k}\subset E.$ Moreover, by $(b)$

$(E(k)-\gamma_{k})\cap(E(j)-\gamma_{j})=\emptyset$ for $k\neq j.$ Therefore
$(c)$ is true.

$(d)$ Now let $E\subset U_{\varepsilon}(S_{\rho}^{^{\prime}}).$ Then by $(b)$
the set $E$ can be divided into a finite number of the pairwise disjoint sets
$E_{1},E_{2},...,E_{n}$ such that there exist the vectors $\gamma_{1}%
,\gamma_{2},...,\gamma_{n}$ satisfying $(E_{k}+\gamma_{k})\subset F^{\star},$
$(E_{k}+\gamma_{k})\cap(E_{j}+\gamma_{j})\neq\emptyset$ for $k,j=1,2,...,n$
and $k\neq j.$ Using $\mu(E_{k}+\gamma_{k})=\mu(E_{k}),$ we get the proof of
$(d),$ since $Tr_{F^{\star}}(E)$ and $E$ are union of the pairwise disjoint
sets $E_{k}+\gamma_{k}$ and $E_{k}$ for $k=1,2,...,n$ respectively
\end{proof}

\begin{theorem}
$(a)$ The set $U_{\varepsilon}(S_{\rho}^{^{\prime}})\backslash Tr(A(\rho))$ is
a subset of $B$. For every connected open subset $E$ of $U_{\varepsilon
}(S_{\rho}^{^{\prime}})\backslash Tr(A(\rho)$ there exists a unique index $N$
such that $\Lambda_{N}(t)=\Lambda(\gamma+t)$ for $\gamma+t\in E,$ where
$\Lambda(\gamma+t)$ is defined in Remark 5. Moreover,
\begin{equation}
\frac{\partial}{\partial t_{j}}\Lambda(\gamma+t)=\frac{\partial}{\partial
t_{j}}\mid\gamma+t\mid^{2}+O(\rho^{1-2\alpha_{1}}),\forall j=1,2,...,d.
\end{equation}

$(b)$ For the part $V_{\rho}\equiv S_{\rho}^{^{\prime}}\backslash
U_{\varepsilon}(Tr(A(\rho)))$ of the approximated isoenergetic surface
$S_{\rho}$ the following holds
\begin{equation}
\mu(V_{\rho})>(1-c_{15}\rho^{-\alpha}))\mu(B(\rho)).
\end{equation}
Moreover, $U_{\varepsilon}(V_{\rho})$ lies in the subset $U_{\varepsilon
}(S_{\rho}^{^{\prime}})\backslash Tr(A(\rho))$ of the simple set $B.$

$(c)$ The isoenergetic surface $I(\rho)$ contains the set $I_{\rho}%
^{^{\prime\prime}},$ which consists of the smooth surfaces and has the
measure
\begin{equation}
\mu(I_{\rho}^{^{\prime\prime}})=\mu(I_{\rho}^{^{\prime}})>(1-c_{16}%
\rho^{-\alpha})\mu(B(\rho)),
\end{equation}
where $I_{\rho}^{^{\prime}}$ is a part of the translated isoenergetic surfaces
of $L(q),$ which is contained in the subset $U_{\varepsilon}(S_{\rho
}^{^{\prime}})\backslash Tr(A(\rho))$ of the simple set $B.$ In particular the
number $\rho^{2}$ for $\rho\gg1$ lies in the spectrum of $L(q),$ that is, the
number of the gaps in the spectrum of $L(q)$ is finite, where $q(x)\in
W_{2}^{s_{0}}(\mathbb{R}^{d}/\Omega),$ $d\geq2,$ $s_{0}=\frac{3d-1}{2}%
(3^{d}+d+2)+\frac{1}{4}d3^{d}+d+6,$ and $\Omega$ is an arbitrary lattice.
\end{theorem}

\begin{proof}
$(a)$ To prove that $U_{\varepsilon}(S_{\rho}^{^{\prime}})\backslash
Tr(A(\rho))\subset B$ we need to show that for each point $\gamma+t$ of
$U_{\varepsilon}(S_{\rho}^{^{\prime}})\backslash Tr(A(\rho))$ the simplicity
conditions (12), (13) hold and $U_{\varepsilon}(S_{\rho}^{^{\prime}%
})\backslash Tr(A(\rho))\subset U(\rho^{\alpha_{1}},p).$ By Lemma 8(a), the
condition (12) holds. Now we prove that (13) holds too. Since $\gamma+t\in
U_{\varepsilon}(S_{\rho}^{^{\prime}}),$ there exists $a\in S_{\rho}^{^{\prime
}}$ such that $\gamma+t\in U_{\varepsilon}(a).$ The inequality (121) and
equality $F(a)=\rho^{2}$ imply
\begin{equation}
F(\gamma+t)\in(\rho^{2}-\varepsilon_{1},\rho^{2}+\varepsilon_{1})
\end{equation}
for $\gamma+t\in U_{\varepsilon}(S_{\rho}^{^{\prime}}).$ On the other hand
$\gamma+t\notin Tr(A(\rho)).$ It means that for any $\gamma^{^{\prime}}%
\in\Gamma,$ we have $\gamma^{^{\prime}}+t\notin A(\rho).$ If $\gamma
^{^{\prime}}\in K$ and $\gamma^{^{\prime}}+t\in E_{k}\backslash E_{k+1},$ then
by definition of $K$ ( see introduction) the inequality $\mid F(\gamma
+t)-\mid\gamma^{^{\prime}}+t\mid^{2}\mid<\frac{1}{3}\rho^{\alpha_{1}}$ holds.
This and (125) imply that $\gamma^{^{\prime}}+t\in(E_{k}\backslash
E_{k+1})\cap K_{\rho}$ ( see (118) for the definition of $K_{\rho}$). Since
$\gamma^{^{\prime}}+t\notin A(\rho),$ we have $\lambda_{i}(\gamma^{^{\prime}%
}+t)\notin(\rho^{2}-3\varepsilon_{1},\rho^{2}+3\varepsilon_{1})$ for
$\gamma^{^{\prime}}\in K$ and $\gamma^{^{\prime}}+t\in E_{k}\backslash
E_{k+1}.$ Therefore (13) follows from (125). Moreover, it is clear that the
inclusion $S_{\rho}^{^{\prime}}\subset U(2\rho^{\alpha_{1}},p)$ ( see
definition of $S_{\rho}$ and $S_{\rho}^{^{\prime}}$) implies that
$U_{\varepsilon}(S_{\rho}^{^{\prime}})\subset U(\rho^{\alpha_{1}},p).$ Thus
$U_{\varepsilon}(S_{\rho}^{^{\prime}})\backslash Tr(A(\rho))\subset B.$

Now let $E$ be a connected open subset of $U_{\varepsilon}(S_{\rho}^{^{\prime
}})\backslash Tr(A(\rho)\subset B.$ By Theorem 5 and Remark 5 for $a\in
E\subset U_{\varepsilon}(S_{\rho}^{^{\prime}})\backslash Tr(A(\rho)$ there
exists a unique index $N(a)$ such that $\Lambda(a)=\Lambda_{N(a)}(a)$,
$\Psi_{a}(x)=\Psi_{N(a),a}(x)$, $\mid(\Psi_{N(a),a}(x),e^{i(a,x)})\mid
^{2}>\frac{1}{2}$ and $\Lambda(a)$ is a simple eigenvalue. On the other hand,
for fixed $N$ the functions $\Lambda_{N}(t)$ and $(\Psi_{N,t}(x),e^{i(t,x)})$
are continuous in a neighborhood of $a$ if $\Lambda_{N}(a)$ is a simple
eigenvalue. Therefore for each $a\in E$ there exists a neighborhood
$U(a)\subset E$ of $a$ such that $\mid(\Psi_{N(a),y}(x),e^{i(y,x)})\mid
^{2}>\frac{1}{2}$, for $y\in U(a).$ Since for $y\in E$ there is a unique
integer $N(y)$ satisfying $\mid(\Psi_{N(y),y}(x),e^{i(y,x)})\mid^{2}>\frac
{1}{2},$ we have $N(y)=N(a)$ for $y\in U(a).$ Hence we proved that%
\begin{equation}
\forall a\in E,\exists U(a)\subset E:N(y)=N(a),\forall y\in U(a).
\end{equation}

Now let $a_{1}$ and $a_{2}$ be two points of $E$ , and let $C\subset E$ be the
arc that joins these points. Let $U(y_{1}),U(y_{2}),...,U(y_{k})$ be a finite
subcover of the open cover $\cup_{a\in C}U(a)$ of the compact $C,$ where
$U(a)$ is the neighborhood of $a$ satisfying (126). By (126), we have
$N(y)=N(y_{i})=N_{i}$ for $y\in U(y_{i}).$ Clearly, if

$U(y_{i})\cap U(y_{j})\neq\emptyset,$ then $N_{i}=N(z)=N_{j},$ where $z\in
U(y_{i})\cap U(y_{j})$. Thus $N_{1}=N_{2}=...=N_{k}$ and $N(a_{1})=N(a_{2}).$

To calculate the partial derivatives of the function $\Lambda(\gamma
+t)=\Lambda_{N}(t)$ we write the operator $L_{t}$ in the form $-\triangle
-(2it,\nabla)+(t,t).$ Then, it is clear that
\begin{align}
\frac{\partial}{\partial t_{j}}\Lambda_{N}(t)  &  =2t_{j}(\Phi_{N,t}%
(x),\Phi_{N,t}(x))-2i(\frac{\partial}{\partial x_{j}}\Phi_{N,t}(x),\Phi
_{N,t}(x)),\\
\Phi_{N,t}(x)  &  =\sum_{\gamma^{^{\prime}}\in\Gamma}b(N,\gamma^{^{\prime}%
})e^{i(\gamma^{^{\prime}},x)},
\end{align}
where $\Phi_{N,t}(x)=e^{-i(t,x)}\Psi_{N,t}(x).$ If $\mid\gamma^{^{\prime}}%
\mid\geq2\rho,$ then using

$\Lambda_{N}\equiv\Lambda(\gamma+t)=\rho^{2}+O(\rho^{-\alpha}),$ ( see (102),
(125)), and the obvious inequality

$\mid\Lambda_{N}-\mid\gamma^{^{\prime}}-\gamma_{1}-\gamma_{2}-...-\gamma
_{k}+t\mid^{2}\mid>c_{17}\mid\gamma^{^{\prime}}\mid^{2}$ for $k=0,1,...,p,$
where $\mid\gamma_{1}\mid<\frac{1}{4p}\mid\gamma^{^{\prime}}\mid,$ and
iterating (18) $p$ times by using the decomposition $q(x)=\sum_{\mid\gamma
_{1}\mid<\frac{1}{4p}\mid\gamma^{^{\prime}}\mid}q_{\gamma_{1}}e^{i(\gamma
_{1},x)}+O(\mid\gamma^{^{\prime}}\mid^{-p}),$ we get
\begin{align}
b(N,\gamma^{^{\prime}})  &  =\sum_{\gamma_{1},\gamma_{2},...}\dfrac
{q_{\gamma_{1}}q_{\gamma_{2}}...q_{\gamma_{p}}b(N,\gamma^{^{\prime}}%
-\sum_{i=1}^{p}\gamma_{i})}{\prod_{j=0}^{p-1}(\Lambda_{N}-\mid\gamma
^{^{\prime}}-\sum_{i=1}^{j}\gamma_{i}+t\mid^{2})}+O(\mid\gamma^{^{\prime}}%
\mid^{-p}),\\
b(N,\gamma^{^{\prime}})  &  =O(\mid\gamma^{^{\prime}}\mid^{-p}),\text{
}\forall\mid\gamma^{^{\prime}}\mid\geq2\rho
\end{align}
By (130) the series in (128) can be differentiated term by term. Hence
\begin{equation}
-i(\frac{\partial}{\partial x_{j}}\Phi_{N,t},\Phi_{N,t})=\sum_{\gamma
^{^{\prime}}\in\Gamma}\gamma^{^{\prime}}(j)\mid b(N,\gamma^{^{\prime}}%
)\mid^{2}=\gamma(j)\mid b(N,\gamma)\mid^{2}+\Sigma_{1}+\Sigma_{2},
\end{equation}
where $\Sigma_{1}=\sum_{\mid\gamma^{^{\prime}}\mid\geq2\rho}\gamma^{^{\prime}%
}(j)\mid b(N,\gamma^{^{\prime}})\mid^{2},$ $\Sigma_{2}=\sum_{\mid
\gamma^{^{\prime}}\mid<2\rho,\gamma^{^{\prime}}\neq\gamma}\gamma^{^{\prime}%
}(j)\mid b(N,\gamma^{^{\prime}})\mid^{2}.$

By (14), $\sum_{2}=O(\rho^{-2\alpha_{1}+1}),$ $\gamma(j)\mid b(N,\gamma
)\mid^{2}=\gamma(j)(1+O(\rho^{-2\alpha_{1}}),$ and by (130), $\sum_{1}%
=O(\rho^{-2\alpha_{1}}).$ Therefore (127) and (131) imply (122).

$(b)$ To prove the inclusion $U_{\varepsilon}(V_{\rho})\subset$
$U_{\varepsilon}(S_{\rho}^{^{\prime}})\backslash Tr(A(\rho))$ we need to show
that if $a\in V_{\rho},$ then $U_{\varepsilon}(a)\subset U_{\varepsilon
}(S_{\rho}^{^{\prime}})\backslash Tr(A(\rho)).$ This is clear, since the
relations $a\in V_{\rho}\subset S_{\rho}^{^{\prime}}$ imply that
$U_{\varepsilon}(a)\subset U_{\varepsilon}(S_{\rho}^{^{\prime}})$ and the
relation $a\notin U_{\varepsilon}(Tr(A(\rho)))$ implies that $U_{\varepsilon
}(a)\cap Tr(A(\rho))=\emptyset.$ To prove (123) first we estimate the measure
of $S_{\rho},S_{\rho}^{^{\prime}},U_{2\varepsilon}(A(\rho))$, namely we prove
\begin{align}
\mu(S_{\rho})  &  >(1-c_{18}\rho^{-\alpha})\mu(B(\rho)),\\
\mu(S_{\rho}^{^{\prime}})  &  >(1-c_{19}\rho^{-\alpha})\mu(B(\rho)),\\
\mu(U_{2\varepsilon}(A(\rho)))  &  =O(\rho^{-\alpha})\mu(B(\rho))\varepsilon
\end{align}
( see below, Estimations 1, 2, 3). The estimation (123) of the measure of the
set $V_{\rho}$ is done in Estimation 4 by using Estimations 1, 2, 3.

$(c)$ In Estimation 5 we prove the formula (124). The Theorem is proved
\end{proof}

In Estimations 1-5 we use the notations: $G(+i,a)=\{x\in G,x_{i}>a\},$
$G(-i,a)=\{x\in G,x_{i}<-a\},$ where $x=(x_{1},x_{2},...,x_{d}),a>0.$ It is
not hard to verify that for any subset $G$ of $U_{\varepsilon}(S_{\rho
}^{^{\prime}})\cup U_{2\varepsilon}(A(\rho))$ , that is, for all considered
sets $G$ in these estimations, and for any $x\in G$ the followings hold
\begin{equation}
\rho-1<\mid x\mid<\rho+1,\text{ }G\subset(\cup_{i=1}^{d}(G(+i,\rho d^{-1})\cup
G(-i,\rho d^{-1}))
\end{equation}
Indeed, if $x\in S_{\rho}^{^{\prime}},$ then $F(x)=\rho^{2}$ and by definition
of $F(x)$ ( see Lemma 8) and (24) we have $\mid x\mid=\rho+O(\rho
^{-1-\alpha_{1}}).$ Hence the inequalities in (135) hold for $x\in
U_{\varepsilon}(S_{\rho}^{^{\prime}}).$ If $x\in$ $A(\rho),$ then by
definition of $A(\rho)$ ( see (117), (118)), we have $x\in K_{\rho},$ and
hence $\mid x\mid=\rho+O(\rho^{-1+\alpha_{1}})$. Thus the inequalities in
(135) hold for $x\in U_{2\varepsilon}(A(\rho))$ too. The inclusion in (135)
follows from these inequalities.

If $G$ $\subset S_{\rho},$ then by (44) we have $\frac{\partial F(x)}{\partial
x_{k}}>0$ for $x\in G(+k,\rho^{-\alpha})$. Therefore to calculate the measure
of $G(+k,a)$ for $a\geq\rho^{-\alpha}$ we use the formula
\begin{equation}
\mu(G(+k,a))=\int\limits_{\Pr_{k}(G(+k,a))}(\frac{\partial F}{\partial x_{k}%
})^{-1}\mid grad(F)\mid dx_{1}...dx_{k-1}dx_{k+1}...dx_{d},
\end{equation}
where $\Pr_{k}(G)\equiv\{(x_{1},x_{2},...,x_{k-1},x_{k+1},x_{k+2}%
,...,x_{d}):x\in G\}$ is the projection of $G$ on the hyperplane $x_{k}=0.$
Instead of $\Pr_{k}(G)$ we write $\Pr(G)$ if $k$ is unambiguous. If $D$ is
$m-$dimensional subset of $\mathbb{R}^{m},$ then to estimate $\mu(D),$ we use
the formula
\begin{equation}
\mu(D)=\int\limits_{\Pr_{k}(D)}\mu(D(x_{1},...x_{k-1},x_{k+1},...,x_{m}%
))dx_{1}...dx_{k-1}dx_{k+1}...dx_{m},
\end{equation}
where $D(x_{1},...x_{k-1},x_{k+1},...,x_{m})=\{x_{k}:(x_{1},x_{2}%
,...,x_{m})\in D\}.$

ESTIMATION 1. Here we prove (132) by using (136). During this estimation the
set $S_{\rho}$ is redenoted by $G.$ First we estimate $\mu(G(+1,a))$ for
$a=\rho^{1-\alpha}$ by using (136) for $k=1$ and \ the relations%
\begin{equation}
\frac{\partial F}{\partial x_{1}}>\rho^{1-\alpha},(\frac{\partial F}{\partial
x_{1}})^{-1}\mid grad(F)\mid=\frac{\rho}{\sqrt{\rho^{2}-x_{2}^{2}-x_{3}%
^{2}-...-x_{d}^{2}}}+O(\rho^{-\alpha}),
\end{equation}%
\begin{equation}
\Pr(G(+1,a))\supset\Pr(A(+1,2a)),
\end{equation}
where $x\in G(+1,a),$ $A=B(\rho)\cap U(3\rho^{\alpha_{1}},p).$ Here (138)
follows from (44), (24), and the definitions of $F(x),$ $S_{\rho}.$ Now we
prove (139). If 

$(x_{2},...,x_{d})\in\Pr_{1}(A(+1,2a)),$ then by definition of $A(+1,2a)$
there exists $x_{1}$ such that
\begin{equation}
x_{1}>2a=2\rho^{1-\alpha},\text{ }x_{1}^{2}+x_{2}^{2}+...+x_{d}^{2}=\rho
^{2},\mid\sum_{i\geq1}(2x_{i}b_{i}-b_{i}^{2})\mid\geq3\rho^{\alpha_{1}}%
\end{equation}
for all $(b_{1},b_{2},...,b_{d})\in\Gamma(p\rho^{\alpha}).$ Therefore for
$h=\rho^{-\alpha}$ we have

$(x_{1}+h)^{2}+x_{2}^{2}+...+x_{q}^{2}>\rho^{2}+\rho^{-\alpha},(x_{1}%
-h)^{2}+x_{2}^{2}+...+x_{q}^{2}<\rho^{2}-\rho^{-\alpha}.$ This and (24) give
$F(x_{1}+h,x_{2},...,x_{d})>\rho^{2}$, $F(x_{1}-h,x_{2},...,x_{d})<\rho^{2}$.
Since $F$ is a continuous function there is $y_{1}\in(x_{1}-h,x_{1}+h)$ such
that (see (140))
\begin{equation}
y_{1}>a,F(y_{1},x_{2},...,x_{d})=\rho^{2},\mid2y_{1}b_{1}-b_{1}^{2}%
+\sum_{i\geq2}(2x_{i}b_{i}-b_{i}^{2})\mid>\rho^{\alpha_{1}},
\end{equation}
since the expression under the absolute value in (141) differ from the
expression under the absolute value in (140) by $2(y_{1}-x_{1})b_{1},$ where
$\mid y_{1}-x_{1}\mid<h=\rho^{-\alpha},$ $\mid b_{1}\mid<p\rho^{\alpha},$
$\mid2(y_{1}-x_{1})b_{1}\mid<2p\rho^{2\alpha}<$ $\rho^{\alpha_{1}}.$ The
relations in (141) means that $(x_{2},...,x_{d})\in\Pr G(+1,a).$ Hence (139)
is proved. Now (136), (138), and the obvious relation $\mu(\Pr G(+1,a))=O(\rho
^{d-1})$ ( see (135)) imply that%
\[
\mu(G(+1,a))=\int\limits_{\Pr(G(+1,a))}\frac{\rho}{\sqrt{\rho^{2}-x_{2}%
^{2}-x_{3}^{2}-...-x_{d}^{2}}}dx_{2}dx_{3}...dx_{d}+O(\frac{1}{\rho^{\alpha}%
})\mu(B(\rho))
\]%
\[
\geq\int\limits_{\Pr(A(+1,2a))}\frac{\rho}{\sqrt{\rho^{2}-x_{2}^{2}-x_{3}%
^{2}-...-x_{d}^{2}}}dx_{2}dx_{3}...dx_{d}-c_{20}\rho^{-\alpha}\mu(B(\rho))
\]%
\[
=\mu(A(+1,2a))-c_{20}\rho^{-\alpha}\mu(B(\rho)).
\]
Similarly, $\mu(G(-1,a))\geq\mu(A(-1,2a))-c_{20}\rho^{-\alpha}\mu(B(\rho)).$
Now using \ the inequality $\mu(G)\geq\mu(G(+1,a))+\mu(G(-1,a))$ we get

$\mu(G)\geq\mu(A(-1,2a))+\mu(A(+1,2a))-2c_{20}\rho^{-\alpha}\mu(B(\rho)).$ On
the other hand it follows from the obvious relation\ 

\ $\mu(\{x\in B(\rho):-2a\leq x_{1}\leq2a\})=O(\rho^{-\alpha})\mu(B(\rho))$ that

$\mu(A(-1,2a))+\mu(A(+1,2a))\geq\mu(A)-c_{20}\rho^{-\alpha}\mu(B(\rho)).$ Therefore

$\mu(G)>\mu(A)-3c_{20}\rho^{-\alpha}\mu(B(\rho)).$ It implies (132), since

$\mu(A))=(1+O(\rho^{-\alpha}))\mu(B(\rho))$ (see (42) ).

ESTIMATION 2. Here we prove (133). For this we estimate the measure of the set
$S_{\rho}\cap P_{b}$ by using (136). During this estimation the set $S_{\rho
}\cap P_{b}$ is redenoted by $G$. We choose the coordinate axis so that the
direction of $b$ coincides with the direction of $(1,0,0,...,0),$ i.e.,
$b=(b_{1},0,0,...,0)$ and $b_{1}>0$. It follows from the definitions of
$S_{\rho},P_{b}$ and $F(x)$ ( see the beginning of this section, (116), and
Lemma 8(a)) that if $(x_{1},x_{2},...,x_{d})\in G,$ then
\begin{align}
x_{1}^{2}+x_{2}^{2}+...+x_{d}^{2}+F_{k_{1}-1}(x) &  =\rho^{2},\\
(x_{1}+b_{1})^{2}+x_{2}^{2}+x_{3}^{2}+...+x_{d}^{2}+F_{k_{1}-1}(x+b) &
=\rho^{2}+h,
\end{align}
where $h\in(-3\varepsilon_{1},3\varepsilon_{1}).$ Subtracting (142) from (143)
and using (24), we get
\begin{equation}
(2x_{1}+b_{1})b_{1}=O(\rho^{-\alpha_{1}}).
\end{equation}
This and the inequalities in (135) imply
\begin{equation}
\mid b_{1}\mid<2\rho+3,\text{ }x_{1}=\frac{b_{1}}{2}+O(\rho^{-\alpha_{1}}%
b_{1}^{-1}),\mid x_{1}^{2}-(\frac{b_{1}}{2})^{2}\mid=O(\rho^{-\alpha_{1}}).
\end{equation}
Consider two cases. Case 1: $b\in\Gamma_{1},$ where $\Gamma_{1}=\{b\in
\Gamma:\mid\rho^{2}-\mid\frac{b}{2}\mid^{2}\mid<3d\rho^{-2\alpha}\}.$ In this
case using the last equality in (145), (142), (24), and taking into account
that $b=(b_{1},0,0,...,0),$ $\alpha_{1}=3\alpha,$ we obtain
\begin{equation}
x_{1}^{2}=\rho^{2}+O(\rho^{-2\alpha}),\mid x_{1}\mid=\rho+O(\rho^{-2\alpha
-1}),x_{2}^{2}+x_{3}^{2}+...+x_{d}^{2}=O(\rho^{-2\alpha}).
\end{equation}
Therefore $G\subset G(+1,a)\cup G(-1,a),$ where $a=\rho-\rho^{-1}$. Using
(136), the obvious relation $\mu(\Pr_{1}(G(\pm1,a))=O(\rho^{-(d-1)\alpha})$
(see (146)) and taking into account that the expression under the integral in
(136) for $k=1$ is equal to $1+O(\rho^{-\alpha})$ (see (146)), we get
$\mu(G(\pm1,a))=O(\rho^{-(d-1)\alpha}).$ Thus $\mu(G)=O(\rho^{-(d-1)\alpha}).$
Since $\mid\Gamma_{1}\mid=O(\rho^{d-1}),$ we have
\begin{equation}
\text{ }\mu(\cup_{b\in\Gamma_{1}}(S_{\rho}\cap P_{b})=O(\rho^{-(d-1)\alpha
+d-1})=O(\rho^{-\alpha})\mu(B(\rho)).
\end{equation}
Case 2: $\mid\rho^{2}-\mid\frac{b}{2}\mid^{2}\mid\geq3d\rho^{-2\alpha}.$
Repeating the proof of (146), we get%
\begin{equation}
\mid x_{1}^{2}-\rho^{2}\mid>2d\rho^{-2\alpha},\text{ }\sum_{k=2}^{d}x_{k}%
^{2}>d\rho^{-2\alpha},\text{ }\max_{k\geq2}\mid x_{k}\mid>\rho^{-\alpha}.
\end{equation}
Therefore $G\subset\cup_{k\geq2}(G(+k,\rho^{-\alpha})\cup G(-k,\rho^{-\alpha
})).$ Now we estimate $\mu(G(+d,\rho^{-\alpha}))$ by using (136). Redenote by
$D$ the set $\Pr_{d}G(+d,\rho^{-\alpha}).$ If $x\in G(+d,\rho^{-\alpha}),$
then according to (142) and (44) the under integral expression in (136) for
$k=d$ is $O(\rho^{1+\alpha}).$ Therefore the first equality in
\begin{equation}
\mu(D)=O(\varepsilon_{1}\mid b\mid^{-1}\rho^{d-2}),\text{ }\mu(G(+d,\rho
^{-\alpha}))=O(\rho^{d-1+\alpha}\varepsilon_{1}\mid b\mid^{-1})
\end{equation}
implies the second equality in (149). To prove the first equality in (149) we
use (137) for $m=d-1$ and $k=1$ and prove the relations $\mu(\Pr_{1}%
D)=O(\rho^{d-2}),$
\begin{equation}
\text{ }\mu(D(x_{2},x_{3},...,x_{d-1}))<6\varepsilon_{1}\mid b\mid^{-1}%
\end{equation}
for $(x_{2},x_{3},...,x_{d-1})\in\Pr_{1}D.$ First relation follows from the
inequalities in (135)). So we need to prove (150). If $x_{1}\in D(x_{2}%
,x_{3},...,x_{d-1})$ then (142) and (143) hold. Subtracting (142) from (143),
we get
\begin{equation}
2x_{1}b_{1}+(b_{1})^{2}+F_{k_{1}-1}(x-b)-F_{k_{1}-1}(x)=h,
\end{equation}
where $x_{2},x_{3},...,x_{d-1}$ are fixed . Hence we have two equations (142)
and (151) with respect two unknown $x_{1}$ and $x_{d}$. Using (44), the
implicit function theorem, and the inequalities $\mid x_{d}\mid>\rho^{-\alpha
},$ $\alpha_{1}>2\alpha$ from (142), we obtain
\begin{equation}
x_{d}=f(x_{1}),\text{ }\frac{df}{dx_{1}}=-\frac{2x_{1}+O(\rho^{-2\alpha
_{1}+\alpha})}{2x_{d}+O(\rho^{-2\alpha_{1}+\alpha})}.
\end{equation}
Substituting this in (151), we get
\begin{equation}
2x_{1}b_{1}+b_{1}^{2}+F_{k_{1}-1}(x_{1}+b_{1},x_{2},...,x_{d-1},f(x_{1}%
))-F_{k_{1}-1}(x_{1},...,x_{d-1},f)=h.
\end{equation}
Using (44), (152), the first equality in (145), and $x_{d}>\rho^{-\alpha}$ we
see that the absolute value of the derivative (w.r.t. $x_{1}$) of the
left-hand side of (153) satisfies the inequality

$\mid2b_{1}+O(\rho^{-2\alpha_{1}+\alpha})+O(\rho^{-2\alpha_{1}+\alpha}%
)\frac{x_{1}+O(\rho^{-2\alpha_{1}+\alpha})}{x_{d}+O(\rho^{-2\alpha_{1}+\alpha
})})\mid>b_{1}$ 

for $x_{1}=\frac{b_{1}}{2}+O(\rho^{-\alpha_{1}})$ (see (145)). Therefore from
(153) by implicit function theorem, we get $\mid\frac{dx_{1}}{dh}\mid<\frac
{1}{\mid b\mid}.$ This inequality and relation $h\in(-3\varepsilon
_{1},3\varepsilon_{1})$ imply (150). Thus (149) is proved. In the same way we
get the same estimation for $G(+k,\rho^{-\alpha})$ and $G(-k,\rho^{-\alpha})$
for $k\geq2$. Hence

$\mu(S_{\rho}\cap P_{b})=O(\rho^{d-1+\alpha}\varepsilon_{1}\mid b\mid^{-1}),$
for $b\notin\Gamma_{1}.$ Since $\mid b\mid<2\rho+3$ ( see (145)) and
$\varepsilon_{1}=\rho^{-d-2\alpha}$, taking into account that the number of
the vectors of $\Gamma$ satisfying $\mid b\mid<2\rho+3$ is $O(\rho^{d}),$ we obtain

$\mu(\cup_{b\notin\Gamma_{1}}(S_{\rho}\cap P_{b}))=O(\rho^{2d-1+\alpha
}\varepsilon_{1})=O(\rho^{-\alpha})\mu(B(\rho)).$ This, (147) and (132) give
the proof of (133).

ESTIMATION 3. Here we prove (134). Denote $U_{2\varepsilon}(A_{k,j}%
(\gamma_{1,}\gamma_{2},...,\gamma_{k}))$ by $G,$ where $\gamma_{1,}\gamma
_{2},...,\gamma_{k}\in\Gamma(p\rho^{\alpha}),k\leq d-1,$ and $A_{k,j}$ is
defined in (117). We turn the coordinate axis so that

$Span\{\gamma_{1,}\gamma_{2},...,\gamma_{k}\}=\{x=(x_{1},x_{2},...,x_{k}%
,0,0,...,0):x_{1},x_{2},...,x_{k}\in\mathbb{R}\}$. Then by (32), we have
$x_{n}=O(\rho^{\alpha_{k}+(k-1)\alpha})$ for $n\leq k,$ $x\in G$. This, (135), and

$\alpha_{k}+(k-1)\alpha<1$ ( see the sixth inequality in (15)) give

$G\subset(\cup_{i>k}(G(+i,\rho d^{-1})\cup G(-i,\rho d^{-1})),$

$\mu(\Pr_{i}(G(+i,\rho d^{-1})))=O(\rho^{k(\alpha_{k}+(k-1)\alpha)+(d-1-k)})$
for $i>k.$ Now using this and (137) for $m=d,$ we prove that
\begin{equation}
\mu(G(+i,\rho d^{-1}))=O(\varepsilon\rho^{k(\alpha_{k}+(k-1)\alpha
)+(d-1-k)}),\forall i>k.
\end{equation}
For this we redenote by $D$ the set $G(+i,\rho d^{-1})$ and prove that
\begin{equation}
\mu((D(x_{1},x_{2},...x_{i-1},x_{i+1},...x_{d}))\leq(42d^{2}+4)\varepsilon
\end{equation}
for $(x_{1},x_{2},...x_{i-1},x_{i+1},...x_{d})\in\Pr_{i}(D)$ and $i>k.$ To
prove (155) it is sufficient to show that if both $x=(x_{1},x_{2}%
,...,x_{i},...x_{d})$ and $x^{^{\prime}}=(x_{1},x_{2},...,x_{i}^{^{\prime}%
},...,x_{d})$ are in $D,$ then $\mid x_{i}-x_{i}^{^{\prime}}\mid\leq
(42d^{2}+4)\varepsilon.$ Assume the converse. Then

$\mid x_{i}-x_{i}^{^{\prime}}\mid>(42d^{2}+4)\varepsilon$. Without loss of
generality it can be assumed that $x_{i}^{^{\prime}}>x_{i}.$ So $x_{i}%
^{^{\prime}}>x_{i}>\rho d^{-1}$ ( see definition of $D$). Since $x$ and
$x^{^{\prime}}$ lie in the $2\varepsilon$ neighborhood of $A_{k,j},$ there
exist points $a$ and $a^{^{\prime}}$ in $A_{k,j}$ such that $\mid
x-a\mid<2\varepsilon$ and $\mid x^{^{\prime}}-a^{^{\prime}}\mid<2\varepsilon.$
It follows from the definitions of the points $x,$ $x^{^{\prime}},a,$
$a^{^{\prime}}$ that the following inequalities hold:%
\begin{align}
\rho d^{-1}-2\varepsilon &  <a_{i}<a_{i}^{^{\prime}},\text{ }a_{i}^{^{\prime}%
}-a_{i}>42d^{2}\varepsilon,\\
(a_{i}^{^{\prime}})^{2}-(a_{i})^{2}  &  >2(\rho d^{-1}-2\varepsilon
)(a_{i}^{^{\prime}}-a_{i}),\nonumber\\
&  \mid\mid a_{s}\mid-\mid a_{s}^{^{\prime}}\mid\mid<4\varepsilon,\forall
s\neq i.\nonumber
\end{align}
On the other hand the inequalities in (135) hold for the points of $A_{k,j}$ ,
that is, we have $\mid a_{s}\mid<\rho+1,\mid a_{s}^{^{\prime}}\mid<\rho+1.$
Therefore these inequalities and the inequalities in (156) imply $\mid\mid
a_{s}\mid^{2}-\mid a_{s}^{^{\prime}}\mid^{2}\mid<12\rho\varepsilon$ for $s\neq
i$, and hence

$\sum_{s\neq i}\mid\mid a_{s}\mid^{2}-\mid a_{s}^{^{\prime}}\mid^{2}%
\mid<12d\rho\varepsilon<\frac{2}{7}\rho d^{-1}(a_{i}^{^{\prime}}-a_{i}),$%

\begin{equation}
\mid\mid a\mid^{2}-\mid a^{^{\prime}}\mid^{2}\mid>\frac{3}{2}\rho d^{-1}\mid
a_{i}^{^{\prime}}-a_{i}\mid.
\end{equation}
Now using the inequality (46), the obvious relation $\frac{1}{2}\alpha_{d}<1$
( see the end of the introduction), the notations $r_{j}(x)=\lambda
_{j}(x)-\mid x\mid^{2}$ ( see Remark 2), $\varepsilon_{1}=7\rho\varepsilon$ (
see Lemma 8(a)), and (157), (156), we get

$\mid r_{j}(a)-r_{j}(a^{^{\prime}})\mid<\rho^{\frac{1}{2}\alpha_{d}}\mid
a-a^{^{\prime}}\mid<\frac{1}{2}\rho d^{-1}\mid a_{i}^{^{\prime}}-a_{i}\mid,$

$\mid\lambda_{j}(a)-\lambda_{j}(a^{^{\prime}})\mid\geq\mid\mid a\mid^{2}-\mid
a^{^{\prime}}\mid^{2}\mid-\mid r_{j}(a)-r_{j}(a^{^{\prime}})\mid>$

$\rho d^{-1}\mid a_{i}^{^{\prime}}-a_{i}\mid>42d\rho\varepsilon>6\varepsilon
_{1}.$

The obtained inequality $\mid\lambda_{j}(a)-\lambda_{j}(a^{^{\prime}}%
)\mid>6\varepsilon_{1}$ contradicts with the inclusions $a$ $\in A_{k,j},$
$a^{^{\prime}}\in A_{k,j},$ since by definition of $A_{k,j}$ ( see (117)) both
$\lambda_{j}(a)$ and $\lambda_{j}(a^{^{\prime}})$ lie in $(\rho^{2}%
-3\varepsilon_{1},\rho^{2}+3\varepsilon_{1}).$ Thus (155), hence (154) is
proved. In the same way we get the same formula for $G(-i,\frac{\rho}{d}).$ So

$\mu(U_{2\varepsilon}(A_{k,j}(\gamma_{1,}\gamma_{2},...,\gamma_{k}%
)))=O(\varepsilon\rho^{k(\alpha_{k}+(k-1)\alpha)+d-1-k}).$ Now taking into
account that $U_{2\varepsilon}(A(\rho))$ is union of $U_{2\varepsilon}%
(A_{k,j}(\gamma_{1,}\gamma_{2},...,\gamma_{k})$ for $k=1,2,..,d-1;$
$j=1,2,...,b_{k}(\gamma_{1},\gamma_{2},...,\gamma_{k}),$ and $\gamma
_{1},\gamma_{2},...,\gamma_{k}\in\Gamma(p\rho^{\alpha})$ \ ( see (117)) and
using that $b_{k}=O(\rho^{d\alpha+\frac{k}{2}\alpha_{k+1}})$ ( see (40)) and
the number of the vectors $(\gamma_{1},\gamma_{2},...,\gamma_{k})$ for
$\gamma_{1},\gamma_{2},...,\gamma_{k}\in\Gamma(p\rho^{\alpha})$ is
$O(\rho^{dk\alpha}),$ we obtain
\[
\mu(U_{2\varepsilon}(A(\rho)))=O(\varepsilon\rho^{d\alpha+\frac{k}{2}%
\alpha_{k+1}+dk\alpha+k(\alpha_{k}+(k-1)\alpha)+d-1-k}).
\]
Therefore to prove (134), it remains to show that

$d\alpha+\frac{k}{2}\alpha_{k+1}+dk\alpha+k(\alpha_{k}+(k-1)\alpha)+d-1-k\leq
d-1-\alpha$ or%
\begin{equation}
(d+1)\alpha+\frac{k}{2}\alpha_{k+1}+dk\alpha+k(\alpha_{k}+(k-1)\alpha)\leq k
\end{equation}
for $1\leq k\leq d-1$. Dividing both side of (158) by $k\alpha$ and using
$\alpha_{k}=3^{k}\alpha,$ $\alpha=\frac{1}{q},$ $q=3^{d}+d+2$ ( see the end of
the introduction) we see that (158) is equivalent to $\frac{d+1}{k}%
+\frac{3^{k+1}}{2}+3^{k}+k-1\leq3^{d}+2.$ The left-hand side of this
inequality gets its maximum at $k=d-1.$ Therefore we need to show that
$\frac{d+1}{d-1}+\frac{5}{6}3^{d}+d\leq3^{d}+4,$ which follows from the
inequalities $\frac{d+1}{d-1}\leq3,d<\frac{1}{6}3^{d}+1$ for $d\geq2.$

ESTIMATION 4. Here we prove (123). During this estimation we denote by $G$ the
set $S_{\rho}^{^{\prime}}\cap U_{\varepsilon}(Tr(A(\rho))$. Since $V_{\rho
}=S_{\rho}^{^{\prime}}\backslash G$ and (133) holds, it is enough to prove
that $\mu(G)=O(\rho^{-\alpha})\mu(B(\rho)).$ For this we use (135) and prove
$\mu(G(+i,\rho d^{-1}))=O(\rho^{-\alpha})\mu(B(\rho))$ for $i=1,2,...,d$ by
using (136) ( the same estimation for $G(-i,\rho d^{-1})$ can be proved in the
same way). By (44), if $x\in G(+i,\rho d^{-1}),$ then the under integral
expression in (136) for $k=i$ and $a=\rho d^{-1}$ is less than $d+1.$
Therefore it is sufficient to prove
\begin{equation}
\mu(\Pr(G(+i,\rho d^{-1}))=O(\rho^{-\alpha})\mu(B(\rho))
\end{equation}
Clearly, if $(x_{1},x_{2},...x_{i-1},x_{i+1},...x_{d})\in\Pr_{i}(G(+i,\rho
d^{-1})),$ then

$\mu(U_{\varepsilon}(G)(x_{1},x_{2},...x_{i-1},x_{i+1},...x_{d}))\geq
2\varepsilon$ and by (137), it follows that
\begin{equation}
\mu(U_{\varepsilon}(G))\geq2\varepsilon\mu(\Pr(G(+i,\rho d^{-1})).
\end{equation}
Hence to prove (159) we need to estimate $\mu(U_{\varepsilon}(G)).$ For this
we prove that
\begin{equation}
U_{\varepsilon}(G)\subset U_{\varepsilon}(S_{\rho}^{^{\prime}}),U_{\varepsilon
}(G)\subset U_{2\varepsilon}(Tr(A(\rho))),U_{\varepsilon}(G)\subset
Tr(U_{2\varepsilon}(A(\rho))).
\end{equation}
The first and second inclusions follow from $G\subset S_{\rho}^{^{\prime}}$
and $G\subset U_{\varepsilon}(Tr(A(\rho)))$ respectively (see definition of
$G$ ). Now we prove the third inclusion in (161). If $x\in U_{\varepsilon
}(G),$ then by the second inclusion of (161) there exists $b$ such that $b\in
Tr(A(\rho)),$ $\mid x-b\mid<2\varepsilon.$ Then by the definition of
$Tr(A(\rho))$ there exist $\gamma\in\Gamma$ and $c\in A(\rho)$ such that
$b=\gamma+c$. Therefore $\mid x-\gamma-c\mid=\mid x-b\mid<2\varepsilon,$

$x-\gamma\in U_{2\varepsilon}(c)\subset U_{2\varepsilon}(A(\rho)).$ This
together with $x\in U_{\varepsilon}(G)\subset U_{\varepsilon}(S_{\rho
}^{^{\prime}})$ (see the first inclusion of (161)) give $x\in
Tr(U_{2\varepsilon}(A(\rho)))$ ( see the definition of $Tr(E)$ in the
beginning of this section), i.e., the third inclusion in (161) is proved. The
third inclusion, Lemma 8(c), and (134) imply that

$\mu(U_{\varepsilon}(G))=O(\rho^{-\alpha})\mu(B(\rho))\varepsilon.$ This and
(160) imply the proof of (159)$\diamondsuit$

ESTIMATION 5 Here we prove (124). Divide the set $V_{\rho}\equiv V$ into
pairwise disjoint subsets $V^{^{\prime}}(\pm1,\rho d^{-1})\equiv V(\pm1,\rho
d^{-1}),$

$V^{^{\prime}}(\pm i,\rho d^{-1})\equiv V(\pm i,\rho d^{-1})\backslash
(\cup_{j=1}^{i-1}(V(\pm j,\rho d^{-1}))),$ for $i=2,3,...,d.$ Take any point
$a\in V^{^{\prime}}(+i,\rho d^{-1})\subset S_{\rho}$ and consider the function
$F(x)$ ( see Lemma 8(a)) on the interval $[a-\varepsilon e_{i},a+\varepsilon
e_{i}],$ where $e_{1}=(1,0,0,...,0)$, $e_{2}=(0,1,0,...,0),...$. By the
definition of \ $S_{\rho}$ we have $F(a)=\rho^{2}$. It follows from (44) and
the definition of $V^{^{\prime}}(+i,\rho d^{-1})$ that $\frac{\partial
F(x)}{\partial x_{i}}>\rho d^{-1}$ for $x\in\lbrack a-\varepsilon
e_{i},a+\varepsilon e_{i}].$ Therefore
\begin{equation}
F(a-\varepsilon e_{i})<\rho^{2}-c_{21}\varepsilon_{1},\text{ }F(a+\varepsilon
e_{i})>\rho^{2}+c_{21}\varepsilon_{1}.
\end{equation}
Since $[a-\varepsilon e_{i},a+\varepsilon e_{i}]\in U_{\varepsilon}(a)\subset
U_{\varepsilon}(V_{\rho})\subset U_{\varepsilon}(S_{\rho}^{^{\prime}%
})\backslash Tr(A(\rho))$ ( see Theorem 7$(b)$), it follows from Theorem
7$(a)$ that there exists index $N$ such that $\Lambda(y)=\Lambda_{N}(y)$ for
$y\in U_{\varepsilon}(a)$ and $\Lambda(y)$ satisfies (102) ( see Remark 5).
Hence (162) implies that
\begin{equation}
\Lambda(a-\varepsilon e_{i})<\rho^{2},\Lambda(a+\varepsilon e_{i})>\rho^{2}.
\end{equation}
Moreover it follows from (122) that the derivative of $\Lambda(y)$ with
respect to $i$th coordinate is positive for $y\in\lbrack a-\varepsilon
e_{i},a+\varepsilon e_{i}].$ So $\Lambda(y)$ is a continuous and increasing
function in $[a-\varepsilon e_{i},a+\varepsilon e_{i}].$ Therefore (163)
implies that there exists a unique point $y(a,i)\in\lbrack a-\varepsilon
e_{i},a+\varepsilon e_{i}]$ such that $\Lambda(y(a,i))=\rho^{2}.$ Define
$I_{\rho}^{^{\prime}}(+i)$ by $I_{\rho}^{^{\prime}}(+i)=\{y(a,i):a\in
V^{^{\prime}}(+i,\rho d^{-1})\}).$ In the same way we define $I_{\rho
}^{^{\prime}}(-i)=\{y(a,i):a\in V^{^{\prime}}(-i,\rho d^{-1})\}$ and put
$I_{\rho}^{^{\prime}}=\cup_{i=1}^{d}(I_{\rho}^{^{\prime}}(+i)\cup I_{\rho
}^{^{\prime}}(-i)).$ To estimate the measure of $I_{\rho}^{^{\prime}}$ we
compare the measure of $V^{^{\prime}}(\pm i,\rho d^{-1})$ with the measure of
$I_{\rho}^{^{\prime}}(\pm i)$ by using the formula (136) and the obvious
relations
\begin{equation}
\Pr(V^{^{\prime}}(\pm i,\rho d^{-1}))=\Pr(I_{\rho}^{^{\prime}}(\pm i)),\text{
}\mu(\Pr(I_{\rho}^{^{\prime}}(\pm i)))=O(\rho^{d-1}),
\end{equation}%
\begin{equation}
(\frac{\partial F}{\partial x_{i}})^{-1}\mid grad(F)\mid-(\frac{\partial
\Lambda}{\partial x_{i}})^{-1}\mid grad(\Lambda)\mid=O(\rho^{-2\alpha_{1}}).
\end{equation}
Here the first equality in (164) follows from the definition of $I_{\rho
}^{^{\prime}}(\pm i)$. The second equality in (164) follows from the
inequalities in (135), since $I_{\rho}^{^{\prime}}\subset U_{\varepsilon
}(S_{\rho}^{^{\prime}})$. Formulas (44), (122) imply (165). Clearly, using
(164), (165), and (136) we get $\mu(V^{^{\prime}}(\pm i,\rho d^{-1}%
))-\mu(I_{\rho}^{^{\prime}}(\pm i))=O(\rho^{d-1-2\alpha_{1}}).$ On the other
hand if

$y\equiv(y_{1},y_{2},...,y_{d})\in I_{\rho}^{^{\prime}}(+i)\cap I_{\rho
}^{^{\prime}}(+j)$ for $i<j$ then there are $a\in V^{^{\prime}}(+i,\rho
d^{-1})$ and $a^{^{\prime}}\in V^{^{\prime}}(+j,\rho d^{-1})$ such that
$y=y(a,i)=y(a^{^{\prime}},j)$ and $y\in\lbrack a-\varepsilon e_{i}%
,a+\varepsilon e_{i}],$ $y\in\lbrack a^{^{\prime}}-\varepsilon e_{j}%
,a^{^{\prime}}+\varepsilon e_{j}].$ These inclusions imply that $\rho
d^{-1}-\varepsilon\leq y_{i}\leq\rho d^{-1}.$ Therefore $\mu(\Pr_{j}(I_{\rho
}^{^{\prime}}(+i)\cap I_{\rho}^{^{\prime}}(+j)))=O(\varepsilon\rho^{d-2}).$
This equality, (136) and (122) imply that $\mu((I_{\rho}^{^{\prime}}(+i)\cap
I_{\rho}^{^{\prime}}(+j)))=O(\varepsilon\rho^{d-2})$ for all $i$ and $j.$ Similarly

$\mu((I_{\rho}^{^{\prime}}(+i)\cap I_{\rho}^{^{\prime}}(-j)))=O(\varepsilon
\rho^{d-2})$ for all $i$ and $j.$ Thus%
\[
\mu(I_{\rho}^{^{\prime}})=\sum_{i}\mu(I_{\rho}^{^{\prime}}(+i))+\sum_{i}%
\mu(I_{\rho}^{^{\prime}}(-i))+O(\varepsilon\rho^{d-2})=
\]
$\ \ \ \ \ \ \ \ \ \ \ \ \ \ \sum_{i}\mu(V^{^{\prime}}(+i,\rho d^{-1}%
))+\sum_{i}\mu(V^{^{\prime}}(-i,\rho d^{-1}))+O(\rho^{d-1-2\alpha_{1}})=$

$\mu(V_{\rho})+O(\rho^{-2\alpha_{1}})\mu(B(\rho)).$ This and (123) yield the
inequality (124) for $I_{\rho}^{^{\prime}}$. Now we define $I_{\rho}%
^{^{\prime\prime}}$ as follows. If $\gamma+t\in$ $I_{\rho}^{^{\prime}}$ then
$\Lambda(\gamma+t)=\rho^{2}$, where $\Lambda(\gamma+t)$ is a unique eigenvalue
satisfying (5) ( see Remark 5). Since

$\Lambda(\gamma+t)=\mid\gamma+t\mid^{2}+O(\rho^{-\alpha_{1}})$ ( see (5) and
(24)), for fixed $t$ there exist only a finite number of vectors $\gamma
_{1},\gamma_{2},...,\gamma_{s}\in\Gamma$ satisfying $\Lambda(\gamma
_{k}+t)=\rho^{2}$. Hence $I_{\rho}^{^{\prime}}$ is the union of pairwise
disjoint subsets $I_{\rho,k}^{^{\prime}}$ $\equiv\{\gamma_{k}+t\in I_{\rho
}^{^{\prime}}:\Lambda(\gamma_{k}+t)=\rho^{2}\}$ for

$k=1,2,...s.$ The translation $I_{\rho,k}^{^{\prime\prime}}=I_{\rho
,k}^{^{\prime}}-\gamma_{k}=\{t\in F^{\ast}:\gamma_{k}+t\in I_{\rho
,k}^{^{\prime}}\}$ of $I_{\rho,k}^{^{\prime}}$ is a part of the isoenergetic
surfaces $I_{\rho}$ of $L(q(x)).$ Put $I_{\rho}^{^{\prime\prime}}=\cup
_{k=1}^{s}I_{\rho,k}^{^{\prime\prime}}.$ If $t\in I_{\rho,k}^{^{\prime\prime}%
}\cap I_{\rho,m}^{^{\prime\prime}}$ for $k\neq m,$ then $\gamma_{k}+t\in
I_{\rho}^{^{\prime}}\subset U_{\varepsilon}(S_{\rho}^{^{\prime}})$ and
$\gamma_{m}+t\in U_{\varepsilon}(S_{\rho}^{^{\prime}}),$ which contradict
Lemma 8(b). So $I_{\rho}^{^{\prime\prime}}$ is union of the pairwise disjoint
subsets $I_{\rho,k}^{^{\prime\prime}}$ for $k=1,2,...s.$ Thus

$\mu(I_{\rho}^{^{\prime\prime}})=\sum_{k}\mu(I_{\rho,k}^{^{\prime\prime}%
})=\sum_{k}\mu(I_{\rho,k}^{^{\prime}})=\mu(I_{\rho}^{^{\prime}})>(1-c_{16}%
\rho^{-\alpha}))\mu(B(\rho))\diamondsuit$

\section{Bloch Functions near the Diffraction Planes}

In this section we obtain the asymptotic formulas for the Bloch function
corresponding to the quasimomentum lying near the diffraction hyperplanes.
Here we assume that $q(x)\in W_{2}^{s}(F),$ where $s\geq6(3^{d}(d+1)^{2})+d.$
In this section we define the number $q$ by $q=4(3^{d}(d+1)).$ The other
numbers $p,\alpha_{k},\alpha,k_{1},p_{1}$ are defined as in the end of
introduction. Clearly these numbers satisfy all inequalities in (15).
\ Therefore the formulas obtained in previous sections hold in this notations
too. Moreover the following relations hold
\begin{equation}
k_{2}<\frac{1}{9}(p-\frac{1}{2}q(d-1),\text{ }k_{2}\alpha_{2}>d+2\alpha,\text{
}4(d+1)\alpha_{d}=1,
\end{equation}
where $k_{2}=[\frac{d}{9\alpha}]+2.$ In this section we construct a subset
$B_{\delta}$ of $V_{\delta}^{^{\prime}}(\rho^{\alpha_{1}})$ such that if
$\gamma+t\equiv\beta+\tau+(j+v)\delta\in B_{\delta}$ ( see Remark 3 for this
notations), then there exists a unique eigenvalue $\Lambda_{N}(\lambda
_{j,\beta}(v,\tau))$ satisfying (99). This is a simple eigenvalue. We say that
the set $B_{\delta}$ is a simple set in the resonance domain $V_{\delta}%
(\rho^{\alpha_{1}}).$ Then we obtain the asymptotic formulas of arbitrary
order for the eigenfunction $\Psi_{N}(x)$ corresponding to the eigenvalue
$\Lambda_{N}(\lambda_{j,\beta}(v,\tau))$. \ At the end of this section we
prove that $B_{\delta}$ has asymptotically full measure on $V_{\delta}%
(\rho^{\alpha_{1}}).$ The construction of \ the simple set $B_{\delta}$ in the
resonance domain $V_{\delta}(\rho^{\alpha_{1}})$ is similar to the
construction of the simple set $B$ in the non-resonance domain (see Step 1 and
Step 2 in introduction). So as in Step 2 we need to find the simplicity
conditions for the eigenvalue $\Lambda_{N}(\lambda_{j,\beta}).$ Since the
first inequality in (166) holds, $\Lambda_{N}(\lambda_{j,\beta})$ satisfies
the formula (99) for $k=k_{2}.$ Therefore it follows from the second
inequality of (166) that $\Lambda_{N}(\lambda_{j,\beta})$ lies in the
$\varepsilon_{1}=\rho^{-d-2\alpha}$ neighbourhood of$\ E(\lambda_{j,\beta
}(v,\tau))$, where $E(\lambda_{j,\beta}(v,\tau))=\lambda_{j,\beta}%
(v,\tau)+E_{k_{2}-1}(\lambda_{j,\beta}(v,\tau)).$ Note that we have the
relations%
\begin{align}
E_{k_{2}-1}(\lambda_{j,\beta})  &  =O(\rho^{-\alpha_{2}}(\ln\rho)),\text{
}\lambda_{j,\beta}(v,\tau)\sim\rho^{2}\\
\lambda_{j,\beta}(v,\tau)  &  =\mid\beta+\tau\mid^{2}+\mu_{j}(v)=\mid
\beta+\tau\mid^{2}+O(\rho^{2\alpha_{1}})\nonumber
\end{align}
( see (100), Lemma 2(b), (52), (51) ). We call $E(\lambda_{j,\beta}(v,\tau))$
the known part of $\Lambda_{N}(\lambda_{j,\beta}(v,\tau))$. Since known parts
of the other eigenvalues are $\lambda_{i}(\gamma^{^{\prime}}+t),$
$F(\gamma^{^{\prime}}+t)$ ( see Step 1), that is, the other eigenvalues lie in
the $\varepsilon_{1}$ neighbourhood of $\lambda_{i}(\gamma^{^{\prime}}+t),$
$F(\gamma^{^{\prime}}+t),$ in order that $\Lambda_{N}(\lambda_{j,\beta}%
(v,\tau))$ does not coincide with any other eigenvalue we use the following
two simplicity conditions
\begin{equation}
\mid E(\lambda_{j,\beta}(v,\tau))-F(\gamma^{^{\prime}}+t)\mid\geq
2\varepsilon_{1},\forall\gamma^{^{\prime}}\in M_{1}%
\end{equation}%
\begin{equation}
\mid E(\lambda_{j,\beta})-\lambda_{i}(\gamma^{^{\prime}}+t)\mid\geq
2\varepsilon_{1},\forall\gamma^{^{\prime}}\in M_{2};i=1,2,...,b_{k},
\end{equation}
where $M$ is the set of $\gamma^{^{\prime}}\in\Gamma$ satisfying $\mid
E(\lambda_{j,\beta}(v,\tau))-\mid\gamma^{^{\prime}}+t\mid^{2}\mid<\frac{1}%
{3}\rho^{\alpha_{1}};$ $M_{1}$ is the set of $\gamma^{^{\prime}}\in M$
satisfying $\gamma^{^{\prime}}+t\in U(\rho^{\alpha_{1}},p);$ $M_{2}$ is the
set of $\gamma^{^{\prime}}\in M$ such that $\gamma^{^{\prime}}+t\notin
U(\rho^{\alpha_{1}},p)$ and $\gamma^{^{\prime}}+t$ has the $\Gamma_{\delta}$ decomposition

$\gamma^{^{\prime}}+t\equiv\beta^{^{\prime}}+\tau+(j^{^{\prime}}%
+v(\beta^{^{\prime}}))\delta$ ( see Remark 3) with $\beta^{^{\prime}}\neq
\beta$.\ Let $B_{\delta}$ be the set of $x\in V_{\delta}^{^{\prime}}%
(\rho^{\alpha_{1}})\cap(R(\frac{3}{2}\rho-\rho^{\alpha_{1}-1})\backslash
R(\frac{1}{2}\rho+\rho^{\alpha_{1}-1}))$ such that $x=\gamma+t,$ where

$\gamma\in\Gamma,t\in F^{\star}$ ( it is $\Gamma$ decomposition of $x$), and
$x=\beta+\tau+(j+v(\beta))\delta,$ where $\beta\in\Gamma_{\delta},$ $\tau\in
F_{\delta},$ $j\in Z,$ $v(\beta)\in W(\rho)$ ( it is $\Gamma_{\delta}$
decomposition of $x$.), and (168), (169) hold. Using this conditions we prove
that $\Lambda_{N}(\lambda_{j,\beta}(v,\tau))$ does not coincide with other
eigenvalues if $\beta+\tau+(j+v)\delta\in B_{\delta}$. The existence and
properties of the sets $B_{\delta},$ will be considered in the end of this
section. Recall that in Section 4 the simplicity conditions (12), (13) implied
the asymptotic formulas \ for the Bloch functions in the non-resonance domain.
Similarly, here the simplicity conditions (168), (169) imply the asymptotic
formula for the Bloch function in the resonance domain $V_{\delta}^{^{\prime}%
}(\rho^{\alpha_{1}})$. For this we use the following lemma.

\begin{lemma}
Let $\Lambda_{N}(\lambda_{j,\beta}(v,\tau))$ be the eigenvalue of the operator
$L_{t}(q(x))$ satisfying (99), where $\beta+\tau+(j+v)\delta\equiv\gamma+t\in
B_{\delta}.$ If for $\gamma^{^{\prime}}+t\equiv\beta^{^{\prime}}%
+\tau+(j^{^{\prime}}+v(\beta^{^{\prime}}))\delta$ at least one of the
following conditions:

Cond.1: $\gamma^{^{\prime}}\in M,$ $\beta^{^{\prime}}\neq\beta,$

Cond.2: $\mid\beta-\beta^{^{\prime}}\mid>(p-1)\rho^{\alpha},$

Cond.3: $\mid\beta-\beta^{^{\prime}}\mid\leq(p-1)\rho^{\alpha},$ $\mid
j^{^{\prime}}\delta\mid\geq h$ hold, then
\begin{equation}
\mid b(N,\gamma^{^{\prime}})\mid\leq c_{4}\rho^{-c\alpha},
\end{equation}
where $h\equiv10^{-p}\rho^{\frac{1}{2}\alpha_{2}},$ $c=p-dq-\frac{1}{4}%
d3^{d}-3,$ $b(N,\gamma^{^{\prime}})=(\Psi_{N,t},e^{i(\gamma^{^{\prime}}%
+t,x)}),$ $\Psi_{N,t}(x)$ is any normalized eigenfunction of $L_{t}(q(x))$
corresponding to the eigenvalue $\Lambda_{N}(\lambda_{j,\beta}(v,\tau))$.
\end{lemma}

\begin{proof}
Repeating the proof of the inequality in(105) and instead of the simplicity
conditions (12), (13) and the set $K$ using the simplicity conditions (168),
(169), and the set $M$ we obtain the proof of (170) under Cond.1. Suppose
Cond.2 holds. Take $n$ vectors $\gamma_{1},\gamma_{2},...,\gamma_{n},$ where
$n\leq p-1,$ from $\Gamma(\rho^{\alpha}).$ Using the decomposition $\gamma
_{i}=\beta_{i}+a_{i}\delta$ ( see (48)), where $\beta_{i}\in\Gamma_{\delta},$
$a_{i}\in\mathbb{R}$ , $1\leq i\leq n\leq(p-1),$ and Cond.2 we have
\begin{align}
\  &  \mid\beta_{i}\mid<\rho^{\alpha},\mid a_{i}\delta\mid<\rho^{\alpha}%
,\beta^{^{\prime}}-\sum_{i=1}^{n}\beta_{i}\neq\beta\\
\gamma^{^{\prime}}+t-\sum_{i=1}^{n}\gamma_{i}  &  =\beta^{^{\prime}}%
-\sum_{i=1}^{n}\beta_{i}+\tau+(j^{^{\prime}}+v(\beta^{^{\prime}})-\sum
_{i=1}^{n}a_{i})\delta.
\end{align}
If $\gamma^{^{\prime}}-\sum_{i=1}^{n}\gamma_{i}\in M,$ then (171) and (170)
under Cond.1. imply that

$b(N,\gamma^{^{\prime}}-\sum_{i=1}^{n}\gamma_{i})=O(\rho^{-c\alpha})$. If
$\gamma^{^{\prime}}\notin M$ then applying (17) $(p-1)$-times and each time
instead of $b(N,\gamma^{^{\prime}}-\sum_{i=1}^{j}\gamma_{i})$ for
$\gamma^{^{\prime}}-\sum_{i=1}^{j}\gamma_{i}\in M$ writing $O(\rho^{-c\alpha
}),$ we obtain
\begin{equation}
b(N,\gamma^{^{\prime}})=\sum_{\gamma_{1},\gamma_{2},...,\gamma_{p-1}}%
\dfrac{q_{\gamma_{1}}q_{\gamma_{2}}...q_{\gamma_{p-1}}b(N,\gamma^{^{\prime}%
}-\sum_{i=1}^{p-1}\gamma_{i})}{\prod_{j=0}^{p-2}(\Lambda_{N}-\mid
\gamma^{^{\prime}}-\sum_{i=1}^{j}\gamma_{i}+t\mid^{2})}+o(\rho^{-c\alpha}),
\end{equation}
where the summation is taken under the conditions $\gamma^{^{\prime}}%
-\sum_{i=1}^{j}\gamma_{i}\notin$ $M,$ for $j\leq p-1$. These conditions, the
definition of $M$ and (99) imply that

$\mid\Lambda_{N}-\mid\gamma^{^{\prime}}-\sum_{i=1}^{j}\gamma_{i}+t\mid^{2}%
\mid>\frac{1}{4}\rho^{\alpha_{1}}$. Therefore (173) and (4) give (170).

Now assume that Cond.3. holds. First we prove that if

$\gamma^{^{\prime}}-\sum_{i=1}^{s}\gamma_{i}\in M$, where $s=0,1,...,n$ and
$n,$ $\gamma_{i}$ are defined in the case of Cond.2., then $\beta^{^{\prime}%
}-\sum_{i=1}^{s}\beta_{i}\neq\beta.$ Assume the converse, i.e., $\beta
^{^{\prime}}-\sum_{i=1}^{s}\beta_{i}=\beta.$ Then the relations $(\beta
+\tau,\delta)=0,\mid a_{i}\delta\mid<\rho^{\alpha},$ (see (171)) $\mid
j^{^{\prime}}\delta\mid\geq h,$ $\mid v(\beta^{^{\prime}})\mid\leq1$ and (172)
imply that%
\begin{equation}
\mid\gamma^{^{\prime}}+t-\sum_{i=1}^{s}\gamma_{i}\mid^{2}\geq\mid\beta
+\tau\mid^{2}+\frac{1}{2}h^{2}.
\end{equation}
On the other hand the definition of $E(\lambda_{j,\beta}(v,\tau)),$ and (167)
yield \
\begin{equation}
E(\lambda_{j,\beta}(v,\tau))=\mid\beta+\tau\mid^{2}+O(\rho^{2\alpha_{1}})
\end{equation}
Therefore using (174), (175) and the relations $\frac{1}{2}h^{2}>c_{22}%
\rho^{\alpha_{2}},$ $\alpha_{2}=3\alpha_{1}$ we get
\[
\mid E(\lambda_{j,\beta}(v,\tau))-\mid\gamma^{^{\prime}}+t-\sum_{i=1}%
^{s}\gamma_{i}\mid^{2}\mid>\rho^{\alpha_{1}}%
\]
which contradicts $\gamma^{^{\prime}}-\sum_{i=1}^{s}\gamma_{i}\in M.$ Thus we
proved that if $\gamma^{^{\prime}}-\sum_{i=1}^{s}\gamma_{i}\in M,$ then
$\beta^{^{\prime}}-\sum_{i=1}^{s}\beta_{i}\neq\beta.$ This relation for $s=0$
means that $\beta^{^{\prime}}\neq\beta$ if $\gamma^{^{\prime}}\in M.$
Therefore if Cond.3. holds and $\gamma^{^{\prime}}\in M,$ then Cond.1. holds
too and hence (170) holds. To prove (170) under Cond.3. in case $\gamma
^{^{\prime}}\notin M$ we repeat the prove of the case of Cond.2, that is, use
(173) and etc.
\end{proof}

\begin{theorem}
If $\gamma+t\equiv\beta+\tau+(j+v(\beta))\delta\in B_{\delta},$ then there
exists a unique eigenvalue $\Lambda_{N}(\lambda_{j,\beta}(v,\tau))$ satisfying
(99). This is a simple eigenvalue and the corresponding eigenfunction
$\Psi_{N}$ satisfies the asymptotic formula
\begin{equation}
\Psi_{N}=\Phi_{j,\beta}(x)+O(\rho^{-\alpha_{2}}\ln\rho).
\end{equation}

\end{theorem}

\begin{proof}
The proof is similar to the proof of Theorem 5. Arguing as in the proof of the
Theorem 5 we see that to prove this theorem it is enough to show that for any
normalized eigenfunction $\Psi_{N}$ corresponding to any eigenvalue$\Lambda
_{N}$ satisfying (99) the following equality holds
\begin{equation}
\sum_{(j^{^{\prime}},\beta^{^{\prime}})\in K_{0}}\mid b(N,j^{^{\prime}}%
,\beta^{^{\prime}})\mid^{2}=O(\rho^{-2\alpha_{2}}(\ln\rho)^{2}),
\end{equation}
where $b(N,j^{^{\prime}},\beta^{^{\prime}})=(\Psi_{N}(x),\Phi_{j^{^{\prime}%
},\beta^{^{\prime}}}(x)),$

$K_{0}=\{(j^{^{\prime}},\beta^{^{\prime}}):j^{^{\prime}}\in Z,\beta^{^{\prime
}}\in\Gamma_{\delta},(j^{^{\prime}},\beta^{^{\prime}})\neq(j,\beta)\}.$ To
prove (177) we divide the set $K_{0}$ into subsets: $K_{1}^{c},$ $K_{1}\cap
S(p-1),$ $K_{1}\cap S^{c}(p-1),$ where

$K_{1}^{c}=K_{0}\backslash$ $K_{1},$ $S^{c}(n)=K_{0}\backslash S(n),K_{1}%
=\{(j^{^{\prime}},\beta^{^{\prime}})\in K_{0}:\mid\Lambda_{N}-\lambda
_{j^{^{\prime}},\beta^{^{\prime}}}\mid<h^{2}\},$

$S(n)=\{(j^{^{\prime}},\beta^{^{\prime}})\in K_{0}:\mid\beta-\beta^{^{\prime}%
}\mid\leq n\rho^{\alpha}$ $,\mid j^{^{\prime}}\delta\mid<10^{n}h\}$ and $h$ is
defined in Lemma 9. If $(j^{^{\prime}},\beta^{^{\prime}})\in K_{1}^{c},$ then
using (53), the definitions of $K_{1}^{c}$ and $h,$ we have
\begin{equation}
\sum_{(j^{^{\prime}},\beta^{^{\prime}})\in K_{1}^{c}}\mid b(N,j^{^{\prime}%
},\beta^{^{\prime}})\mid^{2}=\sup\mid q-q^{\delta}\mid^{2}O(\rho^{-2\alpha
_{2}})=O(\rho^{-2\alpha_{2}}).
\end{equation}
To consider the set $K_{1}\cap S(p-1)$ we prove that
\begin{equation}
K_{1}\cap S(n)=K_{1}\cap\{(j^{^{\prime}},\beta):j^{^{\prime}}\in
Z\}\subset\{(j^{^{\prime}},\beta):\mid j^{^{\prime}}\delta\mid<2h\},
\end{equation}
for $n=1,2,...,p-1.$ Take any element $(j^{^{\prime}},\beta)$ from $K_{1}%
\cap\{(j^{^{\prime}},\beta):j^{^{\prime}}\in Z\}$. Since $\lambda
_{j^{^{\prime}},\beta}(v,\tau)=\mid\beta+\tau\mid^{2}+\mu_{j^{^{\prime}}%
}(v)=\mid\beta+\tau\mid^{2}+\mid(j^{^{\prime}}+v)\delta\mid^{2}+O(1),$ where
$v\in\lbrack0,1]$ ( see Lemma 2(b) and (52)), using the definition of $K_{1}$,
(99) (175) and the relations $h^{2}\sim\rho^{\alpha_{2}}$, $\alpha_{2}%
=3\alpha_{1}$ we obtain

$\mid O(\rho^{2\alpha_{1}})-\mid(j^{^{\prime}}+v)\delta\mid^{2}\mid
<2h^{2},\mid j^{^{\prime}}\delta\mid<2h.$ Hence the inclusion in (179) is
proved and $K_{1}\cap\{(j^{^{\prime}},\beta):j^{^{\prime}}\in Z\}\subset
K_{1}\cap S(n).$ If the inclusion

$K_{1}\cap S(n)\subset K_{1}\cap\{(j^{^{\prime}},\beta):j^{^{\prime}}\in Z\}$
does not hold then there is an element $(j^{^{\prime}},\beta^{^{\prime}})$ of
$K_{1}\cap S(n)$ such that

$0<\mid\beta-\beta^{^{\prime}}\mid\leq n\rho^{\alpha}\leq(p-1)\rho^{\alpha},$
$\mid j^{^{\prime}}\delta\mid<$ $10^{n}h<\frac{1}{2}\rho^{\frac{1}{2}%
\alpha_{2}}$. These inequalities and the inequality
\begin{equation}
\mid\Lambda_{N}-\lambda_{j^{^{\prime}},\beta^{^{\prime}}}\mid>\frac{1}{2}%
\rho^{\alpha_{2}},
\end{equation}
for $0<\mid\beta-\beta^{^{\prime}}\mid\leq(p-1)\rho^{\alpha},$ $\mid
j^{^{\prime}}\delta\mid<$ $\frac{1}{2}\rho^{\frac{1}{2}\alpha_{2}},$ which
follows from (83), (88), imply that $\mid\Lambda_{N}-\lambda_{j^{^{\prime}%
},\beta^{^{\prime}}}\mid>h^{2}.$ This contradicts $(j^{^{\prime}}%
,\beta^{^{\prime}})$ $\in K_{1}.$ So (179) is proved. Therefore
\begin{equation}
\sum_{(j^{^{\prime}},\beta^{^{\prime}})\in K_{1}\cap S(p-1)}\mid
b(N,j^{^{\prime}},\beta^{^{\prime}})\mid^{2}\leq\sum_{\substack{j^{^{\prime}%
}\neq j,\\\mid j^{^{\prime}}\delta\mid<2h}}\mid b(N,j^{^{\prime}},\beta
)\mid^{2}%
\end{equation}
For estimation of $b(N,j^{^{\prime}},\beta)$ for $\mid j^{^{\prime}}\delta
\mid<2h,$ we use (75) as follows. In (75) replacing $\beta^{^{\prime}}$ and
$r$ by $\beta$ and $2h,$ we obtain
\[
(\Lambda_{N}-\lambda_{j^{^{\prime}},\beta})b(N,j^{^{\prime}},\beta
)=O(\rho^{-p\alpha})+
\]%
\begin{equation}
+\sum\limits_{(j_{1},\beta_{1})\in Q(\rho^{\alpha},18h)}A(j^{^{\prime}}%
,\beta,j^{^{\prime}}+j_{1},\beta+\beta_{1})b(N,j^{^{\prime}}+j_{1},\beta
+\beta_{1}).
\end{equation}
By definition of $Q(\rho^{\alpha},18h)$ we have $\mid\beta_{1}\mid
<\rho^{\alpha},$ $\mid j_{1}\delta\mid<18h$ , and hence

$\mid(j^{^{\prime}}+j_{1})\delta\mid<20h<\frac{1}{2}\rho^{\frac{1}{2}%
\alpha_{2}}.$ Therefore in the right-hand side of (182) the multiplicand
$b(N,j^{^{\prime}}+j_{1},\beta+\beta_{1})$ for $(j^{^{\prime}}+j_{1}%
,\beta+\beta_{1})\in D(\beta)$ takes part, where

$D(\beta)=\{(j,\beta+\beta_{1}):\mid j\delta\mid<\frac{1}{2}\rho^{\frac{1}%
{2}\alpha_{2}},0<\mid\beta_{1}\mid<\rho^{\alpha}\}.$ Put
\[
\mid b(N,j_{0},\beta+\beta_{0})\mid=\max_{(j,\beta+\beta_{1})\in D(\beta)}\mid
b(N,j,\beta+\beta_{1})\mid.
\]
By definition of $D(\beta)$ and by (180) we have $\mid\Lambda_{N}%
-\lambda_{j_{0},\beta+\beta_{0}}\mid>\frac{1}{2}\rho^{\alpha_{2}}.$ This
together with (53) gives $\mid b(N,j_{0},\beta+\beta_{0})\mid=$ $O(\rho
^{-\alpha_{2}}).$ Using this, (182) and (71), we get
\begin{equation}
\mid b(N,j^{^{\prime}},\beta)\mid<c_{23}\mid\Lambda_{N}-\lambda_{j^{^{\prime}%
},\beta}\mid^{-1}\rho^{-\alpha_{2}}%
\end{equation}
for $j^{^{\prime}}\neq j,$ $\mid j^{^{\prime}}\delta\mid<2h,$ where

$\Lambda_{N}-\lambda_{j^{^{\prime}},\beta}=\lambda_{j,\beta}-\lambda
_{j^{^{\prime}},\beta}+O(\rho^{-\alpha_{2}})=\mu_{j}(v)-\mu_{j^{^{\prime}}%
}(v)+O(\rho^{-\alpha_{2}})$ \ (see (88) and Lemma 2(b) ) and $v\in W(\rho)$ (
see the definition of $B_{\delta}$). Now using the definition of $W(\rho)$
(see Lemma 3(b)) and (52) we obtain $\sum_{j^{^{\prime}}\neq j}\mid\Lambda
_{N}-\lambda_{j^{^{\prime}},\beta}\mid^{-2}=O(\ln\rho).$ This with (183) and
(181) yield
\begin{equation}
\sum_{(j^{^{\prime}},\beta^{^{\prime}})\in K_{1}\cap S(p-1)}\mid
b(N,j^{^{\prime}},\beta)\mid^{2}=O(\rho^{-2\alpha_{2}}(\ln\rho)^{2}).
\end{equation}
It remains to consider $K_{1}\cap S^{c}(p-1)$. We prove that
\begin{equation}
b(N,j^{^{\prime}},\beta^{^{\prime}})=O(\rho^{-c\alpha})
\end{equation}
for $(j^{^{\prime}},\beta^{^{\prime}})\in K_{1}\cap S^{c}(p-1),$ where the
number $c$ is defined in Lemma 9. For this using the decomposition of
$\varphi_{j^{^{\prime}},v(\beta^{^{\prime}})}(s)$ by $\{e^{i(m+v(\beta
^{^{\prime}}))s}:m\in Z\},$ we get
\begin{equation}
b(N,j^{^{\prime}},\beta^{^{\prime}})=\sum_{m}(\varphi_{j^{^{\prime}}%
,v}(s),e^{i(m+v)s})(\Psi_{N}(x),e^{i(\beta^{^{\prime}}+\tau+(m+v)\delta,x)}).
\end{equation}
If $\mid\beta-\beta^{^{\prime}}\mid>(p-1)\rho^{\alpha}$ then Lemma 9 ( see
Cond. 2) and (73), (186) give the proof (185). So we need to consider the case
$\mid\beta-\beta^{^{\prime}}\mid\leq(p-1)\rho^{\alpha}.$ Then by definition of
$S^{c}(p-1)$ we have $\mid j^{^{\prime}}\delta\mid\geq10^{p_{1}-1}h$. Write
the right-hand side of (186) as $\sum_{1}+\sum_{2},$ where the summations in
$\sum_{1}$ and $\sum_{2}$are taken under conditions $\mid m\delta\mid\geq h$
and $\mid m\delta\mid<h$ respectively. By (73) and Lemma 9 ( see Cond.3 ) we
have $\sum_{1}=O(\rho^{-c\alpha}).$ If $\mid m\delta\mid<h,$ then the
inequality $\mid j^{^{\prime}}\mid>2\mid m\mid$ holds. Therefore using (57),
taking into account that $\mid j^{^{\prime}}\delta\mid\sim\rho^{\alpha_{2}}$
and the number of summand in $\sum_{2}$ is less than $\rho^{\alpha_{2}},$ we
get $\sum_{2}=O(\rho^{-c\alpha}).$ The estimations for $\sum_{1}$, $\sum_{2}$
give (185). Now using $\mid K_{1}\mid=O(\rho^{(d-1)q\alpha}),$ we get
\begin{equation}
\sum_{(j^{^{\prime}},\beta^{^{\prime}})\in K_{1}\cap S^{c}(p-1)}\mid
b(N,j^{^{\prime}},\beta^{^{\prime}})\mid^{2}=O(\rho^{-(2c-(d-1)q)\alpha}).
\end{equation}
This, (178), (184) give the proof of (177), since $(2c-(d-1)q)\alpha
>\alpha_{2}.$ The theorem is proved
\end{proof}

Now using Theorem 8 we obtain the asymptotic formulas of arbitrary order.

\begin{theorem}
The eigenfunction $\Psi_{N}$ defined in Theorem 8, besides formula (175),
satisfies the following asymptotic formulas
\begin{equation}
\Psi_{N}=\Phi_{j,\beta}(x)+\Phi_{k-1}(x)+O(\rho^{-k\alpha_{2}}\ln\rho)
\end{equation}
for $k=1,2,...,n_{1},$ where $n_{1}=[\frac{1}{9}(p-q(\frac{3d-1}{2})-\frac
{1}{4}d3^{d}-3)],$ $\Phi_{0}(x)=0,$ $\Phi_{k-1}(x)$ is a linear combination of
$\Phi_{j,\beta}(x)$ and $\Phi_{j^{^{\prime}},\beta^{^{\prime}}}(x)$ for
$(j^{^{\prime}},\beta^{^{\prime}})\in$ $S(k-1)$ with expressed by
$q(x),\lambda_{j^{^{\prime}},\beta^{^{\prime}}},\Phi_{j^{^{\prime}}%
,\beta^{^{\prime}}}(x)$ coefficients.
\end{theorem}

\begin{proof}
By Theorem 8 the formula (188) for $k=1$ is proved. To prove it for arbitrary
$k$ ( $k\leq n_{1}$) we prove the following equivalent formulas
\begin{align}
\sum_{(j^{^{\prime}},\beta^{^{\prime}})\in S^{c}(k-1)}  &  \mid
b(N,j^{^{\prime}},\beta^{^{\prime}})\mid^{2}=O(\rho^{-2k\alpha_{2}}(\ln
\rho)^{2}),\\
\Psi_{N}  &  =\sum_{(j^{^{\prime}},\beta^{^{\prime}})\in S(k-1)\cup(j,\beta
)}b(N,j^{^{\prime}},\beta^{^{\prime}})\Psi_{j^{^{\prime}},\beta^{^{\prime}}%
}+O(\rho^{-k\alpha_{2}}\ln\rho).
\end{align}
First consider the set $S^{c}(k-1)\cap K_{1}.$ It follows from the relations

$S(k-1)\cap K_{1}=S(p-1)\cap K_{1}$ (see (179)) and $S(k-1)\subset S(p-1)$ for
$0<k<p$ ( see definition of $S(k-1)$) that $(S(p-1))\backslash S(k-1))\cap
K_{1}=\emptyset$ , and hence

$S^{c}(k-1)=S^{c}(p-1)\cup$ $(S(p-1)\backslash S(k-1)),$

$S^{c}(k-1)\cap K_{1}=S^{c}(p-1)\cap K_{1}.$ Therefore using (187), the
equalities $c=p-dq-\frac{1}{4}d3^{d}-3$ ( see Lemma 9), $\alpha_{2}=9\alpha,$
$n_{1}=[\frac{1}{9}(p-q(\frac{3d-1}{2})-\frac{1}{4}d3^{d}-3)]$ ( see Theorem
9) we have
\[
\sum_{(j^{^{\prime}},\beta^{^{\prime}})\in S^{c}(k-1)\cap K_{1}}\mid
b(N,j^{^{\prime}},\beta^{^{\prime}})\mid^{2}=O(\rho^{-2n_{1}\alpha_{2}}).
\]
Thus it remains to prove
\begin{equation}
\sum_{(j^{^{\prime}},\beta^{^{\prime}})\in S^{c}(k-1)\cap K_{1}^{c}}\mid
b(N,j^{^{\prime}},\beta^{^{\prime}})\mid^{2}=O(\rho^{-2k\alpha_{2}}(\ln
\rho)^{2})
\end{equation}
for $k=2,3,...,n_{1}.$ We prove this by induction. By formula (70) and (176)
we have $\Psi_{N}(x)(q(x)-Q(s))=H(x)+O(\rho^{-\alpha_{2}}\ln\rho),$ where
$H(x)$ is a linear combination of $\Phi_{j,\beta}(x)$ and $\Phi_{j^{^{\prime}%
},\beta^{^{\prime}}}(x)$ for $(j^{^{\prime}},\beta^{^{\prime}})\in$ $S(1),$
since $\mid j\delta\mid<r_{1}<h$ (see (51)). Hence $H(x)$ orthogonal to
$\Phi_{j^{^{\prime}},\beta^{^{\prime}}}(x)$ for $(j^{^{\prime}},\beta
^{^{\prime}})\in S^{c}(1)$. Therefore using (75) and the definition of
$K_{1}^{c}$ we have
\begin{align}
\sum_{(j^{^{\prime}},\beta^{^{\prime}})\in S^{c}(1)\cap K_{1}^{c}}  &  \mid
b(N,j^{^{\prime}},\beta^{^{\prime}})\mid^{2}=\sum\mid\dfrac{(O(\rho
^{-\alpha_{2}}\ln\rho),\Phi_{j^{^{\prime}},\beta^{^{\prime}}})}{\Lambda
_{N}-\lambda_{j^{^{\prime}},\beta^{^{\prime}}}}\mid^{2}\\
&  =O(\rho^{-4\alpha_{2}}(\ln\rho)^{2}).\nonumber
\end{align}
Hence (191) for $k=2$ is proved. Assume that this is true for $k=m.$ Then
(190) for $k=m$ holds too. This and (70) for $r=10^{m-1}h$ give

$\Psi_{N}(x)(q(x)-Q(s))=H(x)+O(\rho^{-m\alpha_{2}}\ln\rho),$ where $H(x)$ is a
linear combination of $\Phi_{j,\beta}(x)$ and $\Psi_{j^{^{\prime}}%
,\beta^{^{\prime}}}(x)$ for $(j^{^{\prime}},\beta^{^{\prime}})\in$ $S(m).$
Thus $H(x)$ is orthogonal to $\Psi_{j^{^{\prime}},\beta^{^{\prime}}}(x)$ for
$(j^{^{\prime}},\beta^{^{\prime}})\in$ $S^{c}(m).$ Using this and repeating
the proof of (191) for $k=2$ we obtain the proof of (191) for $k=m+1.$ Thus
(189) and (190) are proved. Here $b(N,j,\beta)$ and $b(N,j^{^{\prime}}%
,\beta^{^{\prime}})$ for $(j^{^{\prime}},\beta^{^{\prime}})\in$ $S(k-1)$ can
be calculated in the same way as we found $b(N,\gamma)$ and $b(N,\gamma
+\gamma^{^{\prime}})$ for $\gamma^{^{\prime}}\in\Gamma((n-1)\rho^{\alpha})$ in
Theorem 5. Namely we apply the formula (75) $2k+2$ times and each time isolate
the terms with multiplicand $b(N,j,\beta).$ Then in the obtained expression
instead of $\Lambda_{N}$ writing the right side of (99) for $k=k_{3},$ where
$k_{3}=[\frac{1}{9}(p-\frac{1}{2}q(d-1)]$ we write $b(N,j^{^{\prime}}%
,\beta^{^{\prime}})$ in term of $b(N,j,\beta)$. Substituting the obtained
formula for $b(N,j^{^{\prime}},\beta^{^{\prime}})$ into (190), taking into
account that $\parallel\Psi_{N}\parallel=1,$ $\arg b(N,j,\beta)=0$ ( it can be
assumed without loss of generality) \ we find $b(N,j,\beta)$ and then
$b(N,j^{^{\prime}},\beta^{^{\prime}})$
\end{proof}

Now we consider the simple set $B_{\delta}$ in the resonance domain
$V_{\delta}(\rho^{\alpha_{1}}).$ As we noted in Remark 3 every vectors $w$ of
$\mathbb{R}^{d}$ has decomposition $w\equiv\beta+\tau+(j+v)\delta$, where
$\beta\in\Gamma_{\delta},$ $\tau\in F_{\delta},$ $j\in Z,$ $v\in\lbrack0,1)$.
Hence the space $\mathbb{R}^{d}$ is the union of the pairwise disjoint sets
$P(\beta,j)\equiv\{\beta+\tau+(j+v)\delta:\tau\in F_{\delta},v\in
\lbrack0,1)\}$ for $\beta\in\Gamma_{\delta},$ $j\in Z.$ To prove that
$B_{\delta\text{ }}$has an asymptotically full measure on $V_{\delta}%
(\rho^{\alpha_{1}}),$ that is,
\begin{equation}
\lim_{\rho\rightarrow\infty}\frac{\mu(B_{\delta})}{\mu(V_{\delta}(\rho
^{\alpha_{1}}))}=1
\end{equation}
we define the following sets: $R_{1}(\rho)=\{j\in Z:\mid j\mid<\frac
{\rho^{\alpha_{1}}}{2\mid\delta\mid^{2}}+\frac{3}{2}\},$

$S_{1}(\rho)=\{j\in Z:\mid j\mid<\frac{\rho^{\alpha_{1}}}{2\mid\delta\mid^{2}%
}-\frac{3}{2}\},$

$R_{2}(\rho)=\{\beta\in\Gamma_{\delta}:\beta\in R_{\delta}(\frac{3}{2}%
\rho+d_{\delta}+1)\backslash R_{\delta}(\frac{1}{2}\rho-d_{\delta}-1))\},$

$S_{2}(\rho)=\{\beta\in\Gamma_{\delta}:\beta\in(R_{\delta}(\frac{3}{2}%
\rho-d_{\delta}-1)\backslash R_{\delta}(\frac{1}{2}\rho+d_{\delta
}+1))\backslash(\bigcup_{b\in\Gamma_{\delta}(\rho^{\alpha_{d}})}V_{b}^{\delta
}(\rho^{\frac{1}{2}}))\},$

where $R_{\delta}(\rho)=\{x\in H_{\delta}:\mid x\mid<\rho\},$ $\Gamma_{\delta
}(\rho^{\alpha_{d}})=\{b\in\Gamma_{\delta}:\mid b\mid<\rho^{\alpha_{d}}\},$

$V_{b}^{\delta}(\rho^{\frac{1}{2}})=\{x\in H_{\delta}:\mid\mid x+b\mid
^{2}-\mid x\mid^{2}\mid<\rho^{\frac{1}{2}}\},$ and

$d_{\delta}=\sup_{x,y\in F_{\delta}}\mid x-y\mid$ is diameter of $F_{\delta}.$

Moreover we define a subset $P^{^{\prime}}(\beta,j)$ of $P(\beta,j)$ as
follows. Introduce the sets $A(\beta,b,\rho)=\{v\in\lbrack0,1):\exists j\in
Z,\mid2(\beta,b)+\mid b\mid^{2}+\mid(j+v)\delta\mid^{2}\mid<4d_{\delta}%
\rho^{\alpha_{d}}\},$

$A(\beta,\rho)=\bigcup_{b\in\Gamma_{\delta}(\rho^{\alpha_{d}})}A(\beta
,b,\rho),$ $S_{3}(\beta,\rho)=W(\rho)\backslash A(\beta,\rho)$ and put

$S_{4}(\beta,j,v,\rho)=\{\tau\in F_{\delta}:\beta+\tau+(j+v)\delta\in
B_{\delta}\}$ for $j\in S_{1},$ $\beta\in S_{2},$ $v\in S_{3}(\beta,j,\rho).$
Then define $P^{^{\prime}}(\beta,j)$ by

$P^{^{\prime}}(\beta,j)=\{\beta+\tau+(j+v)\delta:v\in S_{3}(\beta,\rho
),\tau\in S_{4}(\beta,j,v,\rho)\}.$

It is not hard to see that (193) follows from the following relations:%
\begin{align}
\lim_{\rho\rightarrow\infty}\frac{\mid S_{i}(\rho)\mid}{\mid R_{i}(\rho)\mid}
&  =1,\text{ }\forall i=1,2,\\
B_{\delta}  &  \supset\cup_{j\in S_{1},\beta\in S_{2}}P^{^{\prime}}%
(\beta,j),\\
V_{\delta}(\rho^{\alpha_{1}})  &  \subset\cup_{j\in R_{1},\beta\in R_{2}%
}P(\beta,j),\\
\lim_{\beta\rightarrow\infty}\frac{\mu(P^{^{\prime}}(\beta,j))}{\mu
(P(\beta,j))}  &  =1.
\end{align}
To prove these relations we use the following lemma.

\begin{lemma}
Let $w\equiv\beta+\tau+(j+v)\delta.$ Then the following implications:

(a) \ $w\in V_{\delta}(\rho^{\alpha_{1}})\Rightarrow j\in R_{1},\beta\in
R_{2},$

(b) \ \ $j\in S_{1},\beta\in S_{2}\Rightarrow w\in V_{\delta}(\rho^{\alpha
_{1}})\cap(R(\frac{3}{2}\rho-\rho^{\alpha_{1}-1})\backslash R(\frac{1}{2}%
\rho+\rho^{\alpha_{1}-1})),$ \ 

(c) $\ j\in S_{1},\beta\in S_{2}\Rightarrow w\in V_{\delta}^{^{\prime}}%
(\rho^{\alpha_{1}})\cap(R(\frac{3}{2}\rho-\rho^{\alpha_{1}-1})\backslash
R(\frac{1}{2}\rho+\rho^{\alpha_{1}-1}))$ hold.

The relations (195), (196) and the equality (194) are true.
\end{lemma}

\begin{proof}
Since $(\beta+\tau,\delta)=0$ the inclusion $\omega\in V_{\delta}(\rho
^{\alpha_{1}})$ means that

$\mid\mid(j+v+1)\delta\mid^{2}-\mid(j+v)\delta\mid^{2}\mid<\rho^{\alpha_{1}}$
$,$ where$\mid v\mid<1,$ and

$(\frac{1}{2}\rho)^{2}<\mid\beta+\tau\mid^{2}+\mid(j+v)\delta\mid^{2}%
<(\frac{3}{2}\rho)^{2},$ where $\mid\tau\mid\leq d_{\delta}=O(1)$ ( see the
definition of \ $V_{\delta}(\rho^{\alpha_{1}})$ in introduction). Therefore by
direct calculation we get the proof of the implications $(a)$ and $(b).$

Now we prove $(c).$ It follows from $(b)$ and the definition of $V_{\delta
}^{^{\prime}}(\rho^{\alpha_{1}})$ ( see introduction) that it is enough to
show the relation $w\notin V_{a}(\rho^{\alpha_{1}})$ for $a\in\Gamma
(p\rho^{\alpha})\backslash\delta\mathbb{R}$ holds. Using $a=a_{1}+a_{2}\delta$
( see (48)), where $a_{1}\in\Gamma_{\delta},$ $a_{2}\in\mathbb{R}$ and $\mid
a_{1}\mid<p\rho^{\alpha},$ $\mid a_{2}\delta\mid<p\rho^{\alpha},$ we obtain
$\mid w+a\mid^{2}-\mid w\mid^{2}=d_{1}+d_{2},$ where $d_{1}=\mid\beta
+a_{1}\mid^{2}-\mid\beta\mid^{2},$ $d_{2}=\mid(j+a_{2}+v)\delta\mid^{2}%
-\mid(j+v)\delta\mid^{2}+2(a_{1},\tau).$ The requirements on $j,a_{1},a_{2}$
imply that $d_{2}=O(\rho^{2\alpha_{1}}).$ On the other hand the condition
$\beta\in S_{2}$ gives $\beta\notin V_{a}^{\delta}(\rho^{\frac{1}{2}}),$ i.e.,
$\mid d_{1}\mid\geq\rho^{\frac{1}{2}}$. Since $2\alpha_{1}<\frac{1}{2}$ we
have $\mid\mid w+a\mid^{2}-\mid w\mid^{2}\mid>\frac{1}{2}\rho^{\frac{1}{2}},$
$w\notin V_{a}(\rho^{\alpha_{1}}).$ So $(c)$ is proved.

The inclusion (196) follows from the implication $(a).$ If

$w\equiv\beta+\tau+(j+v)\delta$ belongs to the right-hand side of (195) then
using the implication $(c)$ we obtain $w\in V_{\delta}^{^{\prime}}%
(\rho^{\alpha_{1}}).$ Therefore (195) follows from the definitions of
$P^{^{\prime}}(\beta,j)$ and $S_{4}(\beta,j,v,\rho)$. It remains to prove the
equality (194). Using the definitions of $R_{1},S_{1}$ and inequalities
$\mid\delta\mid<\rho^{\alpha},\alpha_{1}>2\alpha$ we obtain that (194) for
$i=1$ holds. If $\beta\in R_{2}$ then $\beta+F_{\delta}\subset R_{\delta
}(\frac{3}{2}\rho+2d_{\delta}+1)\backslash R_{\delta}(\frac{1}{2}%
\rho-2d_{\delta}-1).$ This implies that,

$\mid R_{2}\mid<(\mu(F_{\delta}))^{-1}\mu(R_{\delta}(\frac{3}{2}%
\rho+2d_{\delta}+1)\backslash R_{\delta}(\frac{1}{2}\rho-2d_{\delta}-1)),$
since the translations $\beta+F_{\delta}$ of $F_{\delta}$ for $\beta\in
\Gamma_{\delta},$ are pairwise disjoint sets having measure $\mu(F_{\delta}).$
Suppose $\beta+\tau\in D(\rho),$ where $D(\rho)=(R_{\delta}(\frac{3}{2}%
\rho-1)\backslash R_{\delta}(\frac{1}{2}\rho+1))\backslash(\bigcup_{b\in
\Gamma_{\delta}(\rho^{\alpha_{d}})}V_{b}^{\delta}(2\rho^{\frac{1}{2}})).$ Then
$\frac{3}{2}\rho-1<\mid\beta+\tau\mid<\frac{1}{2}\rho+1,$ $\mid\mid\beta
+\tau+b\mid^{2}-\mid\beta+\tau\mid^{2}\mid\geq2\rho^{\frac{1}{2}}$ for
$b\in\Gamma_{\delta}(\rho^{\alpha_{d}}).$ Therefore using $\mid\tau\mid\leq
d_{\delta}$ it is not hard to verify that $\beta\in S_{2}.$ Hence the sets
$\beta+F_{\delta}$ for $\beta\in S_{2}$ is cover of $D(\rho).$ Thus $\mid
S_{2}\mid\geq(\mu(F_{\delta}))^{-1}\mu(D(\rho).$ This, the estimation for
$\mid R_{2}\mid,$ and the obvious relations $\mid\Gamma_{\delta}(\rho
^{\alpha_{d}})\mid=O(\rho^{(d-1)\alpha_{d}})),$

$\mu((R_{\delta}(\frac{3}{2}\rho-1)\backslash R_{\delta}(\frac{1}{2}%
\rho+1)))=O(\rho^{d-1}),$

\ $\mu((R_{\delta}(\frac{3}{2}\rho-1)\backslash R_{\delta}(\frac{1}{2}%
\rho+1))\cap V_{b}^{\delta}(2\rho^{\frac{1}{2}}))=O(\rho^{d-2}\rho^{\frac
{1}{2}}),$ $(d-1)\alpha_{d}<\frac{1}{2}$ (see the equality in (166)),%
\[
\lim_{\rho\rightarrow\infty}\frac{\mu((R_{\delta}(\frac{3}{2}\rho-1)\backslash
R_{\delta}(\frac{1}{2}\rho+1)))}{\mu(R(\frac{3}{2}\rho+2d_{\delta
}+1)\backslash R_{\delta}(\frac{1}{2}\rho-2d_{\delta}-1))}=1,
\]
$S_{2}(\rho)\subset R_{2}(\rho)$ imply (194) for $i=2$
\end{proof}

\begin{theorem}
The simple set $B_{\delta}$ has an asymptotically full measure in the
resonance set $V_{\delta}(\rho^{\alpha_{1}})$ in the sense that (193) holds.
\end{theorem}

\begin{proof}
The proof of the Theorem follows from (194)-(197). By Lemma 10 we need to
prove (197). Since the translations $P(\beta,j)-\beta-j\delta$ and
$P^{^{\prime}}(\beta,j)-\beta-j\delta$ of $P(\beta,j)$ and $P^{^{\prime}%
}(\beta,j)$ are $\{\tau+v\delta:v\in\lbrack0,1),\tau\in F_{\delta}\}$ and

$\{\tau+v\delta:v\in S_{3}(\beta,\rho),$ $\tau\in S_{4}(\beta,j,v,\rho)\}$
respectively, it is enough to prove
\begin{equation}
\lim_{\rho\rightarrow\infty}\mu(S_{3}(\beta,\rho))=1,\text{ }\mu(S_{4}%
(\beta,j,v,\rho))=\mu(F_{\delta})(1+O(\rho^{-\alpha})),
\end{equation}
where $j\in S_{1},\beta\in S_{2},v\in S_{3}(\beta,\rho),$ and $O(\rho
^{-\alpha})$ does not depend on $v.$ To prove the first equality in (198) it
is enough to show that
\begin{equation}
\mu(A(\beta,\rho))=O(\rho^{-\alpha}),
\end{equation}
since $W(\rho)\supset A(\varepsilon(\rho))$ and $\mu(A(\varepsilon
(\rho))\rightarrow1$ as $\rho\rightarrow\infty$ (see Lemma 3(b)). Using the
definition of $\ A(\beta,\rho)$ and the obvious relation $\mid\Gamma_{\delta
}(\rho^{\alpha_{d}})\mid=O(\rho^{(d-1)\alpha_{d}})$ we see that (199) holds if
$\mu(A(\beta,b,\rho))=O(\rho^{-d\alpha_{d}}).$ In other word we need to prove
that
\begin{equation}
\mu\{s\in\mathbb{R}:\mid f(s)\mid<4d_{\delta}\rho^{\alpha_{d}}\}=O(\rho
^{-d\alpha_{d}}),
\end{equation}
where $f(s)=2(\beta,b)+\mid b\mid^{2}+s^{2}\mid\delta\mid^{2},$ $\beta\in
S_{2},$ $b\in\Gamma_{\delta}(\rho^{\alpha_{d}}).$ The last inclusions yield
$\mid2(\beta,b)+\mid b\mid^{2}\mid\geq\rho^{\frac{1}{2}}$ for $\mid b\mid
<\rho^{\alpha_{d}}$. This and the inequalities

$\mid f(s)\mid<4d_{\delta}\rho^{\alpha_{d}}$ ( see (200)), $\alpha_{d}%
<\frac{1}{2}$ ( see the equality in (166)) imply that $s^{2}\mid\delta\mid
^{2}>\frac{1}{2}\rho^{\frac{1}{2}}$ from which we obtain $\mid f^{^{\prime}%
}(s)\mid>\mid\delta\mid\rho^{\frac{1}{4}}.$ Therefore (200) follows from the
equality in (166)). Thus (199) and hence the first equality in (198) is proved.

Now we prove the second equality in (198). For this we consider the set
$S_{4}(\beta,j,v,\rho)$ for $j\in S_{1},$ $\beta\in S_{2}$, $v\in S_{3}%
(\beta,\rho).$ By the definitions of $S_{4}$ and $B_{\delta}$ the set
$S_{4}(\beta,j,v,\rho)$ is the set of $\tau\in F_{\delta}$ such that
$E(\lambda_{j,\beta}(v,\tau))$ satisfies the conditions (168), (169). So we
need to consider these conditions. For this we use the decompositions
$\gamma+t=\beta+\tau+(j+v)\delta,$ $\gamma^{^{\prime}}+t=\beta^{^{\prime}%
}+\tau+(j^{^{\prime}}+v(\beta^{^{\prime}},t))\delta,$ (see Remark 3) and the notations

$\lambda_{j,\beta}(v,\tau)=\mu_{j}(v)+\mid\beta+\tau\mid^{2},\lambda
_{i}(\gamma^{^{\prime}}+t)=\mid\gamma^{^{\prime}}+t\mid^{2}+r_{i}%
(\gamma^{^{\prime}}+t).$ ( see Lemma 2(b) and Remark 2). Denoting by $b$ the
vector $\beta^{^{\prime}}-\beta$ we write the decomposition of $\gamma
^{^{\prime}}+t$ in the form $\gamma^{^{\prime}}+t=\beta+b+\tau+(j^{^{\prime}%
}+v(\beta+b,t))\delta.$ Then to every $\gamma^{^{\prime}}\in\Gamma$ there
corresponds $b=b(\gamma^{^{\prime}})\in\Gamma_{\delta}$. For $\gamma
^{^{\prime}}\in M_{1}$ denote by $B^{1}(\beta,b(\gamma^{^{\prime}}),j,v)$ the
set of $\tau$ not satisfying (168). For $\gamma^{^{\prime}}\in M_{2}$ denote
by $B^{2}(\beta,b(\gamma^{^{\prime}}),j,v)$ the set of $\tau$ not satisfying
(169), where $M_{i}$ for $i=1,2$ is defined in (168), (169). Clearly, if
$\tau\in F_{\delta}\backslash(\cup_{s=1,2}(\cup_{\gamma^{^{\prime}}\in M_{s}%
}(B^{s}(\beta,b(\gamma^{^{\prime}}),j,v))$ then the inequalities (168), (169)
hold, that is, $\tau\in S_{4}(\beta,j,v,\rho)$. Therefore using $\mu
(F_{\delta})\sim1$ and proving that
\begin{equation}
\mu(\cup_{\gamma^{^{\prime}}\in M_{s}}B^{s}(\beta,b(\gamma^{^{\prime}%
}),j,v))=O(\rho^{-\alpha}),\text{ }\forall s=1,2,
\end{equation}
we get the proof of the second equality in (198). Now we prove (201). Using
the above notations and the notations of (168), (169) it is not hard to verify
that if $\tau\in B^{s}(\beta,b(\gamma^{^{\prime}}),j,v),$ then
\begin{equation}
\mid2(\beta,b)+\mid b\mid^{2}+\mid(j^{^{\prime}}+v(\beta+b))\delta\mid
^{2}+2(b,\tau)-\mu_{j}(v)+h_{s}(\gamma^{^{\prime}}+t)\mid<2\varepsilon_{1},
\end{equation}
where $h_{1}=F_{k_{1}-1}-E_{k_{2}-1},$ $h_{2}=r_{i}-E_{k_{2}-1},$
$\gamma^{^{\prime}}\in M_{s},s=1,2.$ First we prove that (202) for $s=1,2$ and
$b\equiv b(\gamma^{^{\prime}})\in\Gamma_{\delta}(\rho^{\alpha_{d}})$ does not
hold. The assumption $v\in S_{3}(\beta,\rho)$ implies that $v\notin
A(\beta,\rho).$ This means that

$\mid2(\beta,b)+\mid b\mid^{2}+\mid(j^{^{\prime}}+v(\beta+b))\delta\mid
^{2}\mid\geq$ $4d_{\delta}\rho^{\alpha_{d}}.$ Therefore if
\begin{equation}
\mid2(b,\tau)-\mu_{j}(v)+h_{s}(\gamma^{^{\prime}}+t)\mid<3d_{\delta}%
\rho^{\alpha_{d}},
\end{equation}
then (202) does not hold. Now we prove (203). The relations $b\in
\Gamma_{\delta}(\rho^{\alpha_{d}}),\tau\in F_{\delta}$ imply that
$\mid2(b,\tau)\mid<2d_{\delta}\rho^{\alpha_{d}}.$ The inclusion $j\in S_{1}$
and (52) imply that

$\mu_{j}(v)=O(\rho^{2\alpha_{1}}).$ By (24) and (100), $h_{1}=O(\rho
^{\alpha_{1}}).$ Now we prove that $r_{i}=O(\rho^{\alpha_{1}})$ which implies
that $\mid h_{2}\mid=O(\rho^{\alpha_{1}})$ and hence ends the proof of (203).
The inclusion $\tau\in B^{2}(\beta,b(\gamma^{^{\prime}}),j,v),$ means that
(169) does not holds , that is,

$\mid E(\lambda_{j,\beta}(v,\tau))-\lambda_{i}(\gamma^{^{\prime}}%
+t)\mid<2\varepsilon_{1}.$ On the other hand the inclusion$\ \gamma^{^{\prime
}}\in M_{2}$ implies that $\gamma^{^{\prime}}\in M$ ( see the definitions of
$M_{2},$ and $M$) and hence

$\mid E(\lambda_{j,\beta}(v,\tau))-\mid\gamma^{^{\prime}}+t\mid^{2}\mid
\leq\frac{1}{3}\rho^{\alpha_{1}}$ The last two inequalities imply that
$r_{i}(\gamma^{^{\prime}}+t)=O(\rho^{\alpha_{1}}).$ Thus \ (203) is proved.
Hence (202) for $b\in\Gamma_{\delta}(\rho^{\alpha_{d}})$ does not hold. It
means that the sets $B^{1}(\beta,b,j,v)$ and $B^{^{2}}(\beta,b,j,v)$ for $\mid
b\mid<$ $\rho^{\alpha_{d}}$ are empty.

To estimate the measure of the set $B^{s}(\beta,b(\gamma^{^{\prime}}),j,v)$
for $\gamma^{^{\prime}}\in M_{s}$ ,

$\mid b(\gamma^{^{\prime}})\mid\geq\rho^{\alpha_{d}},$ $b\in\Gamma_{\delta}$
we choose the coordinate axis so that the direction of $b$ coincides with the
direction of $(1,0,0,...,0),$ i.e., $b=(b_{1},0,0,...,0),b_{1}>0$ and the
direction of $\delta$ coincides with the direction of $(0,0,...,0,1).$ Then
$H_{\delta\text{ }}$ and $B^{s}(\beta,b,j,v)$ can be considered as
$\mathbb{R}^{d-1}$ and as a subset of $F_{\delta}\subset\mathbb{R}^{d-1}$
respectively. We estimate the measure of $B^{s}(\beta,b,j,v)$ by using (137) for

$D=B^{s}(\beta,b,j,v),m=d-1,k=1.$ For this we prove that
\begin{equation}
\mu((B^{s}(\beta,b,j,v))(\tau_{2},\tau_{3},...,\tau_{d-1}))<4\varepsilon
_{1}\mid b\mid^{-1},
\end{equation}
for all fixed $(\tau_{2},\tau_{3},...,\tau_{d-1}).$ Assume the converse. Then
there are two points $\tau=(\tau_{1},\tau_{2},\tau_{3},...,\tau_{d-1})\in
F_{\delta},$ $\tau^{^{\prime}}=(\tau_{1}^{^{\prime}},\tau_{2},\tau
_{3},...,\tau_{d-1})\in F_{\delta}$ of $B^{s}(\beta,b,j,v),$ such that
$\mid\tau_{1}-\tau_{1}^{^{\prime}}\mid\geq4\varepsilon_{1}\mid b\mid^{-1}$.
Since (202) holds for $\tau^{^{\prime}}$ and $\tau$ we have
\begin{equation}
\mid2b_{1}(\tau_{1}-\tau_{1}^{^{\prime}})+g_{s}(\tau)-g_{s}(\tau^{^{\prime}%
})\mid<4\varepsilon_{1},
\end{equation}
where $g_{s}(\tau)=h_{s}(\beta^{^{\prime}}+\tau+(j^{^{\prime}}+v(\beta
+b))\delta).$ Using (44), (46), (101), and the inequality$\mid b\mid\geq
\rho^{\alpha_{d}}$, we obtain
\begin{align}
&  \mid g_{1}(\tau)-g_{1}(\tau^{^{\prime}})\mid<\rho^{-\alpha_{1}}\mid\tau
_{1}-\tau_{1}^{^{\prime}}\mid<b_{1}\mid\tau_{1}-\tau_{1}^{^{\prime}}\mid,\\
&  \mid g_{2}(\tau)-g_{2}(\tau^{^{\prime}})\mid<3\rho^{\frac{1}{2}\alpha_{d}%
}\mid\tau_{1}-\tau_{1}^{^{\prime}}\mid<b_{1}\mid\tau_{1}-\tau_{1}^{^{\prime}%
}\mid.
\end{align}
The above inequality $b_{1}\mid\tau_{1}-\tau_{1}^{^{\prime}}\mid
\geq4\varepsilon_{1}$ together with (206) and (207) contradicts (204) for
$s=1$ and for $s=2$ respectively. Hence (204) is proved. Since $B^{s}%
(\beta,b,j,v)\subset F_{\delta}$ and $d_{\delta}=O(1)$, we have $\mu(\Pr
B^{s}(\beta,b,j,v))=O(1).$ Therefore formula (137), the inequalities (204) and
$\mid b\mid\geq\rho^{\alpha_{d}}$ yield

$\mu((B^{s}(\beta,b(\gamma^{^{\prime}}),j,v)=O(\varepsilon_{1}\mid
b(\gamma^{^{\prime}})\mid^{-1})=O(\rho^{-\alpha_{d}}\varepsilon_{1})$ for
$\gamma^{^{\prime}}\in M_{s}\subset M$ and $s=1,2.$ This implies (201), since
$\mid M\mid=O(\rho^{d-1}),$ $\varepsilon_{1}=\rho^{-d-2\alpha}$ and
$O(\rho^{d-1-\alpha_{d}}\varepsilon_{1})=O(\rho^{-\alpha})$
\end{proof}

\end{document}